\documentclass{aastex631}

\usepackage{graphicx}	
\usepackage{amsmath}

\newcommand{\hi}{H{\sc i}}

\shorttitle{A Pilot Study of Magnetic Fields, Star Formation Rates and Gas densities}
\shortauthors{Manna and Roy}

\begin{document}

\title{Magnetic Fields, Star Formation Rates and Gas Densities at Sub-kpc Scales in a Pilot Sample of Nearby Galaxies}

\correspondingauthor{Souvik Manna}
\email{souvik@ncra.tifr.res.in}

\author{Souvik Manna}
\affiliation{National Center for Radio Astrophysics, TIFR, \\ Pune University Campus, Ganeshkhind, Pune 411007, India}

\author{Subhashis Roy}
\affiliation{National Center for Radio Astrophysics, TIFR, \\ Pune University Campus, Ganeshkhind, Pune 411007, India}



\begin{abstract}

 We have estimated the magnetic field strengths of a sample of seven galaxies using their non-thermal synchrotron radio emission at metre wavelengths, and assuming energy equipartition between magnetic fields and cosmic ray particles.
We tested for deviation of magnetic fields from energy equipartition with cosmic ray particles, and found that deviations of $\sim$25\% are typical for the sample galaxies. 
Spatially resolved star formation rates (SFR) were estimated for the seven galaxies along with five galaxies studied previously.
For the combined sample of twelve galaxies, the equipartition magnetic fields (B$_\textrm{eq}$) are correlated with the SFR surface densities ($\Sigma_\textrm{SFR}$) at sub-kpc scales with B$_\textrm{eq}$ $\propto$ $\Sigma_\textrm{SFR}^ {0.31\pm0.06}$, consistent with model predictions.
We estimated gas densities ($\rho_\textrm{gas}$) for a sub-sample of seven galaxies using archival observations of the carbon monoxide (CO) rotational transitions and the atomic hydrogen (\hi) 21 cm line and studied the spatially-resolved correlation between the magnetic fields and $\rho_\textrm{gas}$. 
Magnetic fields and gas densities are found to be correlated at sub-kpc scale as B$_\textrm{eq}$ $\propto$ $\rho_\textrm{gas}^{0.40\pm0.09}$. This is broadly consistent with models, which typically predict B $\propto$ $\rho_\textrm{gas}^{0.5}$.
  
\end{abstract}


\section{Introduction}

\label{introduction}

Magnetic fields are believed to influence several physical processes in a galaxy at almost every scale 
\citep[e.g.][]{Elmegreen1981ApJ,Niklas1997A&A,Groves2003PASA,Price2008MNRAS,Adebahr2013A&A}.
Magnetic fields have been found to consist of two main components: a small-scale turbulent magnetic field up to a few hundred parsecs \citep[e.g.][]{Batchelor1950RSPSA,Groves2003PASA} and a large-scale $``$ordered$"$ or  $``$regular$"$ magnetic field component at scales of a few kpcs \citep[e.g.][]{Moss1996MNRAS,Shukurov2006A&A,Kulsrud2008RPPh}.
Magnetic fields in galaxies can be measured using their effects on different radiation processes like Zeeman splitting of emission lines, polarized emission from dust, the polarization of starlight, Faraday rotation of polarized radio emission, and intensity of synchrotron emission which we use in this work.
Measurement of the line-of-sight component of the magnetic field via the Zeeman effect in galaxies other than the Milky Way has been possible for only a few systems \citep{Kazes1991A&A,Sarma2005AJ,Robishaw2008ApJ}; a significant expansion of such studies is very difficult with current-generation telescopes.

 Magnetic fields in galaxies can be measured and studied using synchrotron emission at radio frequencies, at scales larger than the resolution of the radio observation.
For example, a Very Large Array (VLA) polarization study of NGC 4736 at 8.46 and 4.86 GHz found that the magnetic field in the galaxy was ordered in a spiral shape \citep{Chyzy2008ApJ}.
 An X-shaped structure of the magnetic field in the galactic halo region was observed by stacking the Karl G. Jansky VLA polarized emission maps of 16 nearly edge-on spiral galaxies, obtained as part of the CHANG-ES survey \citep{Krause2020A&A}; such structures had also been observed in  individual spiral galaxies \citep[e.g.][]{Krause2006A&A,Krause2009RM,Heesen2009A&A}.
However, polarized radio emission from external individual galaxies is difficult to study at low radio frequencies due to Faraday depolarization \citep[e.g.][]{Sokoloff1998}.

 The average magnetic field strength can also be estimated from the total intensity of synchrotron radio emission, assuming energy equipartition between magnetic fields and cosmic ray particles \cite[e.g.][]{Miley1980,BeckandKrause2005}.
Equipartition magnetic fields have been studied in several nearby galaxies, but primarily at frequencies $>$1 GHz \citep[e.g.][]{Chyzy2000A&A,Soida2001A&A,Heesen2009A&A,Fletcher2011MNRAS,Adebahr2013A&A}. 
 \cite{Vargas2018ApJ} studied a sample of three nearly edge-on galaxies from the CHANG-ES survey to separate the thermal Bremsstrahlung from the non-thermal synchrotron emission at 1.5 and 6 GHz.
At these frequencies, the thermal component is large and hence the correction for the thermal emission can be as large as $\sim$ 20\%, making the derived magnetic field strengths prone to errors. Conversely, the steep spectral index of synchrotron emission implies that it will dominate the total emission at frequencies $<1$ GHz, with $\sim$ 95\% contribution \citep{Basu2012ApJ,RoyManna2021}. Thus, magnetic field strengths derived using observations at $<$1 GHz are very robust to any correction for thermal emission. 

Magnetic fields are believed to play an important role at various stages of the star-formation process - from the fragmentation of clouds at the few kpc scales to the final collapse of gas into stars \citep[e.g.][]{Elmegreen1981ApJ,Crutcher1999alma,Price2008MNRAS,Van_Loo2015ApJ}.
 To understand the influence of magnetic fields and star-formation activities on different physical processes in the ISM at different physical scales, several studies on radio-infrared correlations have been carried out in the past \citep[e.g.][]{Murphy2006ApJ,Murphy2006ApJb,Murphy2008ApJ,Tabatabaei2013A&A}.
Magnetic fields (B) and star formation rate surface densities (SFRSD) are expected to be correlated \citep{Niklas1997A&A}.
Semi-analytical model also predicts a strong correlation between B and SFRSDs ($\Sigma_\textrm{SFR}$) as B $\propto \Sigma_\textrm{SFR}^{1/3}$ at sub-kpc scales to explain the local radio-FIR correlation \citep{Schleicher2013A&A,Schleicher2016A&A}.
Observational studies of the correlation between B and star formation rates (SFR) have been done primarily in samples of nearby dwarf galaxies. For example, \cite{Chyzy2011A&A} studied 12 local group dwarf galaxies to find that the galaxy-averaged magnetic field and the SFR follow B $\sim$ SFR$^{0.30\pm0.04}$, consistent with the prediction of B $\propto \Sigma_\textrm{SFR}^{1/3}$. However, \cite{Jurusik2014A&A} found the same power-law index in a sample of Magellanic type dwarf galaxies to be 0.25$\pm$0.02, somewhat lower than the expectation. Recently, a study of the dwarf galaxy IC 10 by \cite{Basu2017MNRAS} provides the only study of the correlation between spatially-resolved magnetic fields and SFRSDs; these authors found that the SFRSD is related to the magnetic field as B $\propto \Sigma_\textrm{SFR}^{0.35 \pm 0.03}$.
Therefore, it is important to test such predictions by carrying out systematic spatially-resolved studies of magnetic fields in galaxies and their connection to the star formation rate in nearby large galaxies.


The energy density of magnetic fields and gas in galaxies are expected to be in equipartition, which implies B $\propto \sqrt{\rho_\textrm{gas}}$ \citep[e.g.][]{Chandrasekhar1953,Groves2003PASA}. The observed Radio-FIR correlation can be explained based on such equipartition between the energy density of magnetic fields and gas \citep{Niklas1997A&A}. Several other numerical magnetohydrodynamic (MHD) simulations of the ISM have predicted the coupling constant ($k$) between magnetic fields and gas (B $\propto \rho_\textrm{gas}^{k}$) to be in the range of $\approx$0.4$-$0.6 \citep{Fiedler1993ApJ,Kim2001JKAS,Thompson2006ApJ}. \cite{Niklas1997A&A} studied the correlation between galaxy-integrated equipartition magnetic fields and gas densities for a sample of 43 galaxies to find a power-law index of 0.48 $\pm$ 0.05; the observed correlation is consistent with B $\propto \sqrt{\rho_\textrm{gas}}$.
 Although the correlation between gas surface densities and SRFSDs has been extensively studied in the nearby Universe
\citep[Kennicutt-Schmidt law; e.g.][]{Kennicutt1998ApJ,Onodera2010ApJ,Roychowdhury2015MNRAS}, systematic studies of spatially-resolved correlations between magnetic fields, SFRs and gas densities in nearby galaxies are yet to be carried out.
It is thus important to carry out a systematic investigation of both the B-$\rho$ and the B-SFR correlations, at high-spatial resolutions ($\approx$ sub-kpc scales), using direct estimates of the magnetic fields, gas densities, and star-formation rates, in a sample of nearby galaxies.
 In this paper, we present a pilot study of the connection between spatially resolved magnetic fields, SFRSDs and gas densities in a sample of nearby galaxies.

We have selected a sample of 46 galaxies (Sample 0; Table \ref{different_samples_studied}) from the Spitzer Local Volume Legacy (LVL) sample of 258 galaxies within 11 Mpc \citep{Dale2009ApJ}. As a pilot project, seven (Sample 1; Table \ref{different_samples_studied}) of these 46 galaxies have been observed with the Giant Metrewave Radio Telescope (GMRT) at 0.33 GHz \citep{RoyManna2021}. Six of our seven sample galaxies are spirals and the other one is a dwarf irregular  Magellanic-type galaxy.

In this paper, we present spatially resolved equipartition magnetic field (B$_\textrm{eq}$) maps of the seven galaxies in Sample 1 (Table \ref{different_samples_studied}). We also incorporate the magnetic field maps of five galaxies studied by \citet{Basu2012MNRAS} from previous GMRT observations in our study.
We derived SFRSD maps of all 12 galaxies (Sample 2; Table \ref{different_samples_studied}) using extinction-free diagnostics and used these maps to study the relation between SFRSDs and B$_\textrm{eq}$ at sub-kpc scales in our pilot study.
We used available archival CO and \hi\ 21 cm data to measure the gas densities ($\rho_\textrm{gas}$) of seven (Sample 3; Table \ref{different_samples_studied}) of the combined sample of 12 galaxies and studied the correlation between $\rho_\textrm{gas}$ and B$_\textrm{eq}$ in these galaxies. 
We also studied the magnetic field-gas connection through an indirect measurement of their coupling coefficient using radio$-$FIR correlations of the galaxies in Sample 1.

The paper is organized as follows. The analysis of the data is discussed in Section \ref{data_analysis}. In Section \ref{results}, we present the results of our analysis, including the correlation between magnetic fields, SFRSDs and gas densities of the seven galaxies in Sample 1. In Section \ref{study_basu_2013}, we have extended our study to include a sample of five galaxies of \cite{Basu2012MNRAS} . We discuss the results in Section \ref{discussion}. A summary of this paper is presented in Section \ref{summary}.

\section{Data Analysis}
\label{data_analysis}
As can be seen in Table \ref{sample_details}, six of the seven galaxies in Sample 1 are spirals of varying inclination angles. The seventh galaxy NGC 4449 is a
dwarf irregular galaxy.
Basic information about the seven sample galaxies, including their types, distances, inclination angles, position angles, angular resolutions, spatial resolutions, and RMS noise obtained on the GMRT and VLA images are also listed in Table \ref{sample_details}. The distances, inclination angles, and position angles of the galaxies were taken from \cite{Dale2009ApJ}.
Radio observations and the data reduction procedures are discussed in detail in \cite{RoyManna2021}. Briefly, we used GMRT 0.33~GHz observations (covering 0.309$-$0.342~GHz) and archival VLA
observations at 1.4 and $\sim$6~GHz to derive non-thermal emission maps for each galaxy.
 We used H$\alpha$ and 24$\mu$m observations of the seven galaxies to model free-free emission from them and subsequently, we subtracted the modelled free-free emission from the observed radio emission to get the non-thermal radio maps at 0.33, 1.4 and $\sim$6 GHz \citep{RoyManna2021}.
To generate the non-thermal spectral index maps, we used the non-thermal radio maps at 0.33 and $\sim$6 GHz for NGC 2683, NGC 3627, NGC 4096, and NGC 4449. For the rest of the galaxies (NGC 4490, NGC 4826, and NGC 5194), we used non-thermal images at 0.33 and 1.4 GHz to generate the non-thermal spectral index maps \citep{RoyManna2021}. 
In the following subsections, we present the analysis of other ancillary data and relevant measurements.

\begin{table}
 \caption{Details of the seven sample galaxies. Note that the images at 0.33~GHz were obtained from observations with the GMRT reported in \citet{RoyManna2021} while those at 1.4~GHz were obtained from archival VLA data.
 The distances to the galaxies were taken from \citet{Dale2009ApJ}. Galaxies with an asterisk are those for which spatially-resolved CO data are available.}
\scriptsize
\centering
 \begin{tabular}{||c c c c c c c c c c c||} 
 \hline

Name & Class & Distance & Inclination  & Position  & uv  & Angular  & Spatial  & RMS  & RMS & VLA \\ 
     &          &   (Mpc)  &  angle             &  angle  & range  & resolution  & resolution  & (0.33 GHz)   & (1.4 GHz) & Project ID \\
     
 & & &  (deg)& (deg) & (k$\lambda$) &  $(\textrm{arcsec}^{2})$ & (pc) & ($\mu$Jy/beam) & ($\mu$Jy/beam) & (1.4 GHz) \\
 
 \hline
 
NGC 2683 & Sb        &  7.7 & 83 & 43  & 0.19 - 15  & 19 $\times$ 13 & 670  & 200  &   40    &  AI23 \\
NGC 3627$^{*}$ & SAB       &  10. & 65 & 170 & 0.26 - 25  & 16 $\times$ 11 & 760  & 800  &   370   &  AS541, AP462\\ 
NGC 4096 & SABc      &  8.3 & 76 & 20  & 0.14 - 17  & 14 $\times$ 12 & 730  & 100  &   25    &  16A-013 \\
NGC 4449 & Irregular &  4.2 & 0  & 0   & 0.15 - 15  & 26 $\times$ 15 & 360  & 300  &   180   &  AB167 \\
NGC 4490 & SBm       &  8.0 & 60 & 126 & 0.13 - 14  & 19 $\times$ 18 & 560  & 230  &   100   &  AA181 \\
NGC 4826$^{*}$ & SAab      &  7.5 & 60 & 120 & 0.22 - 20  & 15 $\times$ 14 & 650  & 280  &   70    &  AS541 \\
NGC 5194$^{*}$ &  Sbc      &  8.0 & 20 & 10  & 0.15 - 10  & 23 $\times$ 18 & 740  & 310  &   30    &  AB505, AN57 \\
 
 \hline
 \end{tabular}


\label{sample_details}

\end{table}

 \begin{table}
\centering
\caption{ List of different samples studied in this paper.}
 \begin{tabular}{||c c||} 
 \hline

 Sample Name & Galaxies  \\ 
 \hline\hline
Sample 0 & Full sample containing 46 galaxies from Spitzer LVL survey\\
 \hline
Sample 1 & Pilot sample containing 7 galaxies from Sample 0; galaxies listed in Table \ref{sample_details} \\ 
  \hline
 Sample 2 & Sample 1 + 5 galaxies (NGC 1097, NGC 4736, NGC 5055, NGC 5236 and NGC 6946) \\ & from \cite{Basu2012ApJ} = 12 galaxies; used to probe the B$_\textrm{eq}$-SFRSD correlations \\ 
 \hline
 Sample 3 & A subset of 7 galaxies (NGC 3627, NGC 4826, NGC 5194, NGC 4736, NGC 5055, NGC 5236 and \\ & NGC 6946) from Sample 2 which have archival CO data; used to study the B$_\textrm{eq}$-gas density correlations \\
 
 \hline
 \end{tabular}
\label{different_samples_studied}
\end{table}

\subsection{Magnetic Field
Strengths}
\label{data_analysis_magnetic_field}


The average magnetic field strengths can be estimated from the observed synchrotron flux densities, assuming energy equipartition between cosmic ray particles and magnetic fields                                  \citep[``Classical Equipartition Formula", e.g.][]{Pacholczyk1970,Miley1980,Longair2011}. The equipartition condition is achieved when the total energy in magnetic fields and cosmic ray particles is minimum.
 



The classical equipartition formalism has shortcomings that lead to an overestimation of the magnetic field strength (B) at regions of steep spectral indices and underestimation of B at flat spectral index regions. To overcome these shortcomings of the classical equipartition formula, \cite{BeckandKrause2005} proposed a revised formula to estimate the average magnetic field strength. The formula is expressed as

\begin{equation}
B_\textrm{eq}= [4 \pi (K_\textrm{0} +1) E_\textrm{p}^{1-2 \alpha_{nt}} \frac{f (\alpha_{nt})}{C_{4} (i)} \frac{I_{\nu} \nu^{\alpha_{nt}}}{l}]^{\frac{1}{\alpha_{nt} + 3}}
\label{revised_formula}
\end{equation}

$K_\textrm{0}$, $E_\textrm{p}$, $I_{\nu}$, and $\alpha_{nt}$ are the number density ratio of cosmic ray protons to electrons, the proton rest mass energy, the intensity of the synchrotron emission at frequency $\nu$, and the spectral index of synchrotron emission, respectively. 
$f (\alpha_{nt})$ is a function of $\alpha_{nt}$ given as
$f (\alpha_{nt}) = (2 \alpha_{nt} +1) [2 (\alpha_{nt} -1) c_{2} (\alpha_{nt})
 c_{1}^{\alpha_{nt}}]$
\citep{BeckandKrause2005}. 
 $C_{4} (i)$ is a constant that depends on the inclination angle ($i$) of the galaxy and is expressed as $C_{4} (i)=[cos(i)]^{(\gamma +1)/2}$, where $\gamma = (2 \alpha_{nt} +1)$. $l$ is the path length of the synchrotron emission. The path length was assumed to be 1 kpc for a galaxy with an inclination angle of 0 degree (face-on). For galaxies with low- and moderate- inclination angles ($< 75^\circ$), the assumed path length was corrected for the inclinations of the galaxies as $l$/cos($i$). For the two nearly edge-on galaxies in Sample 1, NGC 2683 and NGC 4096, we have assumed an oblate spheroidal shape of the synchrotron emission, such that the diameter on the plane of the galaxy is equal to its major axis. The path lengths ($l$) were then appropriately calculated, with  the path length being maximum (equal to the galaxy's major axis) at the optical centre of the galaxy and gradually declining to the edge of the galaxy. 
We note that B$_\textrm{eq}$ has only a weak dependence on $l$ as $B_\textrm{eq}(r)= l(r)^\frac{-1}{\alpha_{nt}+3}$ and hence is less sensitive to the exact choice of $l$.
 Values of $K_\textrm{0}$ and $E_\textrm{p}$ were assumed to be 100 and 938.28 MeV, respectively, the same as used by \cite{BeckandKrause2005}.
 Finally, we used non-thermal radio maps at 0.33 GHz ($I_{\nu}$) and spectral index maps ($\alpha_{nt}$) made using 0.33 and 1.4 or $\sim$6 GHz radio observations \citep
 {RoyManna2021} to produce magnetic field maps of the sample galaxies using Equation \ref{revised_formula}.

The revised equipartition formula diverges for spectral index values $\le$ 0.5 because such flat spectra indicate energy loss of electrons through ionizations or Coulomb interactions \citep{Sarazin1999ApJ}.
The central bulge and arm regions have a mostly flatter spectrum due to the association of star-forming regions and the estimates of equipartition magnetic fields in such regions might be affected by systematic uncertainties. This issue affects the derived magnetic field strengths for 8\%, 12\%, 3\%, 70\%, 17\%, 7\%, and 6\% of the projected total surface area of NGC 2683, NGC 3627, NGC 4096, NGC 4449, NGC 4490, NGC 4826, and NGC 5194, respectively. We note that a large fraction of the derived magnetic field values are affected for NGC 4449 due to its non-thermal spectral indices being predominantly flat. This could bias the B$_\textrm{eq}$ for NGC 4449.

 \subsubsection{Uncertainties on Magnetic Field Maps}
 \label{Error_Magnetic_Field}
 The procedure we used to estimate the uncertainties on our magnetic field maps is similar to that of \citet{Basu2013MNRAS}.
We used a Monte Carlo method that generated $10^{4}$ random flux density values for each pixel in a galaxy map at 0.33~GHz and either 1.4 GHz or 6 GHz. These flux density values have Gaussian probability distributions with rms values equal to the measured rms of each of the 0.33 and 1.4/6 GHz maps.
For each of the $10^{4}$ intensity maps, we computed a magnetic field map using the procedure described in the beginning of Section~\ref{data_analysis_magnetic_field}. 
The rms of these $10^4$ magnetic field maps provided us with the magnetic field uncertainty maps for each of the seven galaxies in sample 1.


\subsection{Star Formation Rates}
\label{data_analysis_SFR}
Rest frame H$\alpha$ and ultraviolet (UV) observations are the best tracers of recent SFRs as the radiation from these predominantly
 originate in newly formed massive stars. However, the observations are affected by extinction caused by interstellar dust in both the host galaxy as well as the Milky Way. SFRs estimated from H$\alpha$ and UV observations are therefore corrected for the extinction. Dust-corrected SFRs can be estimated by combining far-ultraviolet (FUV) and H$\alpha$ data with infrared (IR) data to exploit the complementary strengths at different wavelengths \citep[e.g.][]{Kennicutt2012ARA&A,Buat1992A&A,Meurer1995AJ,Meurer1999ApJ,Cortese2008MNRAS,Leroy2012}.
In addition to the FUV+IR and H$\alpha$+IR tracers, the low-frequency radio emission from galaxies, which is predominantly optically thin synchrotron emission, can be used to estimate their dust-unobscured SFRs via the radio-FIR correlation \citep[e.g.][]{Yun2001ApJ}.

We estimated the spatially-resolved star formation rates of our Sample 1 galaxies using FUV+24$\mu$m, H$\alpha$+24$\mu$m, and 1.4 GHz data, which are discussed, respectively, in the following Sections, \ref{SFR_from_FUV}, \ref{SFR_from_halpha}, and \ref{SFR_from_radio}. 
 We used data of these different frequencies as tracers in order to (1) get a fair comparison between different SFR diagnostics and (2) for studying star-formation history at different timescales. All SFRs in this paper assume a Kroupa IMF \citep{Kroupa2001MNRAS}.

\subsubsection{SFRs using FUV and 24$\mu$m Observations}
\label{SFR_from_FUV}

To estimate SFRSD maps of the seven galaxies in Sample 1 (Table \ref{different_samples_studied}) using FUV+24$\mu$m emission, we used SPITZER 24 $\mu$m IR data \citep{Dale2009ApJ} and GALEX FUV data \citep[11HUGS survey;][]{Kennicutt2008ApJS}. 
We first convolved both the 24 $\mu$m and the FUV maps of all galaxies to the same resolutions as our magnetic field maps. 
 The FUV data were corrected for extinction due to dust in the Milky Way (see Section~\ref{Galactic_Extinction_Correction}). 
The FUV images were in units of counts/sec/pixel and were converted to flux-density units of MJy $\textrm{Sr}^{-1}$.
We also converted the 24$\mu$m images to units of MJy $\textrm{Sr}^{-1}$ and used the following calibration from \cite{Leroy2012} to derive SFRSD maps for the sample galaxies:

\begin{equation}
\Sigma_\textrm{SFR} [\textrm{M}_{\odot}\textrm{yr}^{-1} \textrm{kpc}^{-2}]=0.081 \; \textrm{I}_\textrm{FUV} [\textrm{MJy} \;  \textrm{sr}^{-1}] + 0.032 \; \textrm{I}_{24 \mu \textrm{m}} [\textrm{MJy} \; \textrm{sr}^{-1}]
\label{sfr_calibration_FUV}
\end{equation}


The uncertainties of the coefficients are $\sim$10-30\%. Note that the uncertainty in SFR estimates arises from issues such as the error in sampling the stellar IMF of different star-forming regions, determining the contribution of different emission which are not associated with recent star formation, etc. \citep[e.g][]{kennicutt12araa,Leroy2012}. 

\subsubsection{SFRs using H$\alpha$ and 24$\mu$m Observations}
\label{SFR_from_halpha}

To estimate SFRSD maps using H$\alpha$+24$\mu$m as a tracer, we used 24 $\mu$m emission along with H$\alpha$ emission from 11HUGS \citep{Kennicutt2008ApJS}, 
for all but NGC 5194, for which we used data from the SINGS survey \citep{Kennicutt2003}.
All the maps were convolved and regridded to the resolution and pixel size of the magnetic field maps. For the H$\alpha$ maps from 11HUGS and SINGS, the flux density units were converted to erg/s/cm$^{-2}$.
We used the following calibration from \cite{Leroy2012} to estimate SFRSDs of the  galaxies in Sample 1.

\begin{equation}
\Sigma_\textrm{SFR} [\textrm{M}_{\odot}\textrm{yr}^{-1} \textrm{kpc}^{-2}]=634.0 \; \textrm{I}_{\textrm{H}\alpha} [\textrm{erg} \; \textrm{s}^{-1} \; \textrm{sr}^{-1}] + 0.0025 \; \textrm{I}_{24 \mu \textrm{m}} [\textrm{MJy} \; \textrm{sr}^{-1}]
\label{sfr_calibration_halpha}
\end{equation}

%

%
%

\subsubsection{SFRs using 1.4 GHz Observations}
\label{SFR_from_radio}

Our 1.4 GHz non-thermal maps of the galaxies (Sample 1) \citep{RoyManna2021} and an SFR calibration from \cite{Murphy2011} were used to derive SFRSD maps (Equation \ref{SFR_calibration_1420}). The calibration is based on the observed radio-FIR correlation in a sample of nearby star-forming galaxies \citep{Bell2003ApJ} and has a scatter of 0.26 dex. We used this galaxy-integrated calibration (Equation \ref{SFR_calibration_1420}) to derive the formula for spatially-resolved radio-$\Sigma_\textrm{SFR}$ calibration.  
\begin{equation}
\frac{\textrm{SFR}_\textrm{1.4GHz}}{\textrm{M}_{\odot}\textrm{yr}^{-1}}= 6.35 \times 10^{-29} \frac{\textrm{L}_\textrm{1.4GHz}}{\textrm{erg} \: \textrm{Hz}^{-1} \textrm{s}^{-1}}
\label{SFR_calibration_1420}
\end{equation}

The spatially-resolved calibration is consistent with the calibration of \cite{Heesen2014AJ}. We used the above relation to estimate the SFRSD maps of the sample galaxies from the measured 1.4~GHz surface brightness.


\subsubsection{Galactic Extinction Correction for FUV Emission}
\label{Galactic_Extinction_Correction}
 We corrected for the extinction of FUV emission due to dust in the Milky Way using the E(B-V) values along the line of sight to the sample galaxies from \cite{Bianchi2017ApJS}.
 The extinction coefficients (A$_\textrm{FUV}$) of the GALEX FUV bands were measured using Table 1 from \cite{Bianchi2017ApJS} and intrinsic fluxes ($\textrm{F}_\textrm{intrinsic}$) were estimated from the following formula: 

\begin{equation}
\textrm{A}_\textrm{FUV}= -2.5 \times log[\frac{\textrm{F}_\textrm{observed}}{\textrm{F}_\textrm{intrinsic}}]
\end{equation}

 The extinction percentage of the FUV emission is listed in Table \ref{extinction_factor_FUV}.

 \begin{table}
\caption{FUV extinction values of the Sample 1 galaxies due to the Milky Way foreground dust. The extinctions were computed using E(B-V) values along the line of sight to the sample galaxies from \cite{Bianchi2017ApJS}.}
\centering
 \begin{tabular}{||c c||} 
 \hline
 Name & Percentage extinction \\ 
 \hline\hline
NGC 2683 & 22   \\
NGC 3627 & 23   \\ 
NGC 4096 & 21  \\
NGC 4449 & 15   \\
NGC 4490 & 15    \\
NGC 4826 & 13    \\
NGC 5194 & 27   \\
 
 \hline
 \end{tabular}
\label{extinction_factor_FUV}
\end{table}

  \subsection{Gas Densities}
  \label{gas_density_measurement}
 
Atomic hydrogen (\hi) and molecular hydrogen (H$_2$) predominantly contribute to the total gas mass of galaxies.
H$_2$ is best traced using rotational transitions in CO \citep[e.g.][]{Bolatto2013ARA&A}. Spatially-resolved observations of CO transitions exist for only three of our seven galaxies in Sample 1 (Table \ref{different_samples_studied}). We have used CO J=2-1 line data of NGC 3627 and NGC 5194 from the HERA CO-Line Extragalactic Survey \citep[HERACLES;][]{Leroy2009AJ} and CO J=1-0 data of NGC 4826 from the BIMA Survey of Nearby Galaxies \citep[BIMA SONG;][]{Regan2001ApJ}. 
The HERACLES and BIMA survey have a spatial resolution of 13$''$ and 6$''$, respectively. The velocity resolution of the HERACLES and BIMA spectral cubes are $\sim$5 and 6 km/s, respectively.
We restrict our study of the connection between gas densities and magnetic fields to only these three galaxies for which spatially-resolved CO data are available.

The HI Nearby Galaxy Survey (THINGS; \citealp{Walter2008AJ}) used VLA observations to obtain very high spectral ($\le$ 5.2 km/s) and spatial ($\sim6^{''}$) resolution maps of nearby galaxies at 21cm. We used the publicly available 21cm moment maps from this THINGS survey to estimate the distribution of \hi\ in the three galaxies for which CO data are available.
All CO and \hi\ 21 cm maps were convolved and regridded to a common resolution and pixel size of the non-thermal radio maps. Gas  densities were estimated (for NGC 3627, NGC 4826 and NGC 5194) following \cite{Basu2013MNRAS} assuming
CO to H$_{2}$ conversion factor of 2 $\times 10^{20}$ $(\textrm{K} \; \textrm{km} \; \textrm{s}^{-1})^{-1}$ \citep[e.g.][]{Bolatto2013ARA&A}. A line ratio of 0.8 was assumed to convert $\textrm{CO}_\textrm{J=2-1}$ to $\textrm{CO}_\textrm{{J=1-0}}$ \citep[e.g.][]{Leroy2009AJ}.
We accounted for the contribution of helium to the gas density using $\rho_\textrm{gas}$=1.36 $\times$ ($\rho_\textrm{\hi}$ + $\rho_{\textrm{H}_{2}}$). Line of sight depths were assumed to be 300 and 400 pc for molecular and atomic gas, respectively \citep{Basu2013MNRAS}.

 \section{Results}
 \label{results}

\subsection{Magnetic Fields in the Galaxies}
\begin{figure}
 \centering
 \includegraphics[trim={0 0 0 0},clip,scale=0.3,width=0.45\linewidth]{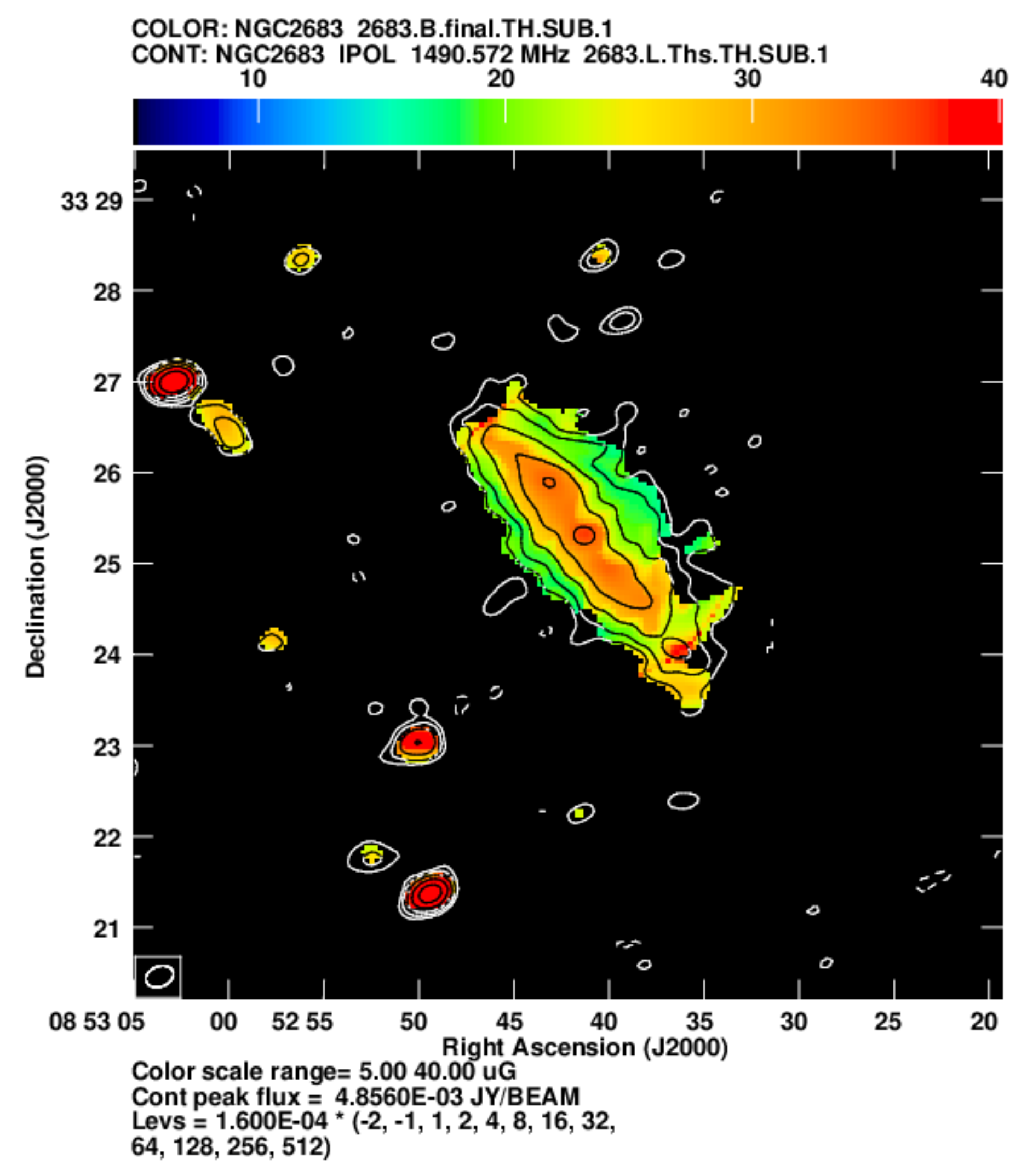}
 \includegraphics[trim={0 0 0 0},clip,scale=0.3,width=0.45\linewidth]{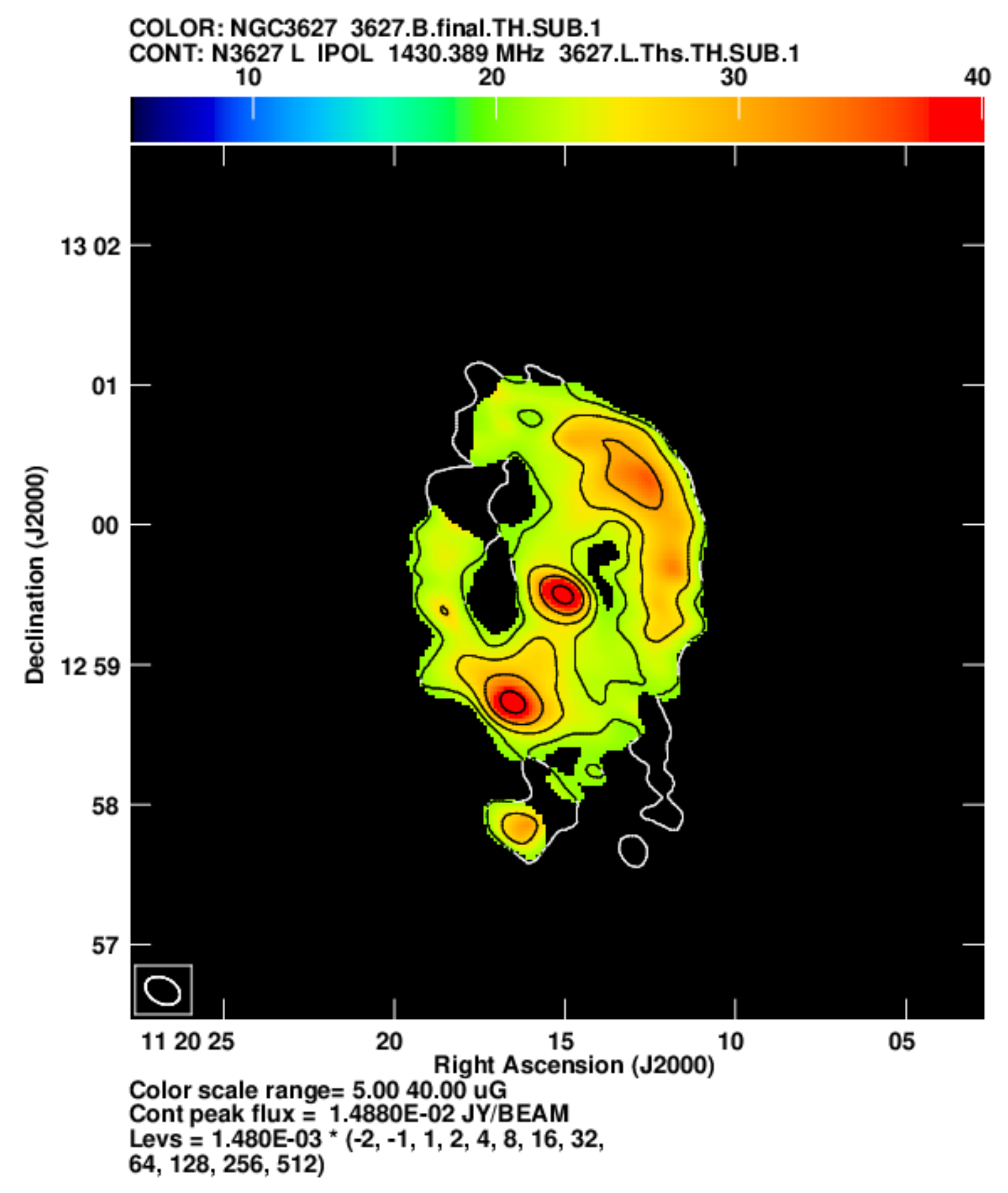}
 \includegraphics[trim={0 0 0 0},clip,scale=0.3,width=0.45\linewidth]{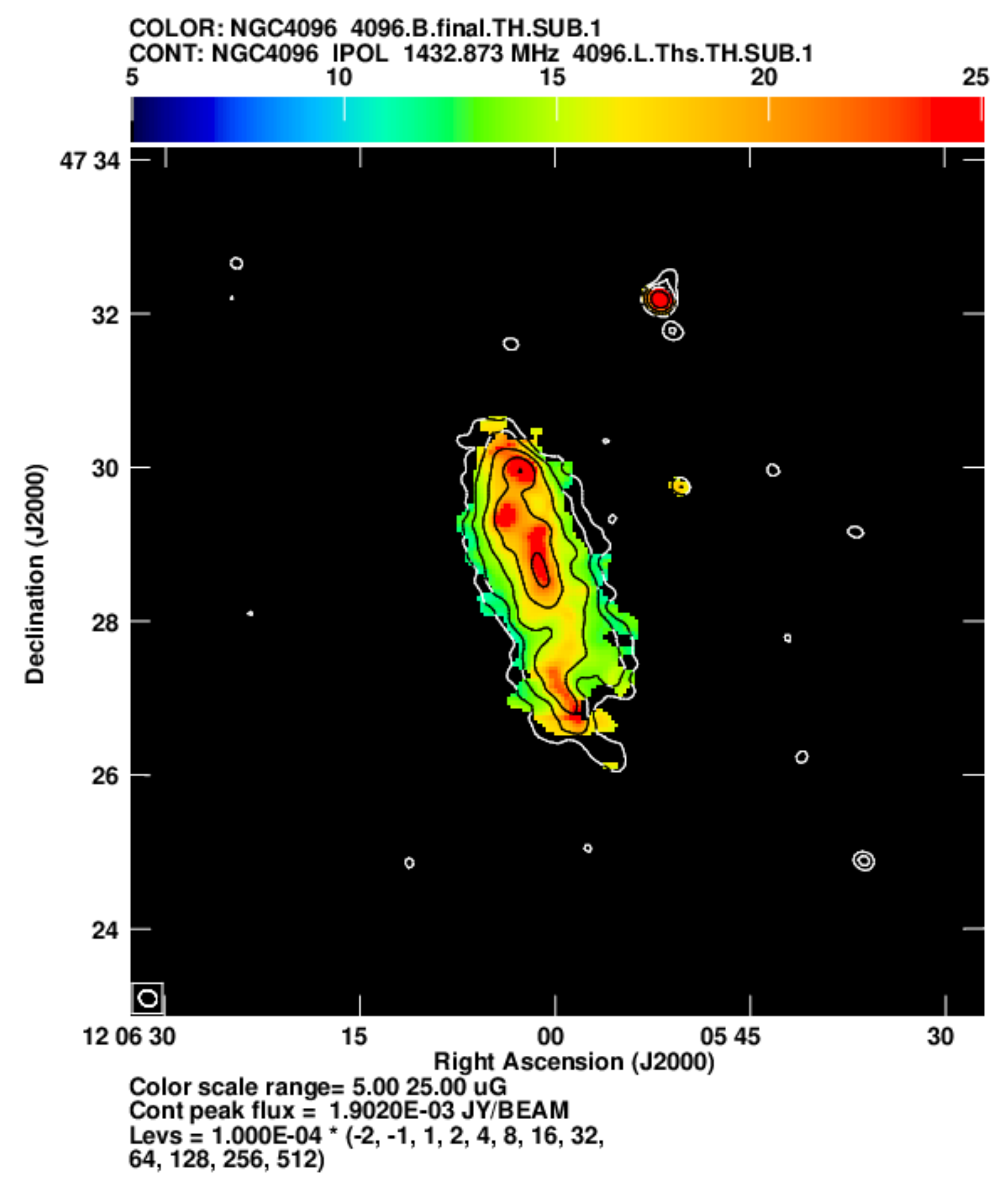}
 \includegraphics[trim={0 0 0 0},clip,scale=0.3,width=0.45\linewidth]{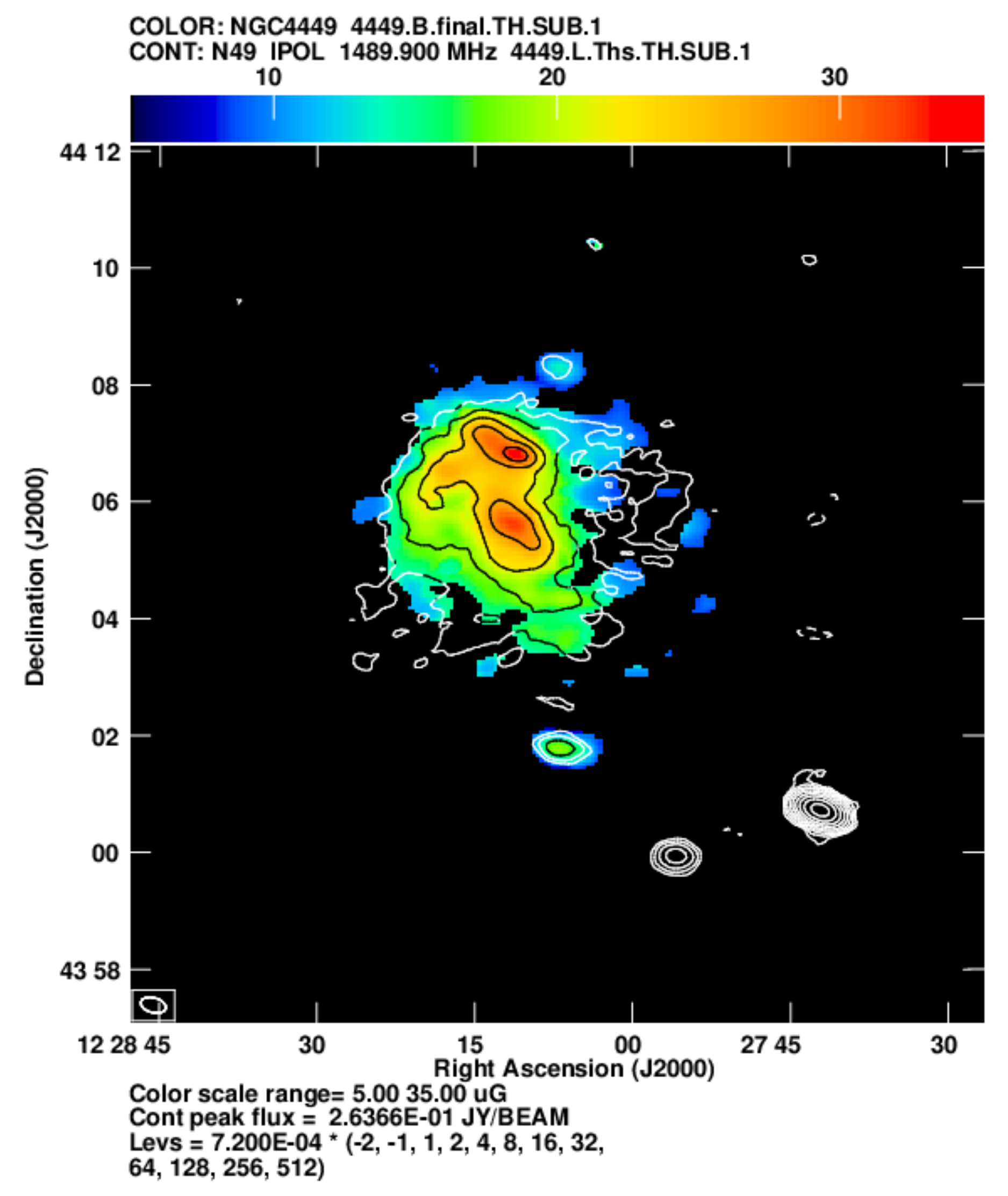}
  \caption{The equipartition magnetic field maps of NGC 2683, NGC 3627, NGC 4449 and NGC 4096 (clockwise from top left) (Sample 1). Non-thermal radio contours at 1.4 GHz are overlaid on magnetic field maps. The magnetic field strengths are shown in color with non-thermal emission at 1.4 GHz shown  as overlaid contours. Contour levels are presented below each panel in the figure. The circle in the bottom-left corner of the panels indicates the angular resolution of the maps. The uncertainties on mean magnetic fields are 0.06$\mu$G, 0.17$\mu$G, 0.04$\mu$G and 0.18$\mu$G for the above galaxies, respectively.}
  
\label{B_cont_1}
  \end{figure}

\begin{figure}
 \centering
 \includegraphics[width=0.45\linewidth]{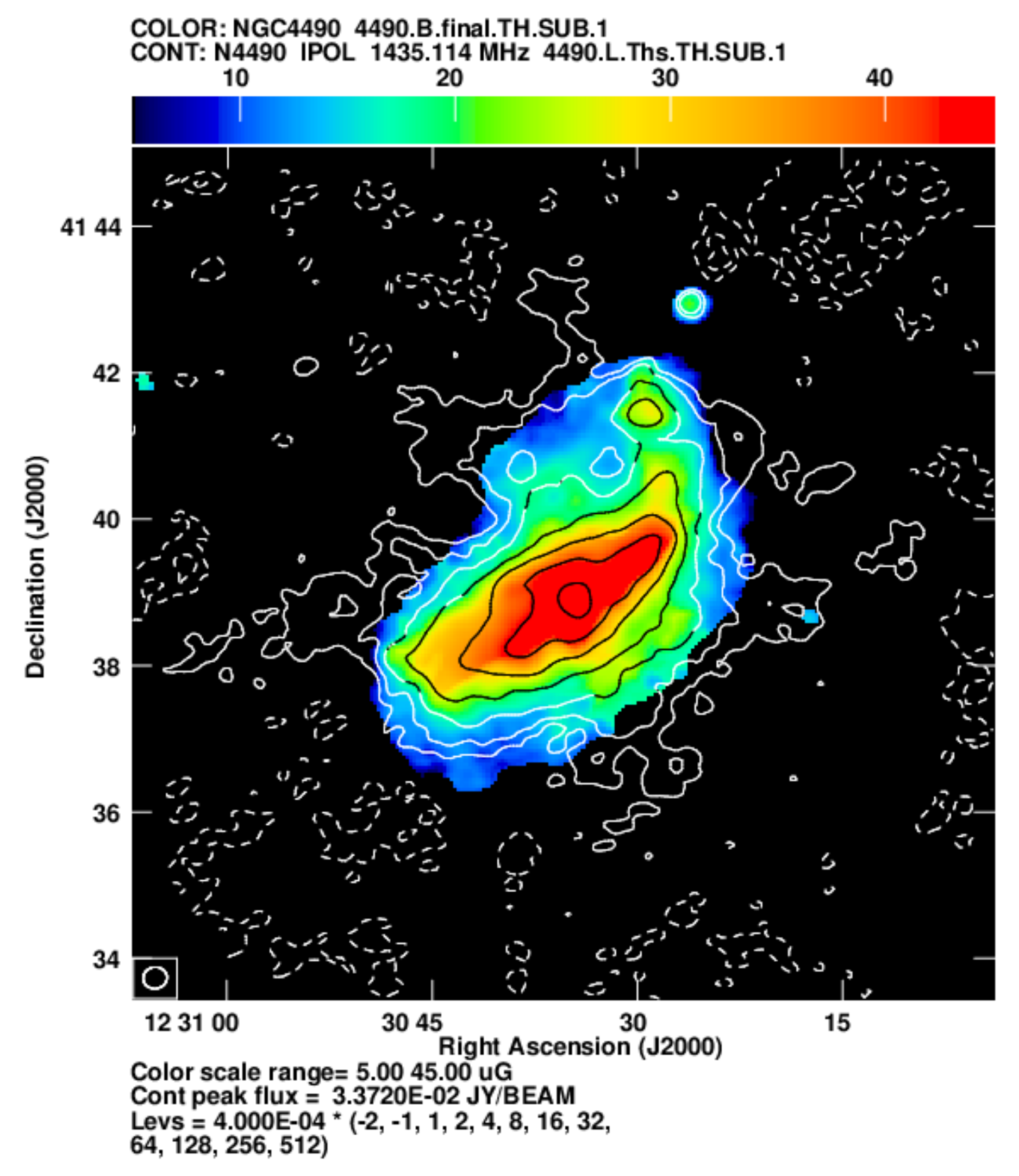}
 \includegraphics[width=0.45\linewidth]{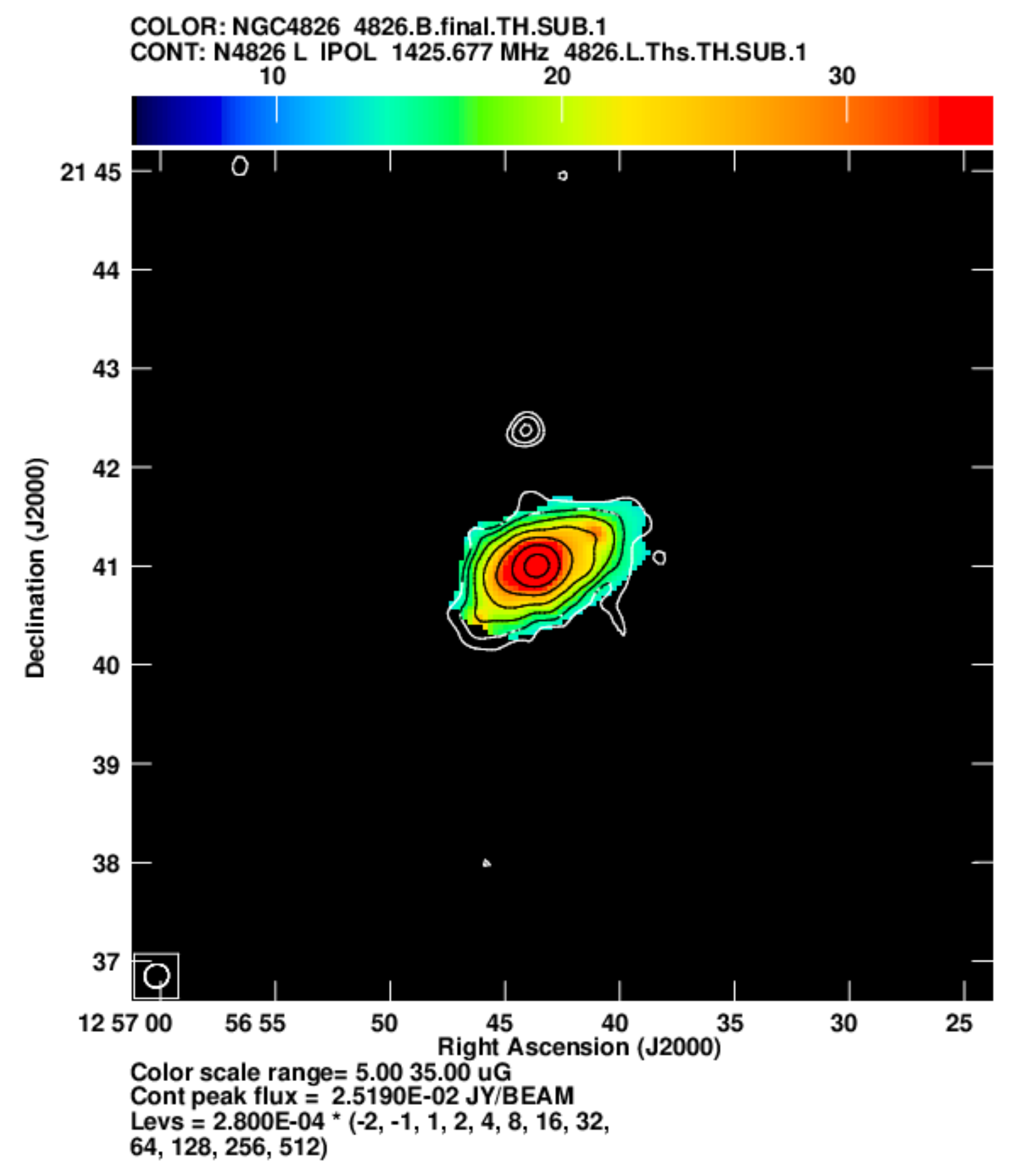}
 \includegraphics[width=0.43\linewidth]{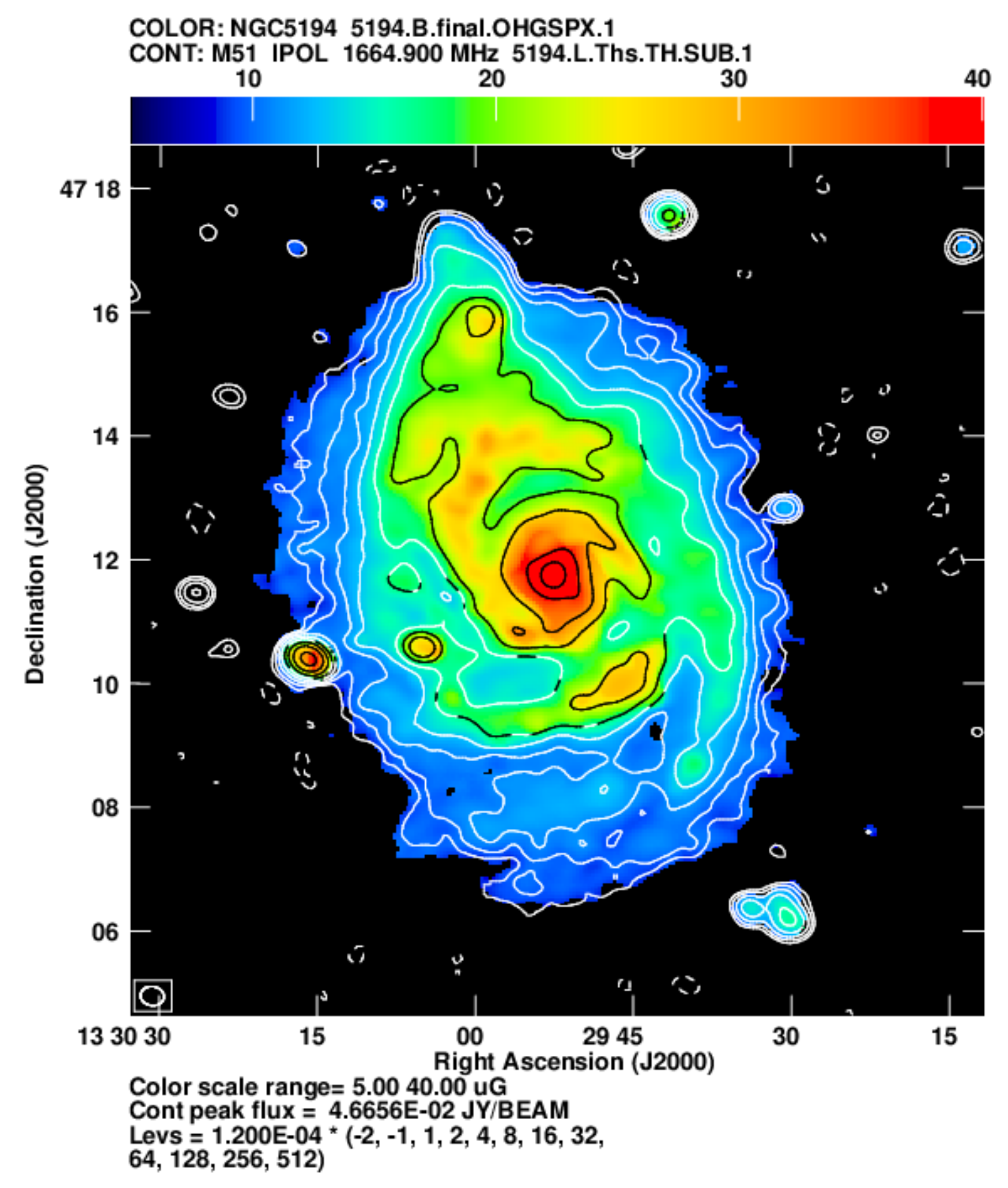}
 \includegraphics[width=0.5\linewidth]{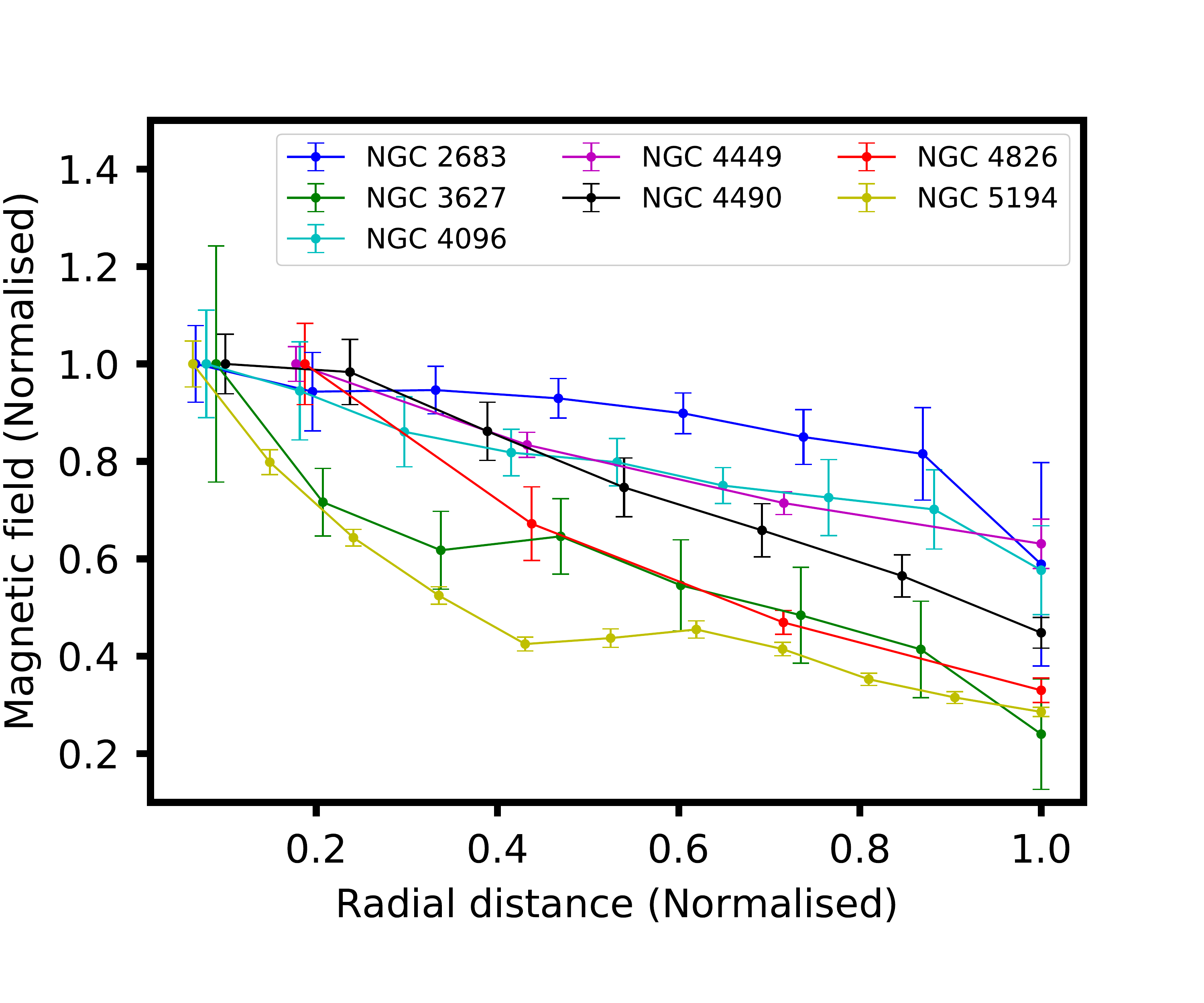}
 

 \caption{The equipartition magnetic field maps of NGC 4490 (top left), NGC 4826 (top right) and NGC 5194 (bottom left). The magnetic field strengths are shown in color with non-thermal emission at 1.4 GHz shown as overlaid contours. Contour levels are presented below each panel in the figure. The circle in the bottom-left corner of the panels indicates the angular resolution of the maps. The uncertainties on mean magnetic fields are 0.06$\mu$G, 0.11$\mu$G and 0.02$\mu$G, respectively. The bottom right panel presents the radial variation of magnetic field strengths with galactocentric distance for all seven galaxies in Sample 1.}
 
\label{B_cont_2}
  \end{figure}

We have estimated spatially resolved revised equipartition magnetic field maps for seven galaxies in Sample 1, using the procedures of Section~\ref{data_analysis_magnetic_field}; these maps are shown in Figures \ref{B_cont_1} \& \ref{B_cont_2}.
Flux density contours of 1.4 GHz observations are overlaid on magnetic field maps. The resolution of these maps corresponds to spatial scales of $\sim$ 0.4$-$0.8 kpc (see Table \ref{sample_details}). The bottom right panel of Figure \ref{B_cont_2} shows the radial variation of the magnetic field with galactocentric radius of all the seven galaxies where both the axes are normalized by their maximum values. Here, we have averaged the magnetic field strengths over an annular elliptical region of width equal to the beam size of the corresponding map. Position and inclination angle (Table \ref{sample_details}) of each galaxy were used while selecting the elliptical regions. 
 We find magnetic fields to be stronger at the central region and at the star formation sites (arm regions) with field strengths up to 50 $\mu$G. Field strengths fall by $\sim$50\% at the edges of the magnetic field maps.
 The Milky Way also shows such a trend in the variation of magnetic field strengths \citep{Beck1996ARA&A}. 
 We note that our analysis was limited to distances where the signal-to-noise ratio in spectral index maps is $>$ 5; the magnetic field strengths at these distances are thus likely to be reliable.

We note that, compared to the magnetic field strengths obtained using the classical equipartition expression, these values are higher by $\sim$ 1.3$-$1.5 for a non-thermal spectral index of -0.6, and 
they match for a spectral index of -0.75 \citep{BeckandKrause2005}.

Figure \ref{B_map_err_1} shows the uncertainties in the magnetic field values for Sample 1 derived using the Monte Carlo method described in Section~\ref{Error_Magnetic_Field}. Statistical uncertainties on mean magnetic fields for these seven galaxies are provided in Table \ref{uncertainties_magnetic_fields}.


 \begin{table}
\centering
 \caption{Statistical uncertainties on mean magnetic fields for galaxies in Sample 1.}
 \begin{tabular}{||c c||} 
 \hline
 Name & Statistical uncertainty \\ 
      & on mean magnetic fields \\
      & ($\mu$G) \\
 \hline\hline
NGC 2683 & 0.06   \\
NGC 3627 & 0.17   \\ 
NGC 4096 & 0.04  \\
NGC 4449 & 0.18   \\
NGC 4490 & 0.06    \\
NGC 4826 & 0.11    \\
NGC 5194 & 0.02   \\
 \hline
 \end{tabular}
\label{uncertainties_magnetic_fields}
\end{table}


 \subsection{Star Formation Rates in the Galaxies}
We have estimated the global, galaxy-averaged SFRs of Sample 1 galaxies
using 1.4 GHz, FUV+24$\mu$m, and H$\alpha$+24$\mu$m emission using calibrations discussed in Section \ref{data_analysis_SFR}.
Globally integrated star formation rates of the sample galaxies
are given in Table \ref{SFR_global}. No systematic offset was found in the SFR values estimated using these tracers. The differences in the SFR values for our galaxies are much less than the calibration uncertainty
except for NGC 4490. 
For NGC 4490, SFR calculated from 1.4 GHz emission is higher than the same from FUV+24$\mu$m emission by a factor of 2.2.

\begin{figure*}
 	\centering
 \includegraphics[width=0.49\linewidth]{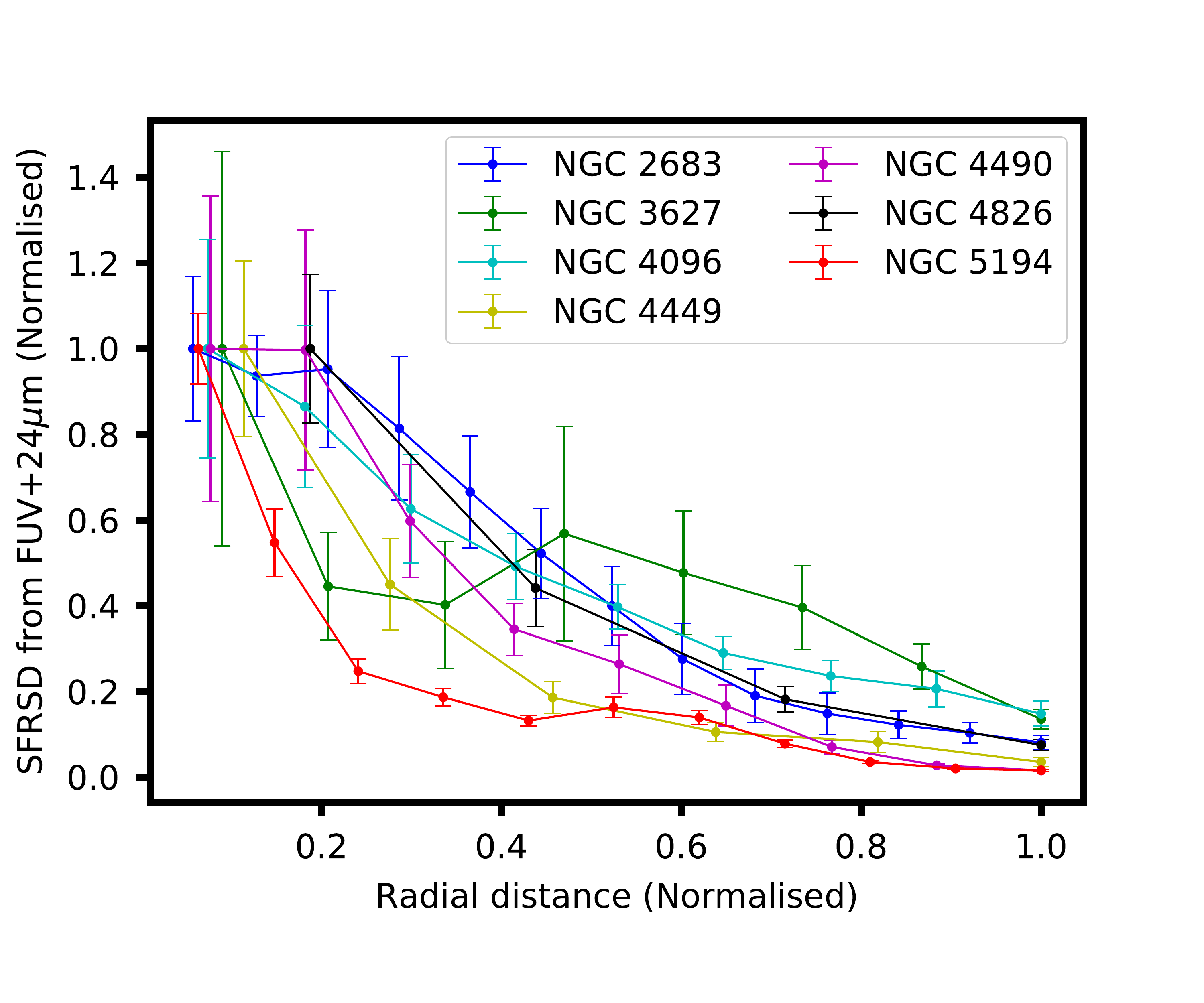}
 \includegraphics[width=0.49\linewidth]{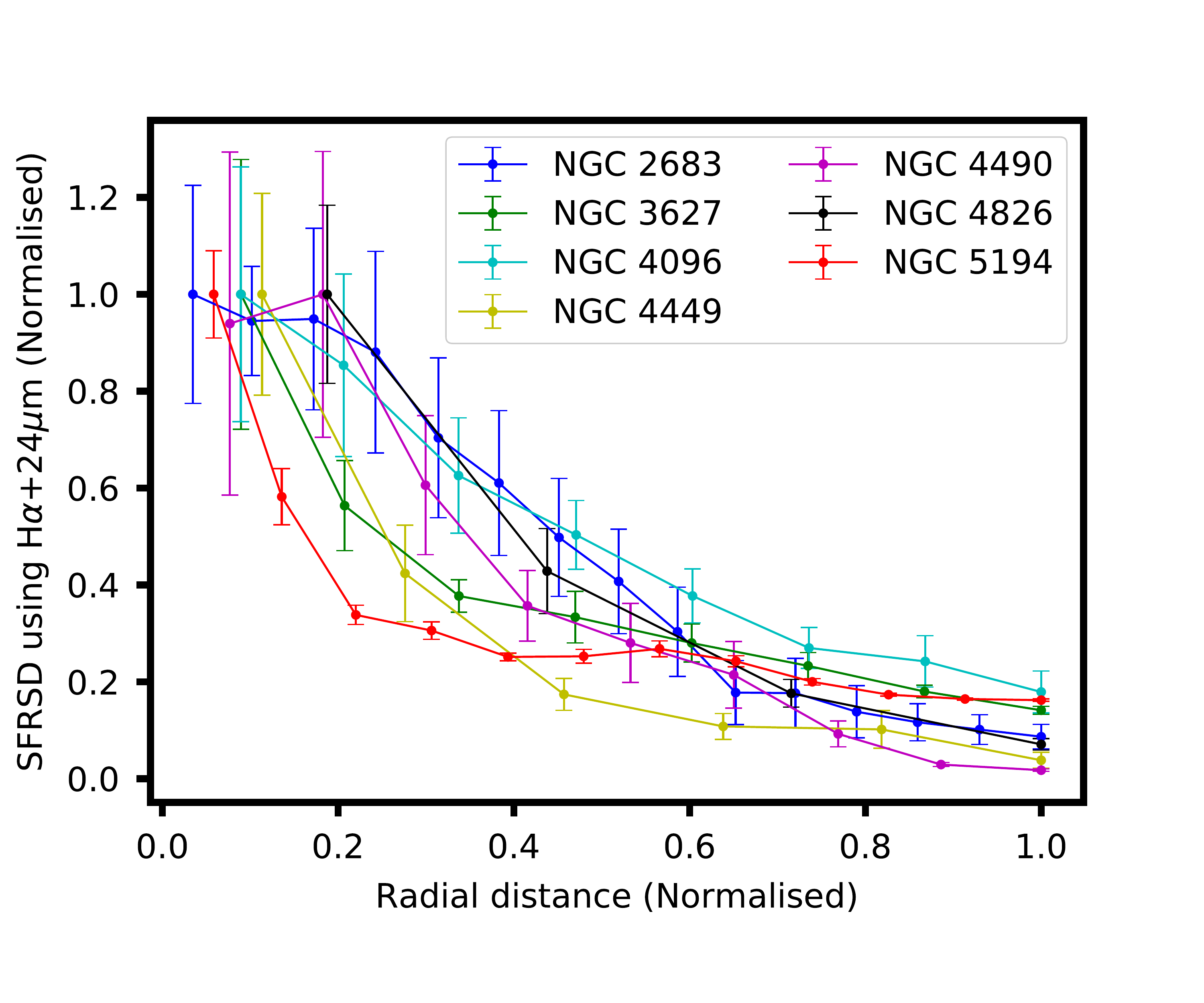}
  \caption{The variation of SFRSDs (normalized),
 estimated using FUV+24$\mu$m (left panel) and H$\alpha$+24$\mu$m (right panel) emission as a function of galactocentric distance (normalized) for all seven galaxies in Sample 1.}
     \label{SFRSD_radial}    
 \end{figure*}

As discussed in Section \ref{data_analysis_SFR}, we have estimated SFRSD maps of the seven galaxies (Sample 1) using FUV+24$\mu$m, H$\alpha$+24$\mu$m and 1.4GHz emission. We show SFRSD maps of the seven galaxies in the Appendix (Figures \ref{SFRSD_cont_1}-\ref{SFRSD_cont_2}), where SFRSDs estimated using 1.4 GHz and FUV+24$\mu$m emission are shown in contours and colors, respectively. In the Appendix (Figures \ref{SFRSD_halpha_1}-\ref{SFRSD_halpha_2}), we also present the SFRSD maps estimated using H$\alpha$+24$\mu$m and 1.4 GHz emission in colors and contours, respectively. The SFRSD maps of each galaxy in Figures~\ref{SFRSD_cont_1}$-$\ref{SFRSD_halpha_2} are shown in the same color scale and contours.
To determine the radial variation of SFRSDs, we have averaged the SFRSD maps
of our sample galaxies over tilted rings centred on the optical centre of each galaxy using their inclinations and position angles. The width of the tilted rings was taken to be equal to the beam size of the corresponding image. Figure \ref{SFRSD_radial} shows the radial variation of the average SFRSD, derived using FUV+24$\mu$m and H$\alpha$+24$\mu$m emission, with galactocentric distance where both the axes are normalized to their maximum values. We also derived the radial variation of SFRSDs for the galaxies using 1.4 GHz emission and 
 it is consistent within $1\sigma$ statistical uncertainties, with those derived using FUV+24$\mu$m and H$\alpha$+24$\mu$m data.
Azimuthally averaged SFRSDs of all the seven galaxies decrease gradually towards the outer region and drop by a factor of 6 to 8 at the edge. 

 \begin{table}
\centering
\caption{Galaxy-averaged star formation rates of the galaxies in Sample 1, using 1.4 GHz, FUV+24$\mu$m, and H$\alpha$+24$\mu$m data. The uncertainties on the SFR values are $\approx$ 30\%.}
 \begin{tabular}{||c c c c||} 
 \hline
 Name & SFR from 1.4 GHz (M$_{\odot}$yr$^{-1}$) & SFR from FUV+24$\mu$m (M$_{\odot}$yr$^{-1}$) & SFR from H$\alpha$+24$\mu$m (M$_{\odot}$yr$^{-1}$) \\ 
 \hline\hline
NGC 2683 & 0.28  & 0.25 & 0.33 \\
NGC 3627 &  1.56 & 2.00 & 1.84\\ 
NGC 4096 &  0.42 & 0.35 & 0.38\\
NGC 4449 & 0.37  & 0.38 & 0.32\\
NGC 4490 &  4.63 & 2.13 & 2.30\\
NGC 4826 & 0.63  & 0.73 & 0.78\\
NGC 5194 & 4.16  & 3.88 & 3.65\\
 \hline
 \end{tabular}
\label{SFR_global}
\end{table}

  \subsection{Details of the Individual Galaxies of Sample 1}

{\bf(i) NGC 2683:} 
In this galaxy, \cite{Krause2020A&A} found very weak linear polarisation using C-band and L-band VLA observations. 
Based on the optical image, we could separate the central region from the disk. The average magnetic field in the central region is found to be $\approx$31 $\mu$G and the outer region of the disk has an average value of $\approx$19 $\mu$G (see Figure \ref{B_cont_1} and Table \ref{B_regions_table}). 

\cite{Wiegert2015AJ} used WISE 22 $\mu$m  data to estimate a galaxy-averaged SFR of $\approx$0.09~M$_\odot$yr$^{-1}$ for NGC 2683.
From our analysis, integrated SFR was measured to be $\sim$0.24 M$_\odot$yr$^{-1}$ and $\sim$0.28 M$_\odot$yr$^{-1}$ using FUV+24$\mu$m and 1.4 GHz radio emission, respectively.
 However, we note that \cite{Wiegert2015AJ} used a distance of 6.27 Mpc for this galaxy, but we have used a distance of 7.7 Mpc. 
The SFR is estimated to be 0.16 M$_\odot$yr$^{-1}$ using FUV+24$\mu$m emission, assuming the same distance as used by \cite{Wiegert2015AJ}. 
Taking the calibration uncertainties and the assumed distance into account, our estimated SFR is hence  consistent with that of \cite{Wiegert2015AJ}.
We note that the contours on the background sources (Figure \ref{SFRSD_cont_1}) are not real SFRSDs, as these are likely to be background AGNs.



 {\bf(ii) NGC 3627:} 
 NGC 3627 was observed at 8.46 GHz and 4.85 GHz using the VLA in its D-configuration \citep{Soida2001A&A}. These authors estimated the magnetic field strengths using the classical equipartition formula \citep{Longair2011} and found an average equipartition magnetic field strength of 11$\pm$2 $\mu$G, assuming a constant non-thermal spectral index of 0.9 and a disk thickness of 2 kpc.
 \citet{Soida2001A&A} also studied the polarized emission at these frequencies to find a regular magnetic field of 4$\pm$1 $\mu$G. They suggested two distinct magnetic field components of NGC 3627: one for the spiral arms and another for the inter-arm regions. We have separately studied equipartition magnetic fields in the arm and interarm regions of the galaxy. We find that the central region and the edges of the extended bar have magnetic field strengths of $\approx$ 34 $\mu$G  (see Figure \ref{B_cont_1}). The arm region has a field strength of $\approx$28 $\mu$G (see Table \ref{B_regions_table}). However, the magnetic field strength in the interarm regions has values $\approx$21 $\mu$G. We note that our estimates of the equipartition magnetic field strengths in the galaxy are higher than those found by \cite{Soida2001A&A}; this difference likely arises from the fact that \citet{Soida2001A&A} estimated the magnetic field strengths using the classical equipartition formula, which is known to significantly underestimate the magnetic field in the star-forming regions.

We measured a galaxy-averaged SFR of $\approx$2.0 M$_\odot$yr$^{-1}$ and $\approx$1.56 M$_\odot$yr$^{-1}$ from FUV+24$\mu$m and 1.4 GHz emission, respectively. Our measurements of spatially resolved SFRs in different regions are consistent, within calibration uncertainties, with the SFR estimates of \cite{Watanabe2011MNRAS}.

{\bf(iii) NGC 4096:} 
Our estimate of the equipartition magnetic field in NGC 4096 varies from $\approx$21 $\mu$G at the centre to $\approx$12 $\mu$G at the edge (table \ref{B_regions_table}). 
The magnetic field strength in both the central region and northern periphery is quite similar, with typical field strengths of $\approx$ 20 $\mu$G; this is presumably due to its high inclination. The outer part of the galaxy has an average field strength of $\approx$14 $\mu$G. NGC 4096 was observed \citep{Irwin2012AJ,Wiegert2015AJ} with its B-field and further studied by \cite{Krause2020A&A} 
who found very little polarized emission from the galaxy.

\cite{Wiegert2015AJ} used the 22 $\mu$m$-$SFR calibration to measure a galaxy-averaged SFR of 0.27$\pm$0.02 M$_\odot$yr$^{-1}$. Our measurement of the galaxy-averaged SFR is $\approx$0.35 M$_\odot$yr$^{-1}$ and $\approx$0.43 M$_\odot$yr$^{-1}$ using FUV+24$\mu$m and 1.4 GHz emission, respectively. Considering the calibration uncertainties, our estimates are consistent with that of \cite{Wiegert2015AJ}.
 

{\bf(iv) NGC 4449:} This is an optically bright irregular starburst galaxy.
\cite{Chyzy2000A&A} used  VLA 4.86 and 8.46
GHz observations to find a galaxy-averaged equipartition magnetic field of $\approx$14 $\mu$G. These authors also used polarization emission to estimate a regular field of $\approx$8 $\mu$G. The equipartition magnetic field map of NGC 4449 from our study is
shown in Figure \ref{B_cont_1}. As noted in Section~\ref{data_analysis_magnetic_field}, about 70 \% of the total projected area of this galaxy has spectral index values of less than 0.55. We have replaced the pixel values with $\alpha_{\rm nt}<0.55$ with $\alpha_{\rm nt}=0.55$ while computing the magnetic field for NGC 4449 (see Section \ref{data_analysis_magnetic_field}). The average magnetic field strength is $\approx$17 $\mu$G in this galaxy, which is comparable to the findings of \cite{Chyzy2000A&A}. 

Our measurements of the galaxy-averaged SFR are $\approx$0.38 M$_\odot$yr$^{-1}$  and $\approx$0.37 M$_\odot$yr$^{-1}$ using FUV+24$\mu$m and 1.4 GHz emission, respectively, which are consistent with the SFR of 0.47 M$_\odot$yr$^{-1}$ estimated by \cite{Chyzy2011A&A}. 

{\bf(v) NGC 4490:} 
\cite{Nikiel2016MNRAS} observed NGC 4490 at 0.61 GHz using the GMRT, and at 4.86 $\&$ 8.44 GHz using VLA + Effelsberg. The authors used these observations to find a mean equipartition magnetic field of 21.9$\pm$2.9 $\mu$G, with typical field strengths in the range of 18 $\mu$G to 40 $\mu$G. We have found a typical equipartition magnetic field strength of $\approx$40 $\mu$G in the central region, which decreases to $\approx$17 $\mu$G in the outer region (see Figure \ref{B_cont_2}); these values are consistent with the estimates of \cite{Nikiel2016MNRAS}. We find a relatively lower magnetic field strength of $\approx$15 $\mu$G in both the interacting region and the companion galaxy NGC 4485. Therefore, a gradual decrease in the average magnetic field strength occurs from the center to the outer region.

\cite{Clemens1999MNRAS} used radio observations to find a galaxy-averaged SFR of 4.7 M$_\odot$yr$^{-1}$. We found a similar SFR ($\approx$4.63 M$_\odot$yr$^{-1}$) using 1.4 GHz radio emission but a factor of $\sim$2 lower SFR (2.13 M$_\odot$yr$^{-1}$) using the  FUV+24$\mu$m emission (Table \ref{SFR_global}). Extinction corrections for NGC 4490 are believed to be higher than those typically assumed and this may lead to an underestimation of the SFR while using the FUV+24$\mu$m diagnostics \citep{Clemens1999MNRAS}. 


{\bf(vi) NGC 4826:} 
No spatially resolved maps of magnetic fields and SFRSDs are available in the literature. 
 We measure the central and outer regions of the galaxy to have an average
equipartition magnetic field strength of $\approx$38 $\mu$G and $\approx$20 $\mu$G, respectively (see Figure \ref{B_cont_2} and Table \ref{B_regions_table}). 
We find galaxy-averaged SFR of $\approx$0.73  M$_\odot$yr$^{-1}$ and $\approx$0.63 M$_\odot$yr$^{-1}$ using FUV+24$\mu$m and 1.4 GHz data, respectively.

{\bf(vii) NGC 5194:} \cite{Fletcher2011MNRAS} used VLA C-band observations of the galaxy and assumed a constant thermal and non-thermal spectral index of 0.1 and 1.1 to find an average equipartition magnetic field strength of 20 $\mu$G using the revised formula by \cite{BeckandKrause2005}. They found a magnetic field of 20$-$25 $\mu$G in the spiral arms, higher than the 15$-$20 $\mu$G typical in the interarm regions. Using VLA observations at S-band (2$-$4~GHz) frequencies, \cite{Kierdorf2020A&A} found the field strength of turbulent and regular components of the magnetic field in the arm regions of 18$-$24~$\mu$G and 8$-$16~$\mu$G, respectively. 
We find an equipartition magnetic field strength of $\approx$25 $\mu$G in the arm region and $\approx$18 $\mu$G in the interarm region (see Table \ref{B_regions_table}). The peripheral region has a magnetic field of $\approx$12 $\mu$G, while the overlapping region between NGC 5194 and NGC 5195 has an average B$_\textrm{eq}$ of $\approx$16 $\mu$G. Considering our use of Equation \ref{revised_formula} \citep{BeckandKrause2005}, measurements are roughly consistent with the earlier study of \cite{Fletcher2011MNRAS} and \cite{Kierdorf2020A&A}.


Spatially resolved SFRs were measured in several star-forming regions of NGC 5194 using H$\alpha$+24$\mu$m and 
H$\alpha$+Pa$\alpha$ emission \citep{Kennicutt2007ApJ}. SFRSDs in different regions were found to be in the range of 0.10 to 0.46 M$_\odot$yr$^{-1}$kpc$^{-2}$. Our estimates using the two tracers are consistent with the estimates of \citet{Kennicutt2007ApJ} (See Figures \ref{SFRSD_cont_2} \& \ref{SFRSD_halpha_2}). Furthermore, we find that the galaxy-integrated SFR derived using FUV+24$\mu$m ($\approx$3.88  M$_\odot$yr$^{-1}$) and 1.4 GHz data ($\approx$4.16 M$_\odot$yr$^{-1}$) are consistent
 with each other, within 1-sigma statistical uncertainty.


\begin{table}
 \caption{Magnetic field strengths in different regions of the galaxies in Sample 1. For the irregular galaxy NGC 4449, we could only measure the galaxy-integrated magnetic field. We have separated the two nearly face-on galaxies (NGC 3627 and NGC 5194) into arm and inter-arm regions. For the rest of the galaxies, we could not separate the arm and inter-arm region due to their higher inclinations.}
\footnotesize
 \centering 
  \begin{tabular} {||c c c c c c||}
\hline
    Galaxy  &  Galaxy-average & B$_\textrm{eq}$ in & B$_\textrm{eq}$ in & B$_\textrm{eq}$ in & B$_\textrm{eq}$ in \\
    
      name  & B$_\textrm{eq}$ & central region    & disk region         & arm region         &  inter-arm region \\
    
            & ($\mu$G)       &   ($\mu$G)        &  ($\mu$G)           &   ($\mu$G)         &   ($\mu$G)    \\
\hline
    NGC 2683 & 24$\pm$6  & 31$\pm$3  & 19$\pm$5 & --       & -- \\
    NGC 3627 & 25$\pm$4  & 34$\pm$8  & --       & 28$\pm$5 & 21$\pm$4   \\
    NGC 4096 & 16$\pm$4  & 21$\pm$5  & 14$\pm$3 & --       & -- \\
    NGC 4449 & 17$\pm$6  &  -- & --  & --       & -- \\
    NGC 4490 & 23$\pm$10 & 40$\pm$6  & 17$\pm$7 & --       & -- \\
    NGC 4826 & 23$\pm$9  & 38$\pm$8  & 20$\pm$5 & --       & -- \\
    NGC 5194 & 16$\pm$6  & 34$\pm$6  &  --      & 25$\pm$5 & 18$\pm$4  \\
\hline
  \end{tabular}
\label{B_regions_table}
\end{table}

\subsection{Is the Minimum Energy Condition  Valid for the Sample Galaxies?}
 \label{validity_minimum_energy_condition}
We have estimated magnetic fields for the galaxies in Sample 1 assuming the ``minimum energy condition$"$ or  ``equipartition condition$"$, i.e. by assuming that the energy density in the magnetic field is approximately equal to the energy density in cosmic ray particles. Therefore, it is important to verify the validity of this assumption in our sample galaxies. The tightness of the spatially-resolved radio$-$FIR correlation can be used to estimate the deviation of the energy densities from the minimum energy condition \citep{Hummel1986A&A,Basu2013MNRAS}. According to the simplified model of \cite{Hummel1986A&A}, when the minimum energy condition is satisfied, the distribution of $I_\textrm{nt}/I_\textrm{FIR}$ will be similar to the distribution of $\textrm{B}^{1+\alpha_{nt}}$. The model assumes the following to be constant across galaxies: (a) the ratio of the number densities of relativistic electrons and dust-heating stars, (b) the volume ratio of radio and FIR emitting regions, and (c) the ratio of efficiency factors for both the radio and FIR emission.  
In this model, the cumulative distribution function (CDF) of the quantity $I_\textrm{nt}/I_\textrm{FIR}$ and $\textrm{B}_\textrm{eq}^{1+\alpha_{nt}}$ is expected to follow each other if $\textrm{B}_\textrm{eq}$ is close to $\textrm{B}$. 

\begin{figure*}
 	\centering
 \includegraphics[width=0.49\linewidth]{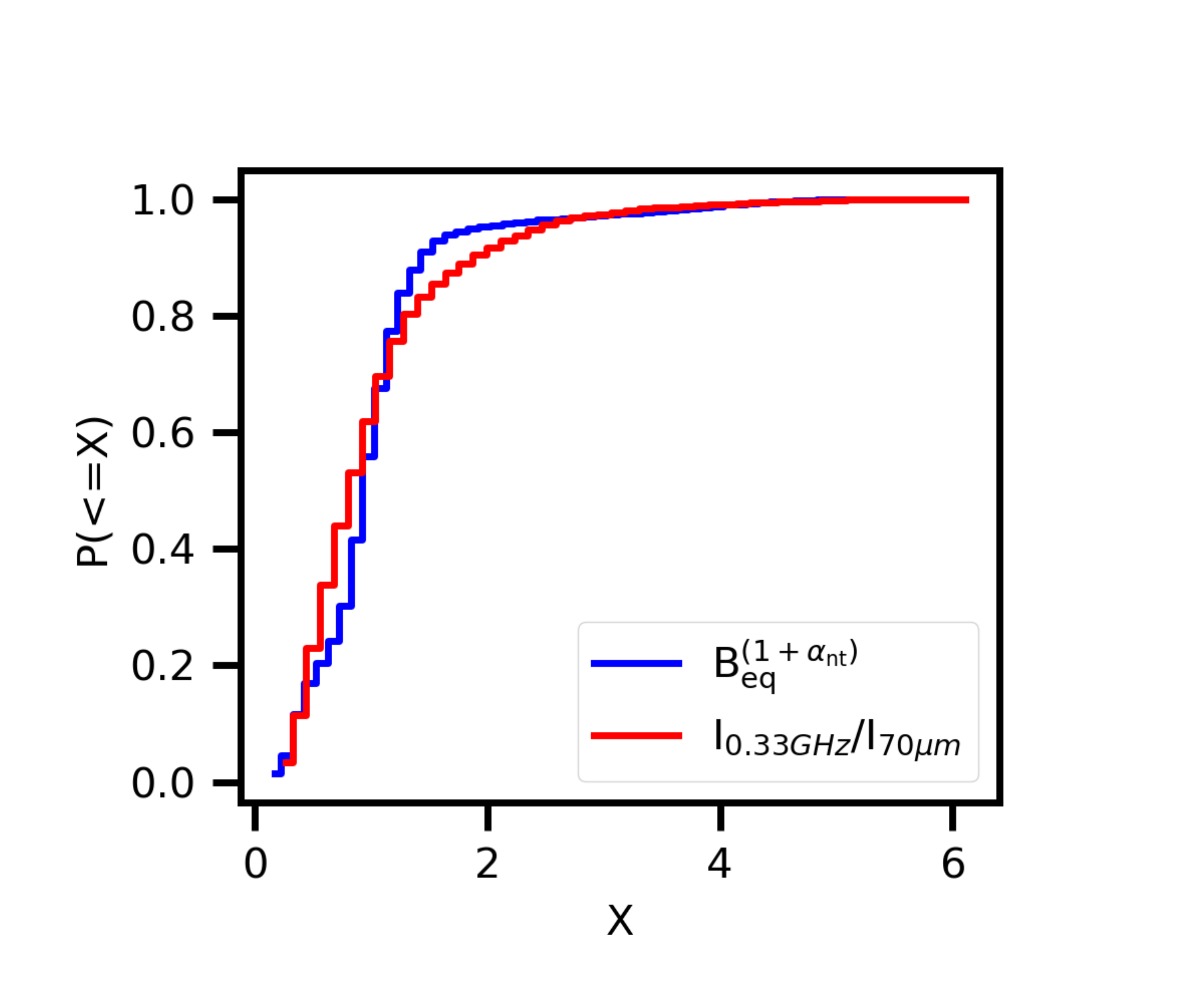}
 \includegraphics[width=0.49\linewidth]{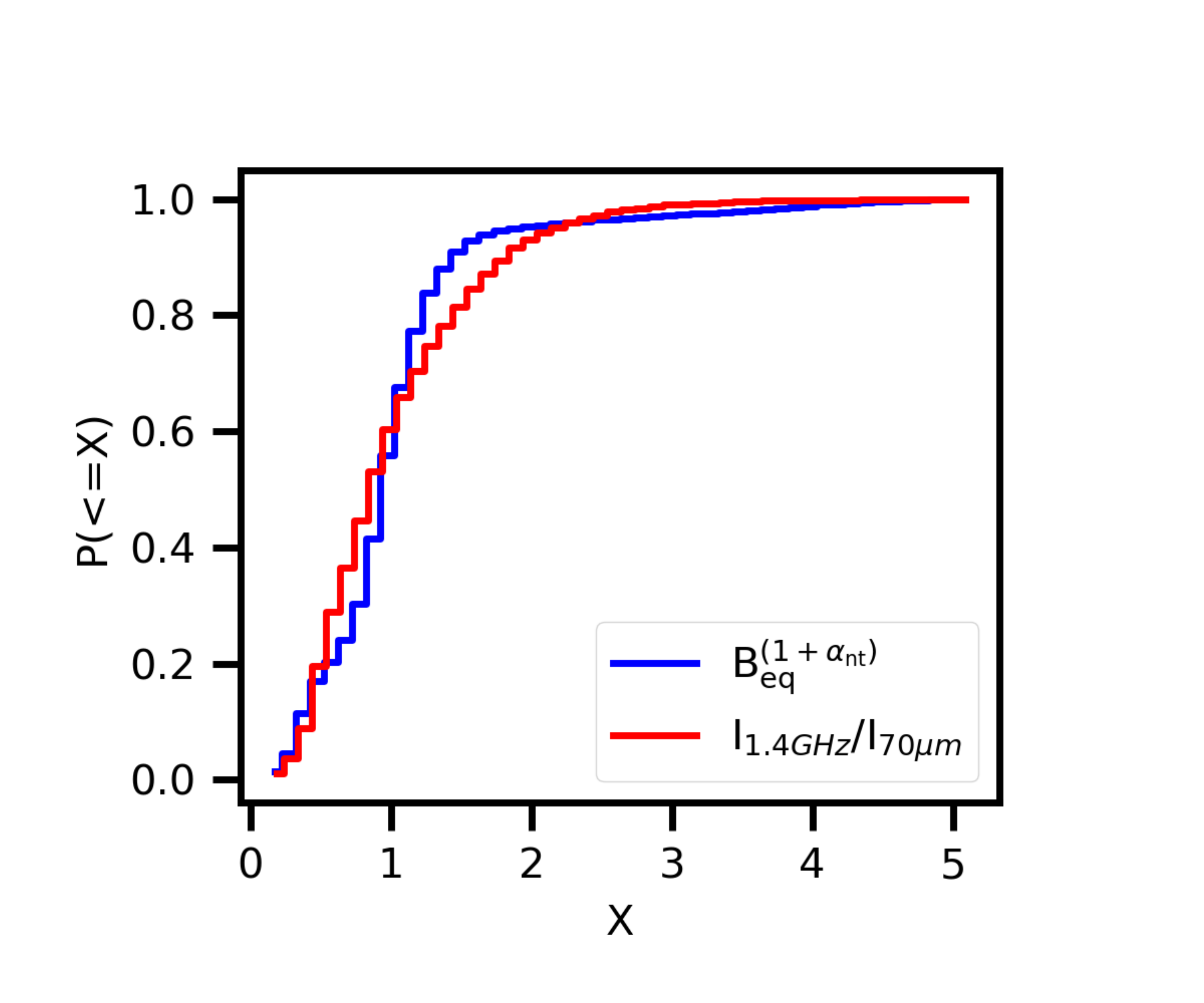}
 \includegraphics[width=0.49\linewidth]{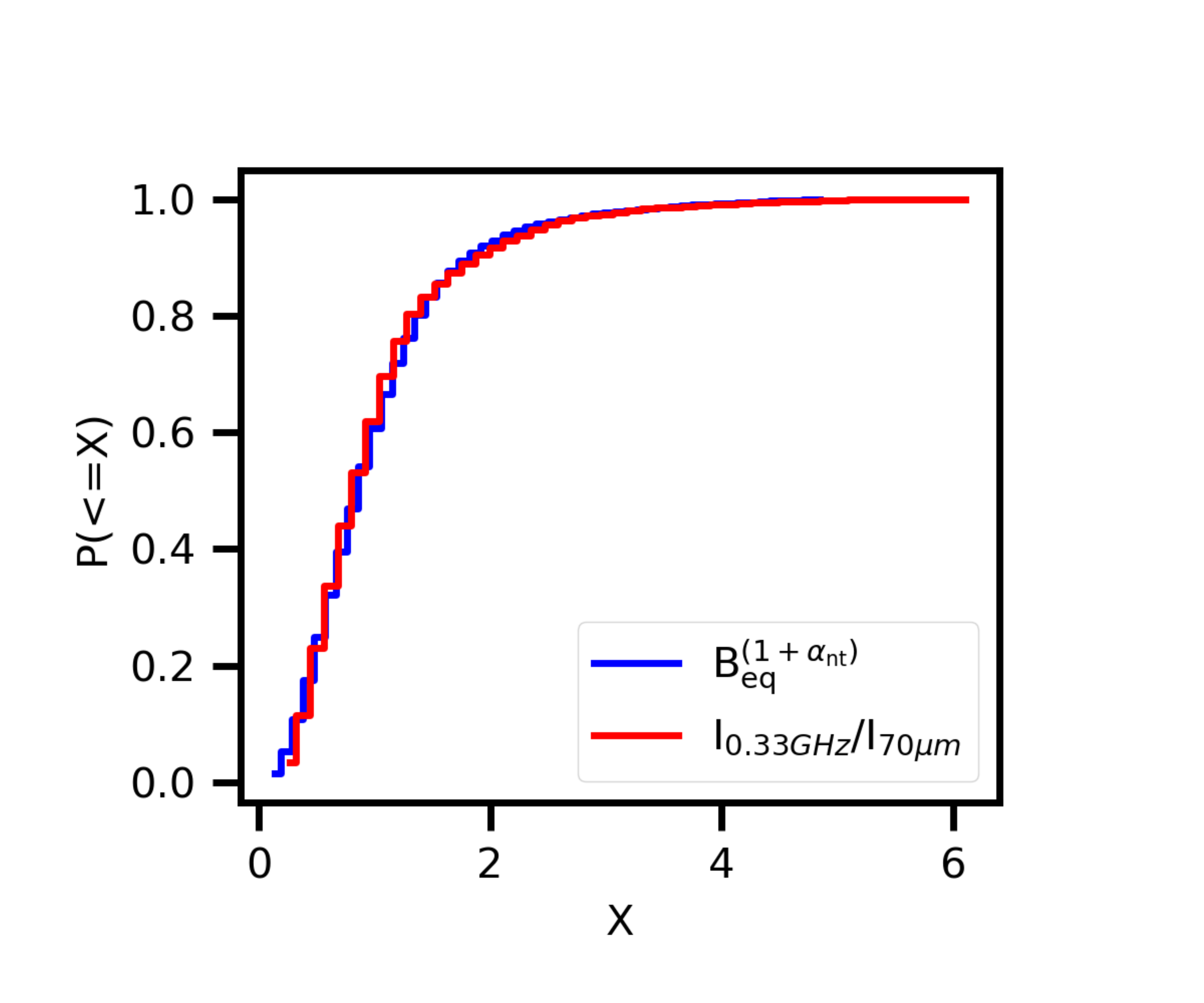}
 \includegraphics[width=0.49\linewidth]{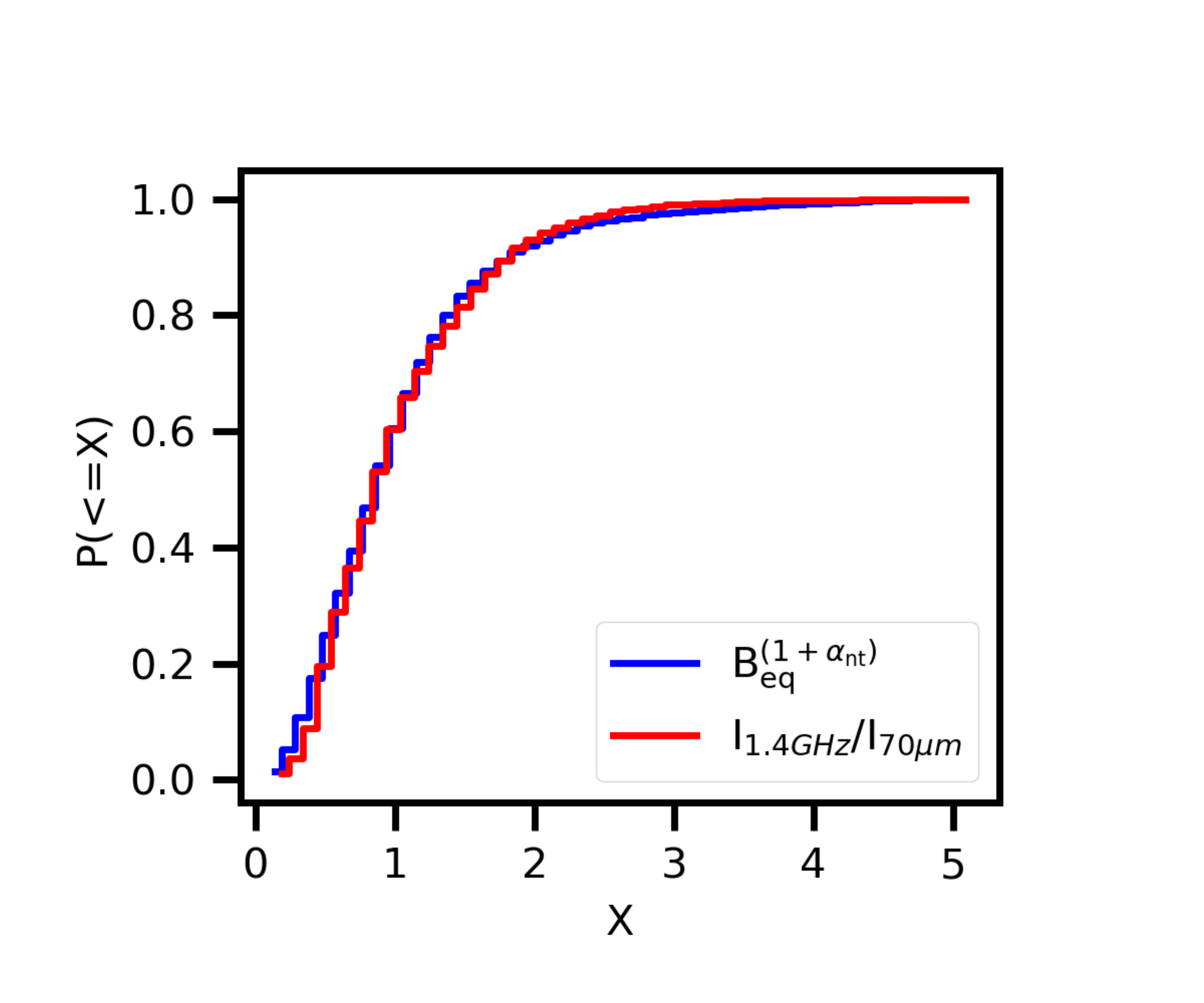}
  \caption{The top panels show the cumulative distribution function (CDF) of $\textrm{I}_\textrm{nt,radio}/\textrm{I}_{70\mu m}$ (in red) and $\textrm{B}_\textrm{eq}^{1+\alpha_{nt}}$ (in blue), where $\textrm{I}_\textrm{nt}$ is the non-thermal emission at 0.33 GHz (top left) and 1.4 GHz (top right) (Sample 1). The variables are normalized by their median values. The bottom panels show the same but now with the magnetic field perturbed from its measured value using $\sigma$=0.1 (see Section \ref{validity_minimum_energy_condition}); the CDFs of the $\textrm{I}_\textrm{nt,radio}/\textrm{I}_{70\mu m}$ and $\textrm{B}_\textrm{eq}^{1+\alpha_{nt}}$ are now consistent with being derived from the same distribution.}
  
     \label{CDF_sample}
 \end{figure*}

To verify the validity of the minimum energy condition in our sample galaxies, we have followed the procedure as in \cite{Hummel1986A&A} and \cite{Basu2013MNRAS}. The CDF of $I_\textrm{nt}/I_\textrm{FIR}$ and $B_\textrm{eq}^{1+\alpha_{nt}}$ were estimated using our radio maps of the sample galaxies at both 0.33 and 1.4 GHz. We used an ensemble of spatially-resolved values of $\alpha_{nt}$, $I_\textrm{nt}$ (both at 0.33 and 1.4 GHz), $I_\textrm{FIR}$ (70 $\mu$m) and magnetic fields (B$_\textrm{eq}$), which are averaged over the beam size from all the galaxies in Sample 1 (Table \ref{different_samples_studied}) to generate these distributions. The CDFs of all quantities were normalized by their median values. The top panels in Figure \ref{CDF_sample} show the median-normalized CDFs of $I_\textrm{nt}/I_\textrm{FIR}$ and $B_\textrm{eq}^{1+\alpha_{nt}}$ at both 0.33 and 1.4 GHz.

We find that the CDFs of $I_\textrm{nt}/I_\textrm{FIR}$ and $B_\textrm{eq}^{1+\alpha_{nt}}$ at both 0.33 and 1.4 GHz broadly follow each other but with slight deviations at high and low ends (see top panels in Figure \ref{CDF_sample}). This implies that the minimum energy condition is broadly valid and is consistent with earlier findings. For example,
\cite{Hummel1986A&A} found the distribution of the two quantities is similar in a sample of Sbc galaxies while \cite{Basu2013MNRAS} reached similar conclusions in a study of 5 nearby large spiral galaxies, but with slight deviations observed in the CDFs of $I_\textrm{nt}/I_\textrm{FIR}$ and $B_\textrm{eq}^{1+\alpha_{nt}}$ in the interarm regions of the galaxies.

The observed deviation in the CDFs of $I_\textrm{nt}/I_\textrm{FIR}$ and $B_\textrm{eq}^{1+\alpha_{nt}}$ for our sample galaxies imply a corresponding deviation from the minimum-energy condition. In order to quantify this deviation, we performed a Monte Carlo simulation originally proposed by \cite{Hummel1986A&A}. In this simulation, 
random numbers (X) were drawn from a Gaussian distribution with standard deviation $\sigma$. Thereafter, we multiplied $10^{X}$ with the observed equipartition magnetic fields to introduce deviations from the minimum-energy condition. We thus constructed the CDF of $B_\textrm{eq}^{1+\alpha_{nt}}$ using the deviated magnetic field values.
The CDF of $B_\textrm{eq}^{1+\alpha_{nt}}$ were then compared to the observed CDF of $I_\textrm{nt}/I_\textrm{FIR}$ via a Kolmogorov-Smirnov (KS) test. This procedure was repeated for a range of $\sigma$ from 0 to 0.2. We find that the p-values for the KS test comparing the distributions are maximized when $\sigma$ = 0.1. Indeed, $B_\textrm{eq}^{1+\alpha_{nt}}$ derived after deviating the magnetic field using $\sigma$ = 0.1 and $I_\textrm{nt}/I_\textrm{FIR}$ are consistent with being derived from the same distribution, with a KS test p-value of 0.41 and 0.55, when using  $I_\textrm{nt}$ at 0.33 and 1.4 GHz, respectively. The bottom panels in Figure \ref{CDF_sample} show the CDFs of the two quantities for $\sigma$ = 0.1 at 0.33 and 1.4 GHz; it is clear that the CDFs follow each other.
This implies the actual magnetic field values may deviate from the equipartition values by $\sim$ 25\% in our galaxies in Sample 1. We note that any violation of the assumptions made by \cite{Hummel1986A&A} may also lead to the observed deviation in the CDFs.

 \subsection{Correlation Between Magnetic Fields and SFRSDs}
 \label{results_B_SFRSD_corr}

\begin{figure}
\includegraphics[width=0.95\linewidth]{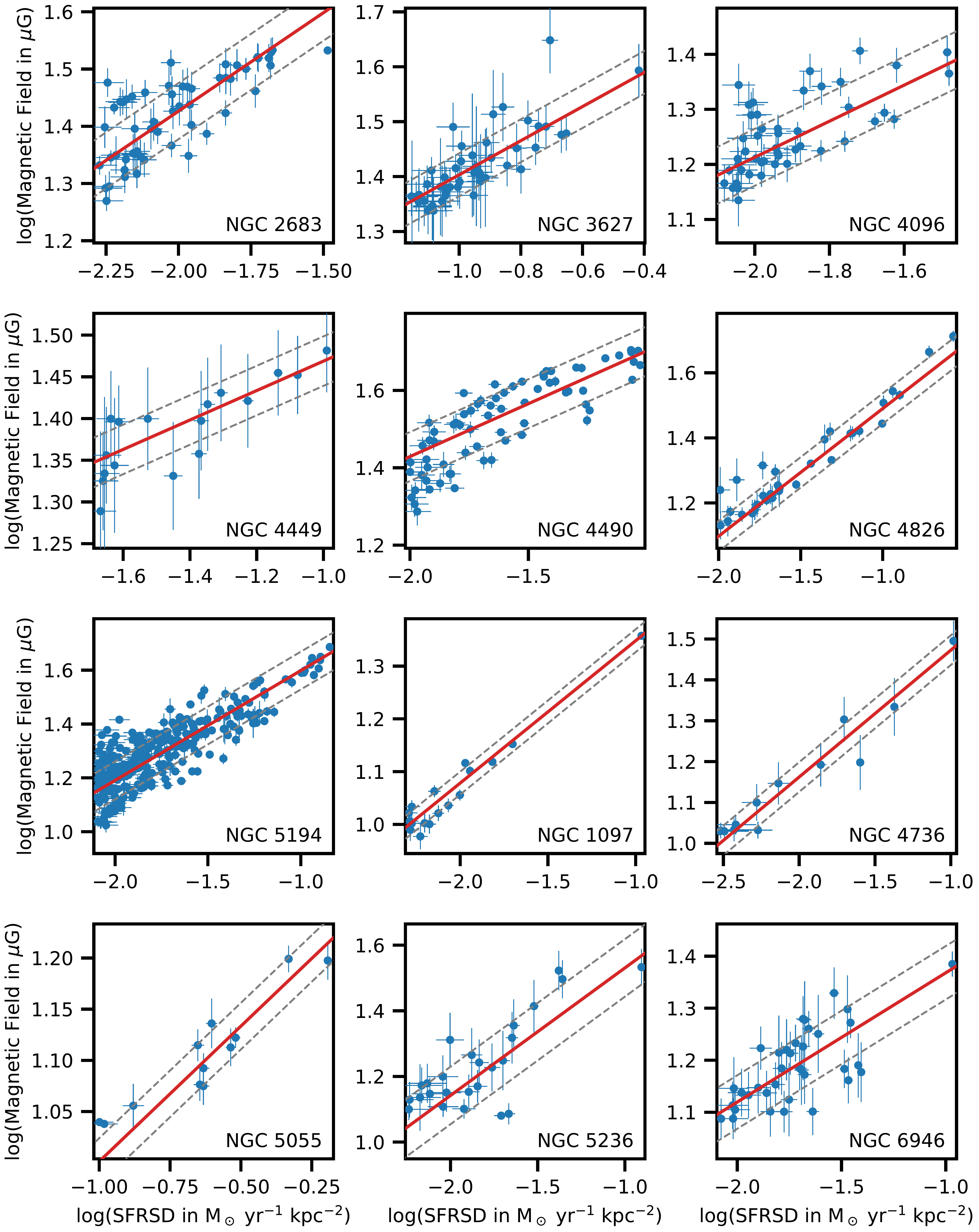}
\caption{The correlation between magnetic fields and SFRSD for the combined sample of 12 galaxies (Sample 2, Table \ref{different_samples_studied}). For the seven galaxies in Sanple 1, the SFRSD estimates shown in the plots were derived using $\textrm{FUV}+24\mu m$ (Section \ref{results_B_SFRSD_corr}). The SFRSD estimates for the five galaxies from \cite{Basu2012MNRAS} (Sample 2) were derived using $\textrm{H}\alpha+24\mu m$ (Section \ref{study_basu_2013}) The red line shows a linear fit to the data points. The black dashed lines show the $\pm 1\sigma$ vertical scatter.}

\label{B_SFRSD_corr}
\end{figure}

We have studied the correlation between the spatially-resolved equipartition magnetic field and SFRSDs for the galaxies in Sample 1 (Table \ref{different_samples_studied}) at scales of $\approx$360$-$760 pc (Table \ref{sample_details}). For the seven sample galaxies, we used the SFRSD maps estimated using the FUV+24$\mu$m emission.
The correlations between magnetic fields and SFRSDs for the seven galaxies are shown in Figure \ref{B_SFRSD_corr}.
Each point represents the logarithms of equipartition magnetic fields and SFRSD values that are averaged over the beam size of the corresponding maps. \cite{daSilva2014MNRAS} found that SFR calibrations could be biased and strongly affected by stochasticity at small spatial scales where the star formation rate is low ($\le$ 10$^{-2.5}$ M$\odot \textrm{yr}^{-1}$); we have therefore excluded regions of low star formation rates from the correlation study. 

We find that the equipartition magnetic field and the SFSRD are correlated in all seven sample galaxies. We use \emph{orthogonal distance regression} in Scipy \citep{2020SciPy-NMeth} to fit a power law of the form $\textrm{B}= \textrm{B}_0 \  \left(\Sigma_\textrm{SFR}\right)^{\eta}$ to the magnetic field $-$ SFRSD data points; the spatially-resolved uncertainty maps of equipartition magnetic fields and rms noise on the SFRSD maps were used to estimate the uncertainties on each data point during the fitting procedure. The best-fit parameters of the power-law are given in Table \ref{B_SFRSD_corr_parameter}. We have also estimated the scatter (rms of the data points along the y-axis) of the correlations which are presented in Table \ref{B_SFRSD_corr_parameter} and are shown in dashed lines in the corresponding plots (Figure \ref{B_SFRSD_corr}). 
We find that six of the seven galaxies have slopes ($\eta$) in the range of $\approx 0.27-0.40$ but that the slope is relatively lower for NGC 4449 with $\eta \approx 0.18$. Averaging over the slope of all galaxies in Sample 1, we find a mean slope of $0.32\pm0.06$.

%


\begin{table}

 \caption{Best-fit parameters and the scatter of the correlation between magnetic fields and SFRSDs for the seven galaxies in Sample 1. The data were fitted with a power law of the form B=B$_{0}(\Sigma_\textrm{SFR})^{\eta}$.}
\centering
 \begin{tabular}{||c c c c c||} 
 \hline
 Name & Slope ($\eta$) & Intercept (B$_{0}$) (log($\mu$G)) & Intercept (B$_{0}$) ($\mu$G) & Scatter \\ 
 \hline\hline
 NGC 2683 & 0.34 $\pm$ 0.04 & 2.10 $\pm$ 0.07 & 125 $\pm$ 1.2 & 0.05 \\
 NGC 3627 & 0.31 $\pm$ 0.03 & 1.71 $\pm$ 0.03 & 51 $\pm$ 1.1 & 0.04 \\
 NGC 4096 & 0.33 $\pm$ 0.04 & 1.80 $\pm$ 0.08 & 63 $\pm$ 1.2 & 0.05 \\
 NGC 4449 & 0.18 $\pm$ 0.03 & 1.64 $\pm$ 0.04 & 43 $\pm$ 1.1 & 0.03 \\  
 NGC 4490 & 0.27 $\pm$ 0.02 & 1.90 $\pm$ 0.03 & 79 $\pm$ 1.1 & 0.06 \\
 NGC 4826 & 0.38 $\pm$ 0.02 & 1.80 $\pm$ 0.02 & 63 $\pm$ 1.0 & 0.05\\
 NGC 5194 & 0.40 $\pm$ 0.01 & 2.00 $\pm$ 0.02 & 100 $\pm$ 1.0 & 0.07\\
 \hline
 \end{tabular}
 \label{B_SFRSD_corr_parameter}
 
\end{table}

\subsection{Correlation Between Magnetic Fields and Gas Densities}
\label{results_B_gas-corr}
\begin{figure*}
 \includegraphics[width=\linewidth]{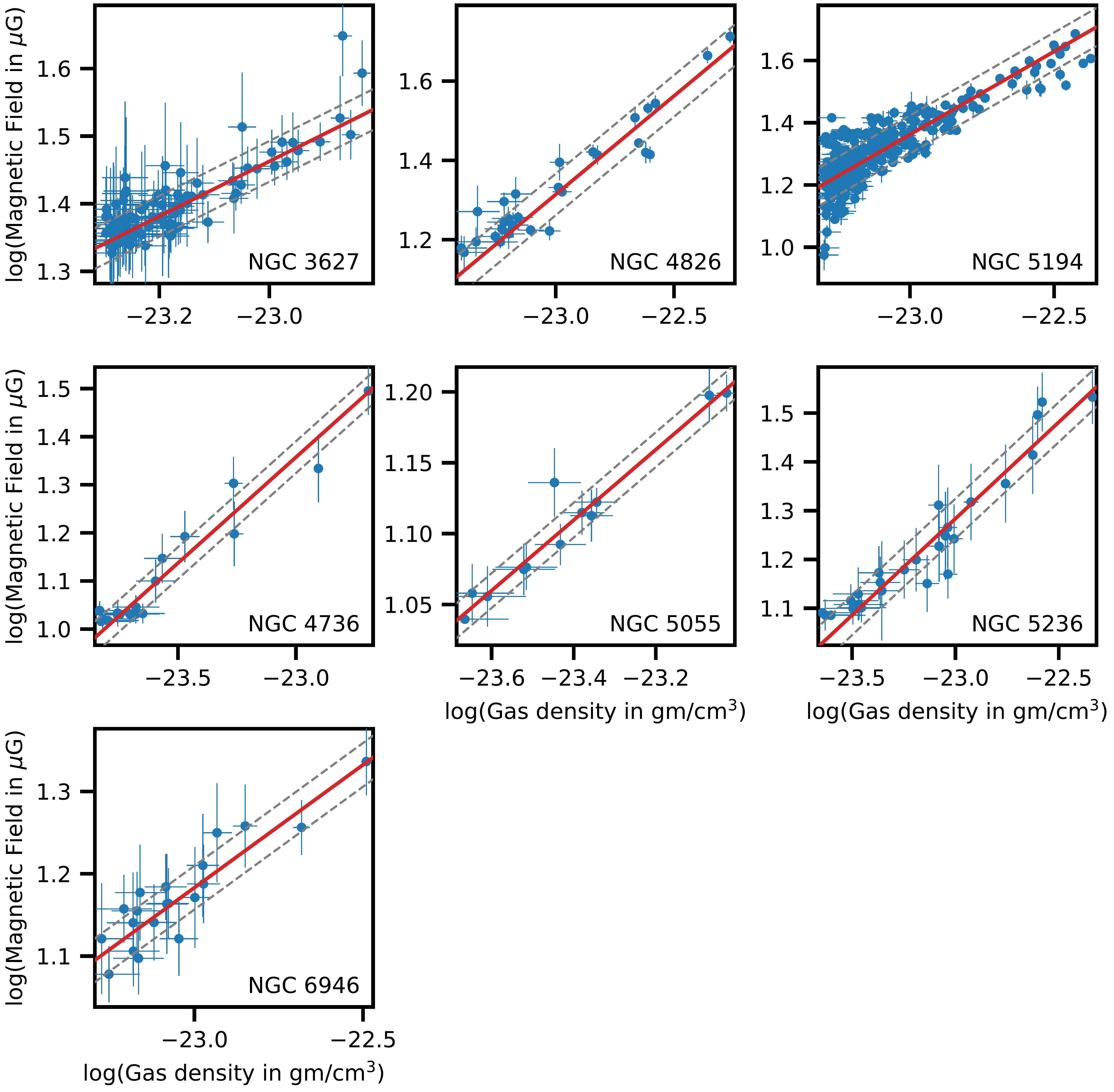}
 
 \caption{ The correlations between magnetic fields ($\mu$G) and gas densities (gm/cm$^{-3}$) for seven galaxies of Sample 3 (Table~\ref{different_samples_studied}). The red line shows a linear fit to the data points. The black dashed lines show the $\pm 1\sigma$ vertical scatter.}
\label{B_gas_density_corr}
  \end{figure*}

We have studied the correlation between spatially-resolved equipartition magnetic fields and gas densities for three of the galaxies in Sample 1, NGC 3627, NGC 4826, and NGC 5194, for which spatially resolved CO observations were available (see Section \ref{gas_density_measurement}).
Similar to the study of correlations between B$_\textrm{eq}$ and SFRSDs, we have studied the correlations between B$_\textrm{eq}$ and gas density values, both averaged over the beam size of the corresponding maps. The correlations between magnetic fields and gas densities of NGC 3627, NGC 4826, and NGC 5194 are shown in Figure \ref{B_gas_density_corr}. We have again used \emph{orthogonal distance regression} in Scipy \citep{2020SciPy-NMeth} to fit a power-law to the B$_\textrm{eq}$ and gas density data points.
The scatters of the three correlations
 are shown in  dashed lines in all the figures. 
  
The measured best-fit power-law indices are 0.40$\pm$0.02,\; 0.49$\pm$0.03 and 0.53$\pm$0.02 (Table \ref{B_gas_corr_parameter}) for NGC 3627, NGC 4826 and NGC 5194, respectively.
The mean of the power-law indices is 0.47$\pm$0.05.

\section{Extending the sample with 5 galaxies from existing GMRT observations}

\label{study_basu_2013}

As mentioned earlier, a study of B$_\textrm{eq}$ and radio-FIR correlations for a sample of five large nearly face-on galaxies was carried out by \citet{Basu2012MNRAS,Basu2012ApJ,Basu2013MNRAS}, using low-radio frequency observations at 0.33 and 1.4 GHz at sub-kpc linear resolutions. In this paper, we expand our study of spatially-resolved correlations between magnetic fields, gas densities, and SFRSDs by including these five galaxies.

We refer readers to \cite{Basu2012MNRAS} for a detailed discussion of their sample, GMRT observations, data reduction procedures, and estimation of non-thermal spectral indices. It is to be noted that the modelling of the thermal free-free emission from these galaxies is performed in the same way as was done for our seven galaxies in Sample 1.

We have estimated the SFRSD maps of these five galaxies using H$\alpha$ data along with 24 $\mu$m IR data. We obtained H$\alpha$ maps of four of the galaxies, NGC 1097, NGC 4736, NGC 5055, and NGC 6946 from the ancillary data at the SINGS website\footnote{https://irsa.ipac.caltech.edu/data/SPITZER/SINGS/} and obtained the H$\alpha$ map of NGC 5236 from 11HUGS \citep{Kennicutt2008ApJS}. We used H$\alpha$ and MIPS 24 $\mu$m data in combination to derive the SFRSD maps of these galaxies using the calibration from \cite{Leroy2012} (Equation \ref{sfr_calibration_halpha}, Section \ref{data_analysis_SFR}). To estimate the equipartition magnetic field strengths of these five galaxies, we have used the non-thermal radio maps at 0.33 and 1.4 GHz
from \cite{Basu2012MNRAS}. The correlations between equipartition magnetic fields and SFRSDs are shown in Figure \ref{B_SFRSD_corr} where, similar to the previous correlation studies, each point represents the logarithms of magnetic fields and SFRSD values that are averaged over the beam size. Similar to the previous correlations (Section \ref{results_B_SFRSD_corr}),  we used \emph{orthogonal distance regression} in Scipy to fit a power law to the data. We have provided the best-fit parameters of the power-law fit in Table \ref{B_SFRSD_corr_parameter_basu_sample}. The scatters of all five correlations (presented in Table \ref{B_SFRSD_corr_parameter_basu_sample}) are shown in  dashed lines in all the figures. We find a mean exponent of 0.30$\pm$0.05 for the five galaxies where the exponent of individual galaxies varies from $\approx$0.25 to $\approx$0.38. 

We have computed maps of cold gas densities of four out of the five galaxies; NGC 4736, NGC 5055, NGC 5236 and NGC 6946, using the atomic and molecular gas surface density maps from \citet{Basu2013MNRAS}. The assumed parameters are taken to be the same as described in Section \ref{gas_density_measurement}. 
 For the remaining galaxy, NGC 1097, we could not measure gas densities as there are no archival CO data available for the galaxy.
Following the procedures of Section~
\ref{results_B_SFRSD_corr}, we have also studied the spatially-resolved correlation between equipartition magnetic fields and gas densities for the four sample galaxies, which are shown in Figure \ref{B_gas_density_corr}. The best-fit parameters are presented in Table \ref{B_gas_corr_parameter}. The exponents of the individual galaxies vary between $\approx$0.25 to $\approx$0.44 where the mean exponent is found to be 0.35$\pm$0.07.

\begin{table}

 \caption{Best-fit parameters and the scatter of the correlation between magnetic fields and SFRSDs for the five galaxies in \cite{Basu2012MNRAS} (Sample 2). The data were fitted with a power law of the form B=B$_{0}(\Sigma_\textrm{SFR})^{\eta}$.}
\centering
 \begin{tabular}{||c c c c c||} 
 \hline
 Name & Slope ($\eta$) & Intercept (B$_{0}$) (log($\mu$G)) & Intercept (B$_{0}$) ($\mu$G) & Scatter \\ 
 \hline\hline
 NGC 1097 & 0.27 $\pm$ 0.01 & 1.61 $\pm$ 0.01 & 41 $\pm$ 1.0 & 0.02 \\
 NGC 4736 & 0.32 $\pm$ 0.02 & 1.78 $\pm$ 0.05 & 60 $\pm$ 1.1 & 0.04 \\
 NGC 5055 & 0.27 $\pm$ 0.04 & 1.26 $\pm$ 0.02 & 18 $\pm$ 1.0 & 0.02 \\
 NGC 5236 & 0.38 $\pm$ 0.07 & 1.91 $\pm$ 0.02 & 81 $\pm$ 1.0 & 0.08 \\  
 NGC 6946 & 0.25 $\pm$ 0.03 & 1.62 $\pm$ 0.05 & 42 $\pm$ 1.1 & 0.05 \\
 \hline
 \end{tabular}
 \label{B_SFRSD_corr_parameter_basu_sample}
\end{table}

\begin{table}

 \caption{Best-fit parameters and the scatter of the correlation between spatially-resolved magnetic fields and gas densities for the seven galaxies in Sample 3. Galaxies with an asterisk are from the sample of \cite{Basu2012MNRAS}.}
\centering
 \begin{tabular}{||c c c||} 
 \hline
 Name & Exponent & Scatter \\ 
 \hline\hline
 NGC 3627       & 0.40 $\pm$ 0.02  & 0.03 \\
 NGC 4826       & 0.49 $\pm$ 0.03  & 0.05 \\
 NGC 5194       & 0.53 $\pm$ 0.02  & 0.06 \\
 NGC 4736$^{*}$ & 0.44 $\pm$ 0.03  & 0.03 \\
 NGC 5055$^{*}$ & 0.25 $\pm$ 0.02  & 0.02 \\
 NGC 5236$^{*}$ & 0.40 $\pm$ 0.03  & 0.04 \\  
 NGC 6946$^{*}$ & 0.31 $\pm$ 0.03  & 0.04 \\
 \hline
 \end{tabular}
 \label{B_gas_corr_parameter}
 
\end{table}

\section{Discussion}
\label{discussion}
Understanding the relationship between the physical condition of the interstellar medium (ISM) and the star formation process is crucial to understand galaxy evolution. Gas and magnetic fields are key constituents of the ISM and therefore it is important to study the interrelations between gas, magnetic fields, and SFRs. Though the Kennicutt$-$Schmidt relation, i.e. the relation between gas densities and SFRs, has been extensively studied at high spatial resolutions in various types of nearby galaxies \citep[e.g.][]{Onodera2010ApJ,Roychowdhury2015MNRAS,Filho2016ApJ,Miettinen2017A&A}, similar high-resolution observations of how the magnetic fields are related to SFRs and gas densities are yet to be systematically investigated. Such observations are critical to understand the validity of several models that predict strong correlations between the magnetic fields and gas densities \citep[e.g.][]{Chandrasekhar1953,Fiedler1993ApJ,Cho2000ApJ,Groves2003PASA} as well as magnetic fields and SFRSDs \citep[e.g.][]{Niklas1997A&A,Schleicher2013A&A,Schleicher2016A&A}. 
Here, 
 we have studied these correlations in a sample of twelve galaxies (Sample 3) at sub-kpc scales (see Sections \ref{results_B_SFRSD_corr}, \ref{results_B_gas-corr} \& \ref{study_basu_2013}). To our knowledge, this is the first spatially resolved study of the above correlations in nearby large galaxies.
In this section, we place these findings in the light of predictions made by various models and in the process attempt to provide physical insights into the interrelation between magnetic fields, gas densities, and star formation rates at sub-kpc scales.

\subsection{Magnetic Fields and SFRSDs}

Several Magneto-Hydrodynamical simulations find that  galactic magnetic fields are amplified by gas turbulence in very short timescales (i.e. $\sim$100 Myr) \citep[e.g.][]{Brandenburg2005PhR,Beresnyak2012PhRvL,Schober2012ApJ,Schleicher2013A&A,Bovino2013NJPh}. The primary driver of gas-turbulence in the ISM of galaxies is supernova explosion \citep{Bacchini2020A&A}, the rate of which is in turn directly coupled to the SFR in the galaxy. Therefore, it is expected that the star formation rates and the magnetic fields in a galaxy will be correlated. Indeed, using semi-analytical models, \cite{Schleicher2013A&A,Schleicher2016A&A} found that
in order to explain the radio-FIR correlation at sub-kpc scales, magnetic fields and SFRSDs, again at sub-kpc scales, must be related as B $\propto \Sigma_\textrm{SFR}^{1/3}$.

Studies in the literature on the correlation between magnetic fields and SFRSDs have focused on dwarf galaxies and those studies were carried out using galaxy-integrated magnetic fields and SFRSDs.
As mentioned in Section \ref{introduction}, to our knowledge, there is only one published work of the spatially-resolved study of the correlation between magnetic fields and SFRSDs \citep{Basu2017MNRAS}.

For the 12 galaxies in Sample 2 (Table \ref{different_samples_studied}), we find that the mean value of the power-law index of the correlation between B$_\textrm{eq}$ and SFRSDs is  0.31$\pm$0.06, i.e B$_\textrm{eq}$ $\propto$ $\Sigma_\textrm{SFR}^{0.31\pm0.06}$(\footnote{The uncertainty quoted is the scatter of the measured value of $\eta$ across the galaxies in Sample 2.}), consistent (at $<$ 1$\sigma$ error) with the model of \cite{Schleicher2013A&A,Schleicher2016A&A}. Thus, it appears that the semi-analytical models that are based on the amplification of magnetic fields due to supernova-driven gas turbulence work remarkably well for the pilot sample, in predicting the correlation between magnetic fields and SFRSDs down to sub-kpc scales. 

We note that the power-law index for the correlation between B$_\textrm{eq}$ and SFRSDs for NGC 4449 was
found to be $0.18\pm0.03$, significantly lower than for the remaining galaxies (Table \ref{B_SFRSD_corr_parameter}) as well as lower than the model prediction of B $\propto \Sigma_\textrm{SFR}^{1/3}$ \cite{Schleicher2013A&A} (at $>5\sigma$ significance).
For the case of NGC 4449, the relatively flat spectral index values ($\alpha_{nt} \le 0.55$) in $\approx 70\%$ of the galaxy meant that the magnetic field values could not be estimated reliably for a large part of the galaxy (see Section \ref{data_analysis_magnetic_field} and \ref{results_B_SFRSD_corr}).
This could lead to biases in the correlation and therefore, the low value of the power-law index for NGC 4449 should be taken with caution.


%
 
\subsubsection{Intercept of the Correlation}
According to the model proposed by \cite{Schleicher2013A&A}, the intercept of the B-$\Sigma_\textrm{SFR}$ correlation depends on several ISM parameters such as gas density ($\rho_{0}$), the fraction of turbulent kinetic energy converted into magnetic energy ($f_{sat}$), the injection rate of turbulent supernova energy (C) and the intercept of Kennicutt-Schmidt (KS) relation ($C_{1}$) (Equation 5).
\begin{equation}
\textrm{B}\sim \sqrt{f_{sat}8\pi} \; \rho_{0}^{1/6} \; (\frac{C}{C_{1}})^{1/3}\; \Sigma_\textrm{SFR}^{1/3}.
\label{B_SFRSD_corr_eqn} 
\end{equation}

\cite{Schleicher2013A&A} predicted the intercept of the B-$\Sigma_\textrm{SFR}$ correlation to be $\sim$ 26 $\mu$G assuming $\rho_{0}$ = 10$^{-24}$ g cm$^{-3}$ and $f_{sat}$ $\sim$ 5 percent.
 We have found an average intercept at 65$\pm$25 $\mu$G of the  B$_\textrm{eq}$-$\Sigma_\textrm{SFR}$ correlation of the 12 galaxies in sample 2 (see Table \ref{B_SFRSD_corr_parameter} \& \ref{B_SFRSD_corr_parameter_basu_sample}). Although the mean value is a factor of $\approx$2.5 higher than the value predicted by \cite{Schleicher2013A&A}, this value is consistent with the predicted value, within the scatter (at $\approx$1.6$\sigma$).
 Future follow-up studies, such as using our full survey (Sample 0 which consists of 46 galaxies), are required to draw statistically robust conclusions about the value of the intercept.
 
 If the value of $f_{sat}$ is indeed higher, this would imply a higher than assumed value of one or more of $\rho_{0}$, C, and $f_{sat}$. 
 The intercept is broadly insensitive to the assumed value of $\rho_{0}$ (Equation \ref{B_SFRSD_corr_eqn}) and therefore, in order to explain a factor of $\approx2.5$ higher value of the intercept, the actual value $\rho_{0}$ has to be higher than the assumed value of $10^{-24}$ g cm$^{-3}$ by a factor of $\approx$ 240; such high gas densities are unphysical and are not observed in typical regions of a galaxy. 
 The other possibility that the assumed value of the injection rate of turbulent supernova energy (C) is higher by a factor of $\approx16$ is also contrary to expectation; \citet{Basu2017MNRAS} found that under reasonable conditions the value of C can be higher by at most a factor of 1.4.
 Therefore, $f_{sat}$ must be higher than 0.05 to explain a significantly higher value of the intercept.
 An understanding of how galaxies can achieve such efficient amplification of magnetic fields with $f_{sat}$ much greater than 5\% requires detailed MHD simulations. We note that \cite{Basu2017MNRAS} found that the value of the intercept for B-$\Sigma_\textrm{SFR}$ for the dwarf galaxy IC 10 is 51 $\mu$G, similar to our findings of a higher than predicted value of the intercept.

\subsection{Magnetic Fields and Gas}

Magnetic fields and gas are expected to be correlated as B $\propto \sqrt{\rho_\textrm{gas}}$ \citep[e.g.][]{Chandrasekhar1953, Groves2003PASA}.
We find that equipartition magnetic fields are correlated with gas densities for the seven galaxies (Sample 3) with an average power-law index, $k$=0.40$\pm$0.09  (see Section \ref{results_B_gas-corr} \& \ref{study_basu_2013})\footnote{The uncertainty quoted is the scatter of the measured value of k across the galaxies in Sample 3.}. 
This value of $k$ is 
consistent with the numerical simulations that predict $k$ $\approx$ 0.4$-$0.6 and also consistent with the theories that predict B $\propto \rho_\textrm{gas}^\textrm{0.5}$.
The power-law index of the correlation between B$_\textrm{eq}$ and gas densities is found to be 0.25$\pm$0.02 and 0.31$\pm$0.03 for NGC 5055 and NGC 6946 respectively, significantly lower than the model predictions and as compared to the other galaxies
in Sample 3.
A lower value of $k$ could mean that either the efficiency of the amplification of the magnetic field is less or that the magnetic field strengths derived assuming the $``$minimum energy condition$"$ are underestimated \citep{Dumas2011AJ}. Strong synchrotron or inverse Compton losses of cosmic-ray electrons could suppress the radio synchrotron emission which would then cause the equipartition magnetic fields to be underestimated.

\subsubsection{Magnetic Fields, Gas Densities and the Radio-FIR Correlations}
\label{k_value_indirect}
Energy equipartition between the magnetic field (B) and the gas density ($\rho_\textrm{gas}$), and between magnetic fields and cosmic ray particles implies that the non-thermal emission is related to the gas density as $\textrm{I}_{nt} \propto$ $\rho_\textrm{gas}^{k(3+\alpha_{nt})}$ where k is the power-law index relating magnetic fields and gas densities ($\textrm{B}_\textrm{eq} \propto$ $\rho_\textrm{gas}^{k}$) \citep{Niklas1997A&A}. Further, the Kennicutt-Schmidt law and the radio-FIR correlation imply that $\textrm{I}_{nt}$ is related to gas densities as (1) $\textrm{I}_{nt} \propto$ $\rho_\textrm{gas}^{m(n+1)}$ for optically thin dust to UV photons and (2) $\textrm{I}_{nt} \propto$ $\rho_\textrm{gas}^{mn}$ for optically thick dust to UV photons, where m is the power-law index of the radio-FIR correlation and n is the power-law index of the Kennicutt-Schmidt law.  Therefore, we can obtain the following relation between the power-law index of all four correlations \citep{Dumas2011AJ}:

\begin{equation}
k=\frac{(n+1) m}{3+ \alpha_{nt}} ;\textrm{Optically thin dust}
\label{B_gas_indirect_eqn_thin}
\end{equation}

\begin{equation}
k=\frac{nm}{3+ \alpha_{nt}} ;\textrm{Optically thick dust}
\label{B_gas_indirect_eqn_thick}
\end{equation}

We can use the above equations to indirectly estimate the power-law index, k, of the correlation between magnetic fields and gas densities. 
For the three galaxies, NGC 3627, NGC 4826, and NGC 5194 \citep{RoyManna2021}, we have estimated gas densities using CO and \hi\ observations. Now we can compare the direct measurement of k with an indirect estimate of k using Equations \ref{B_gas_indirect_eqn_thin} and \ref{B_gas_indirect_eqn_thick}; this will provide additional information on the validity of both the minimum energy conditions that were assumed between magnetic fields and the gas densities as well as the magnetic fields and cosmic ray particles. For the galaxies from \cite{Basu2012MNRAS}, this study was already presented and discussed in \cite{Basu2012ApJ}.

We have estimated $k$ for all the seven sample galaxies from \cite{RoyManna2021} (Sample 1), using the assumption of optically thin dust to UV photons, using (i) the slope of radio-FIR correlation (m) as derived in \cite{RoyManna2021}, (ii) the measured galaxy-averaged spectral index ($\alpha_{nt}$) from \cite{RoyManna2021}, and (iii) a Kennicutt-Schmidt power-law index of 1.4$\pm$0.15 \citep[][]{Kennicutt1998ARA&A}. 

Table \ref{B_gas_indirect_table} provides the relevant values as well as estimated values of $k$ derived using the measured value of m using radio emission at both 0.33 and 1.4 GHz.
For two of the galaxies, NGC 3627 \& NGC 5194, the value of k estimated using Equation \ref{B_gas_indirect_eqn_thin} is comparable to the direct measurement of k. This broadly validates the assumption of energy equipartition between magnetic fields and cosmic ray particles in these two galaxies.

For the optically thin case, the mean of indirectly-estimated $k$ values of the sample of seven galaxies are $0.59\pm0.16$ and $0.53\pm0.19$ at 1.4 and 0.33 GHz, respectively. However, this includes the galaxy NGC 4826, which shows an anomalously high value of k=1.0 and 0.95 derived at 1.4 and 0.33 GHz, respectively. Excluding this galaxy from the mean calculation, we find that $k$=0.52$\pm$0.04 and 0.47$\pm$0.09 at 1.4 and 0.33 GHz, respectively. Remarkably, for all the galaxies except NGC 4826, the $k$ value at 1.4 GHz, for the optically thin case, is consistent with 0.5 within error bars. Thus, the indirectly estimated values of $k$ are consistent with equipartition between magnetic fields and gas energy densities \citep{Chandrasekhar1953,Fiedler1993ApJ,Cho2000ApJ,Groves2003PASA}. This is similar to the findings of \cite{Niklas1997A&A} for their sample of 43 galaxies and \cite{Basu2012ApJ} for their sample of four galaxies.

 

The value of k derived for NGC 4826, for the optically thin case, is a consequence of the anomalously high value of the power-law index of the radio-FIR correlation ($\approx$1.39 and $\approx$1.47 for 0.33 and 1.4 GHz respectively, Table \ref{B_gas_indirect_table}) which is different from the other six galaxies in the sample. NGC 4826 has been classified as a Seyfert 2 galaxy in the past \citep{Malkan2017ApJ} and therefore the emission from the core contributes to the observed power-law index of the radio-FIR correlation \citep{RoyManna2021}. It is likely that the significant contribution of the AGN to the radio emission makes the estimate of $k$ for NGC 4826 unreliable. 

\begin{table}
 \caption{Power law index ($k$) of the relation between magnetic fields and gas densities (B $\propto \rho^{k})$ of galaxies in Sample 1, indirectly estimated using the slope of radio-FIR correlation (m) and the slope of the Kennicutt$-$Schmidt law. See Section \ref{k_value_indirect} for a discussion on these.}
\scriptsize
\centering
 \begin{tabular}{||c c c c c c ||} 
 \hline
 Name & m        & m        & $\alpha_{nt}$ & k (Optically thin)   & k (Optically thin)    \\ 
      & 0.33~GHz & 1.4~GHz  &               & 0.33~GHz & 1.4~GHz   \\
 
 \hline\hline
 NGC 2683 & 0.54$\pm$0.06 & 0.91$\pm$0.07 & -0.84$\pm$0.08  & 0.33 $\pm$ 0.04 & 0.57 $\pm$ 0.06   \\
 NGC 3627 & 0.55$\pm$0.03 & 0.85$\pm$0.13 & -1.10$\pm$0.07 & 0.32 $\pm$ 0.03  & 0.50 $\pm$ 0.08   \\
 NGC 4096 & 0.74$\pm$0.05 & 0.90$\pm$0.04 & -0.78$\pm$0.06  & 0.47 $\pm$ 0.04 & 0.57 $\pm$ 0.05   \\
 NGC 4449 & 0.77$\pm$0.05 & 0.65$\pm$0.04 & -0.48$\pm$0.06  & 0.53 $\pm$ 0.05 & 0.45 $\pm$ 0.04  \\  
 NGC 4490 & 0.68$\pm$0.02 & 0.75$\pm$0.02 & -0.59$\pm$0.07  & 0.45 $\pm$ 0.03 & 0.50 $\pm$ 0.04  \\
 NGC 4826 & 1.39$\pm$0.1  & 1.47$\pm$0.08 & -0.49$\pm$0.06 & 0.95 $\pm$ 0.09  & 1.00 $\pm$ 0.09   \\
 NGC 5194(arm)       & 0.50$\pm$0.05 & 0.65$\pm$0.04 & -0.63$\pm$0.05  & 0.33$\pm$ 0.04 & 0.43 $\pm$ 0.04   \\
  NGC 5194(interarm) & 0.73$\pm$0.11 & 1.03$\pm$0.05 & -0.85$\pm$0.10 & 0.46$\pm$ 0.08  & 0.64 $\pm$ 0.05   \\
 \hline
 \end{tabular}
 \label{B_gas_indirect_table}

\end{table}

\section{Summary}
\label{summary}
\begin{enumerate}
	\item We made spatially resolved maps of equipartition magnetic fields in seven galaxies (Sample 1): NGC 2683, NGC 3627, NGC 4096, NGC 4449, NGC 4490, NGC 4826, and NGC 5194 and find that the magnetic fields are strongest near the central region and go down by a factor of $\sim$2 at the edge of the magnetic field maps.
	\item We have used the tightness of the spatially-resolved radio-FIR correlations to verify the validity of the equipartition condition between magnetic fields and cosmic ray particles for the sample galaxies. We find that the magnetic field values may deviate from the equipartition values by $\sim$25\%.
	\item We have estimated spatially resolved maps of SFRSDs of the galaxies in Sample 1 using FUV+24$\mu$m, H$\alpha$+24$\mu$m, and 1.4 GHz data. Azimuthally averaged SFRSDs drop by a factor of 6 to 8 at the edge of the galaxies,  where SFRSD values are 5 times the rms of the maps.
	\item We also included five additional galaxies: NGC 1097, NGC 4736, NGC 5055, NGC 5236, and NGC 6946 from previous GMRT observations of \cite{Basu2012MNRAS} and estimated their equipartition magnetic field, SFRSD and gas density maps. 
	\item We studied the spatial correlation between magnetic fields and star formation rates at $<$ 1 kpc resolution for the 12 galaxies (Sample 2) and find that magnetic field strengths and SFRSDs are correlated  with an average power-law index of 0.31$\pm$0.06. This result is in remarkable agreement (at $<$ 1$\sigma$ error) with semi-analytical model predictions of $\textrm{B} \propto \Sigma_\textrm{SFR}^{1/3}$  \citep{Schleicher2013A&A,Schleicher2016A&A}.
	\item We measure an average intercept of $\approx65$ $\mu$G from the B-$\Sigma_\textrm{SFR}$ correlations of our galaxies in Sample 2. This is higher than the predictions of \cite{Schleicher2013A&A} by a factor of $\approx2.5$, and, if confirmed with a larger sample, would imply a significantly 
     higher efficiency of magnetic field amplification than what is typically assumed.
	
	\item We used spatially resolved gas density maps for seven (Sample 3) of the 12 galaxies, for which archival CO data was available, to find that magnetic fields are correlated with gas densities as B $\propto \rho_\textrm{gas}^{0.40\pm0.09}$. This result is consistent with numerical simulations that predict $k$ $\approx$ 0.4$-$0.6 and broadly consistent (within $\approx$1 sigma uncertainty)
	with theories that predict B $\propto \rho_\textrm{gas}^\textrm{0.5}$.
	\item We have indirectly estimated the power-law index ($k$) of the correlation between the magnetic fields and the gas densities using the slope of the radio-FIR correlation, the slope of the Kennicutt-Schmidt law, and the non-thermal spectral index. The mean value of $k$, for optically thin dust, was found to be 0.52$\pm$0.04 and 0.47$\pm$0.09 at 1.4 and 0.33 GHz respectively for the six galaxies in Sample 1, with NGC 4826 excluded due to its high value of $k$. This is consistent with the equipartition between magnetic fields and gas. 
The anomalously high values of $k$ (1.0 and 0.95 at 1.4 and 0.33 GHz respectively) for NGC 4826 are possibly due to the contribution of the central AGN to the radio emission.  
\end{enumerate}

 We have started to follow up these pilot study results with a survey of
 a much larger sample of galaxies (Sample 0, Table \ref{different_samples_studied}). For this, we have already observed another 24 galaxies using the upgraded GMRT (uGMRT), a Square Kilometer Array (SKA) pathfinder facility. Sensitivities of the images from these uGMRT observations are significantly better ($\approx3$ times) than those of the observations presented here and the result will be part of a future publication.
In addition, SKA precursors such as the MeerKAT will also provide very deep images of the diffuse radio-continuum emission around nearby galaxies. Eventually, the dramatic increase in sensitivity and $\sim$arc-sec resolution of the SKA has the potential to significantly advance our understanding of magnetic fields in nearby galaxies. For example, the SKA is expected to provide sensitive images of polarised synchrotron emission from nearby galaxies at a few GHz frequencies which would provide information on the large-scale ordered fields on the plane of the sky \citep[e.g.][]{Johnston2015aska}.
Further, polarised emission from nearby galaxies at $<\sim$1 GHz, where significant depolarisations take place, could be modelled through Faraday tomography 
\citep[e.g.][]{Heald2015aska}. A combination of the two approaches could eventually allow us to infer the three-dimensional structure of the magnetic fields in nearby galaxies. 
SKA observations will also provide detailed images of star formation with resolutions of tens of parsecs. These will help to identify any dependence of SFR and IMF on galaxy type, evolution and environment within the local volume \citep{Beswick2015aska}.



\begin{acknowledgments}
We would like to thank Aditya Chowdhury for his help at various stages of this research. We thank Yogesh Wadadekar, Preeti Kharb, and Dipanjan Mitra for reading the manuscript and providing useful comments. Aritra Basu provided their earlier published images and also suggested checking the B vs SFRSD relation for our sample galaxies. We thank him for the above. We also thank the anonymous referee whose comments helped significantly improve the presentation of the paper.
We thank the staff of GMRT that allowed these observations to be made. GMRT is run by National Centre for Radio Astrophysics of the Tata Institute of
fundamental research. We acknowledge the support of the Department of Atomic Energy, Government of India, under project no. 12-R$\&$D-TFR-5.02-0700.
\end{acknowledgments}

\bibliography{galaxyrefs.bib}{}
\bibliographystyle{aasjournal}


\appendix
\section{Magnetic Field Uncertainty Maps}
\begin{figure*}
   	\centering
 \includegraphics[width=0.31\linewidth]{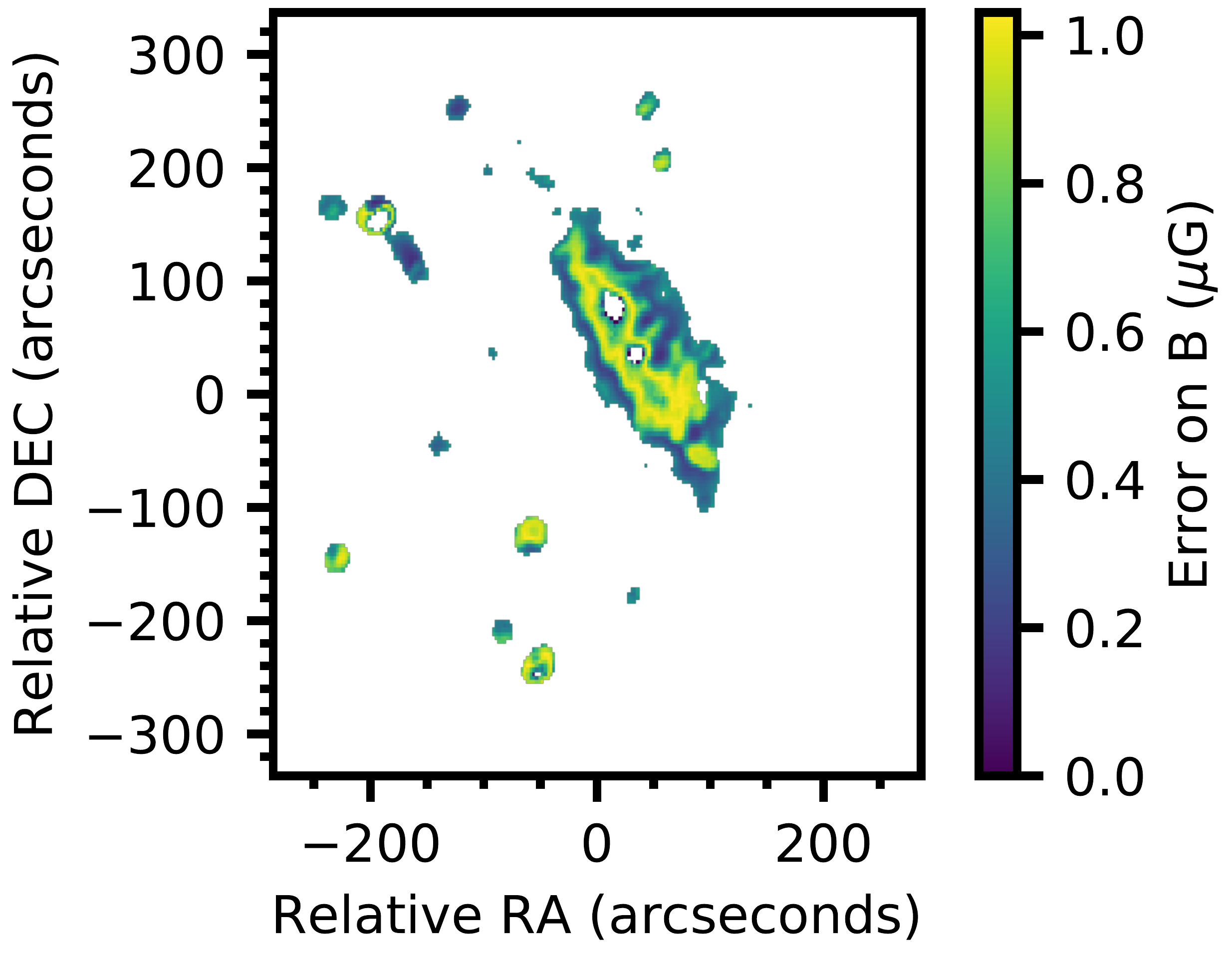}
 \includegraphics[width=0.33\linewidth]{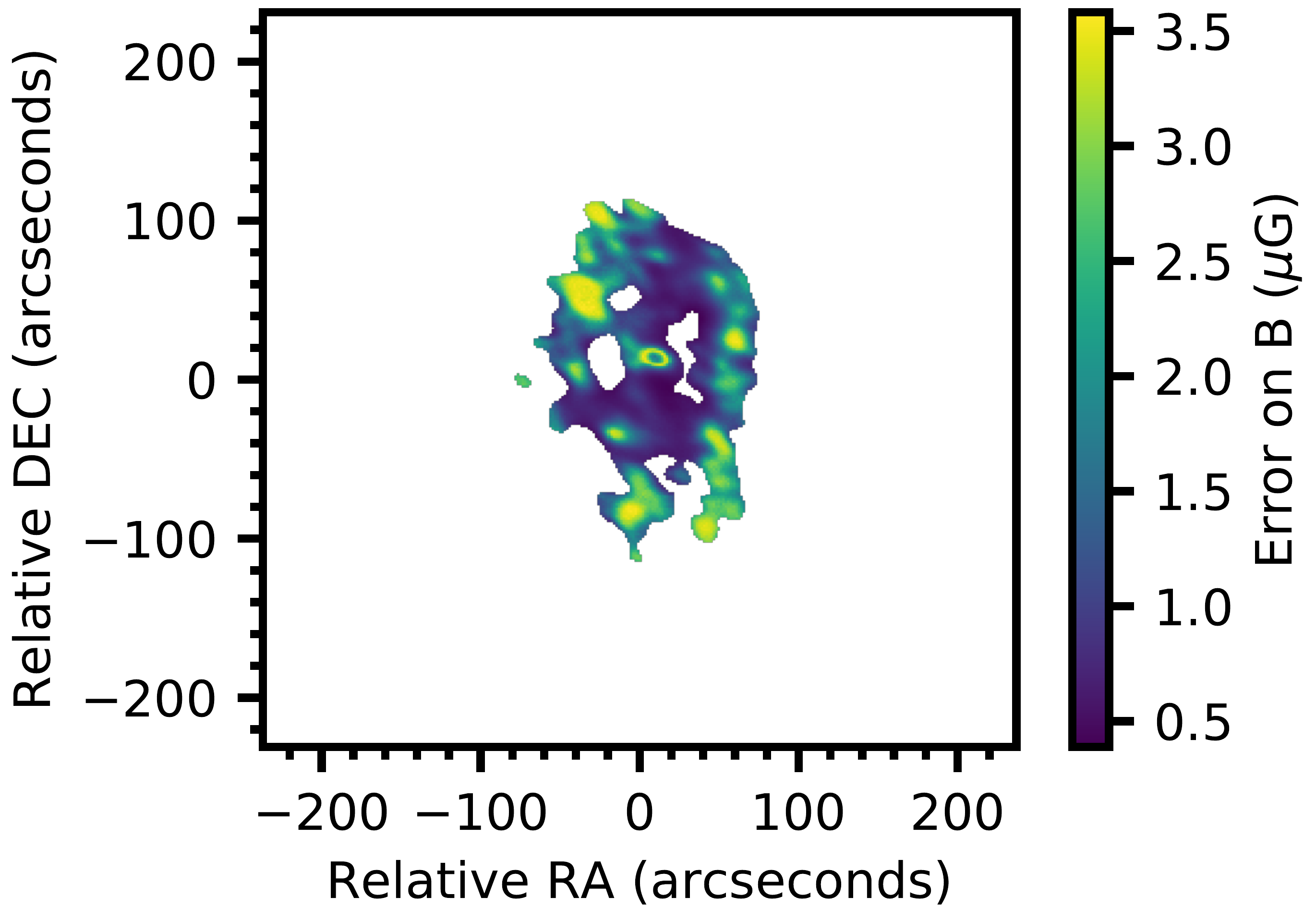}
 \includegraphics[width=0.33\linewidth]{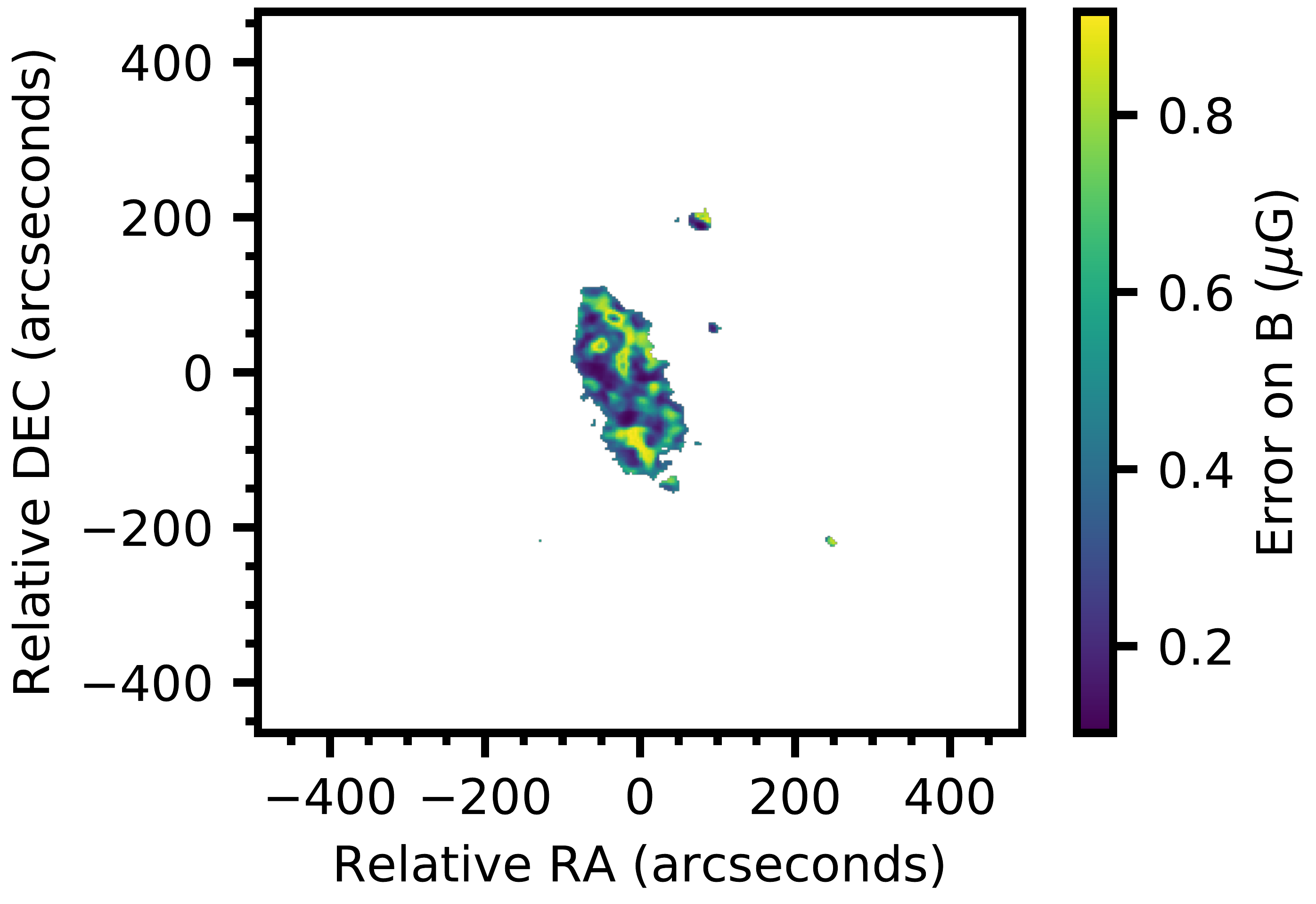}
 \includegraphics[width=0.33\linewidth]{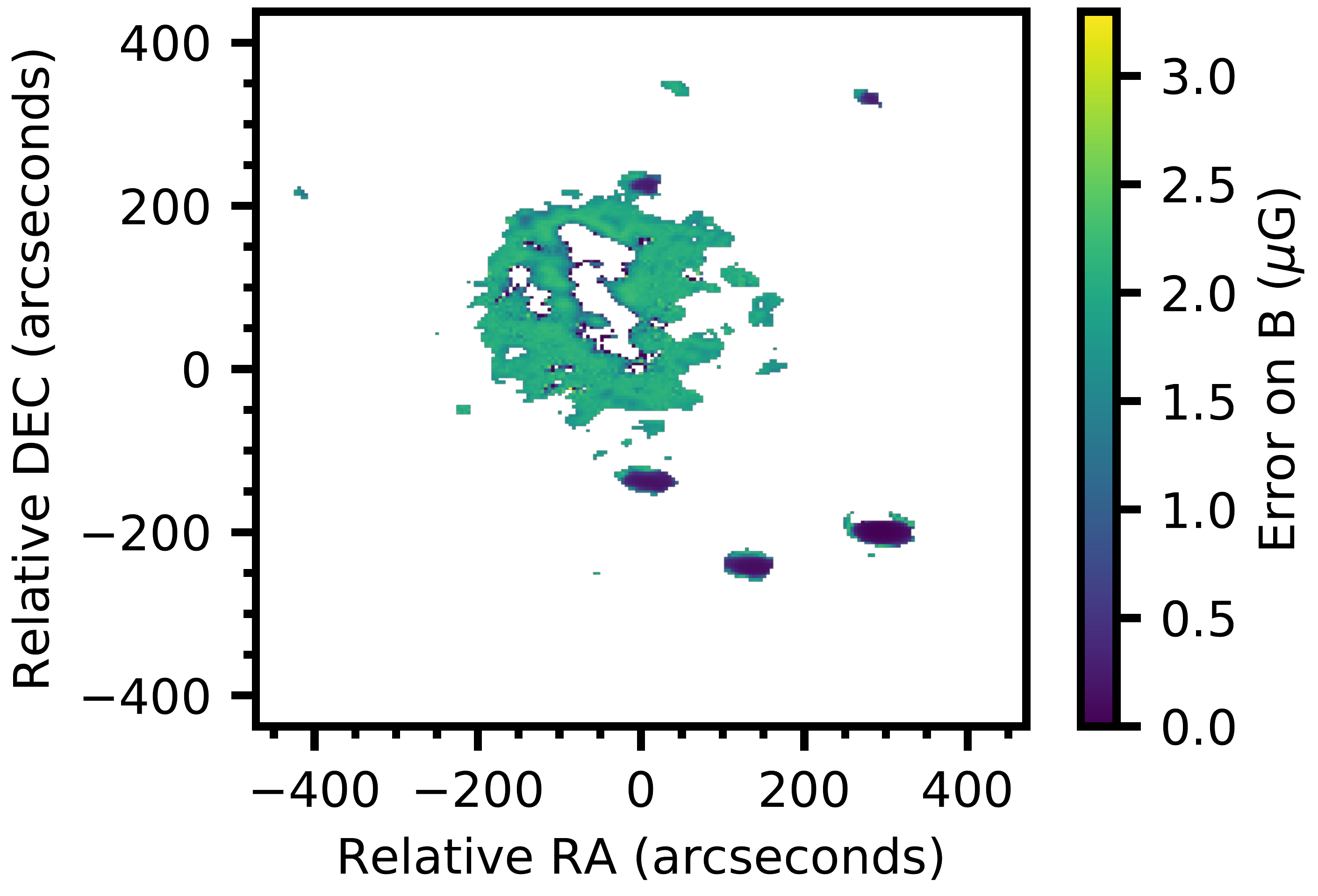}
 \includegraphics[width=0.32\linewidth]{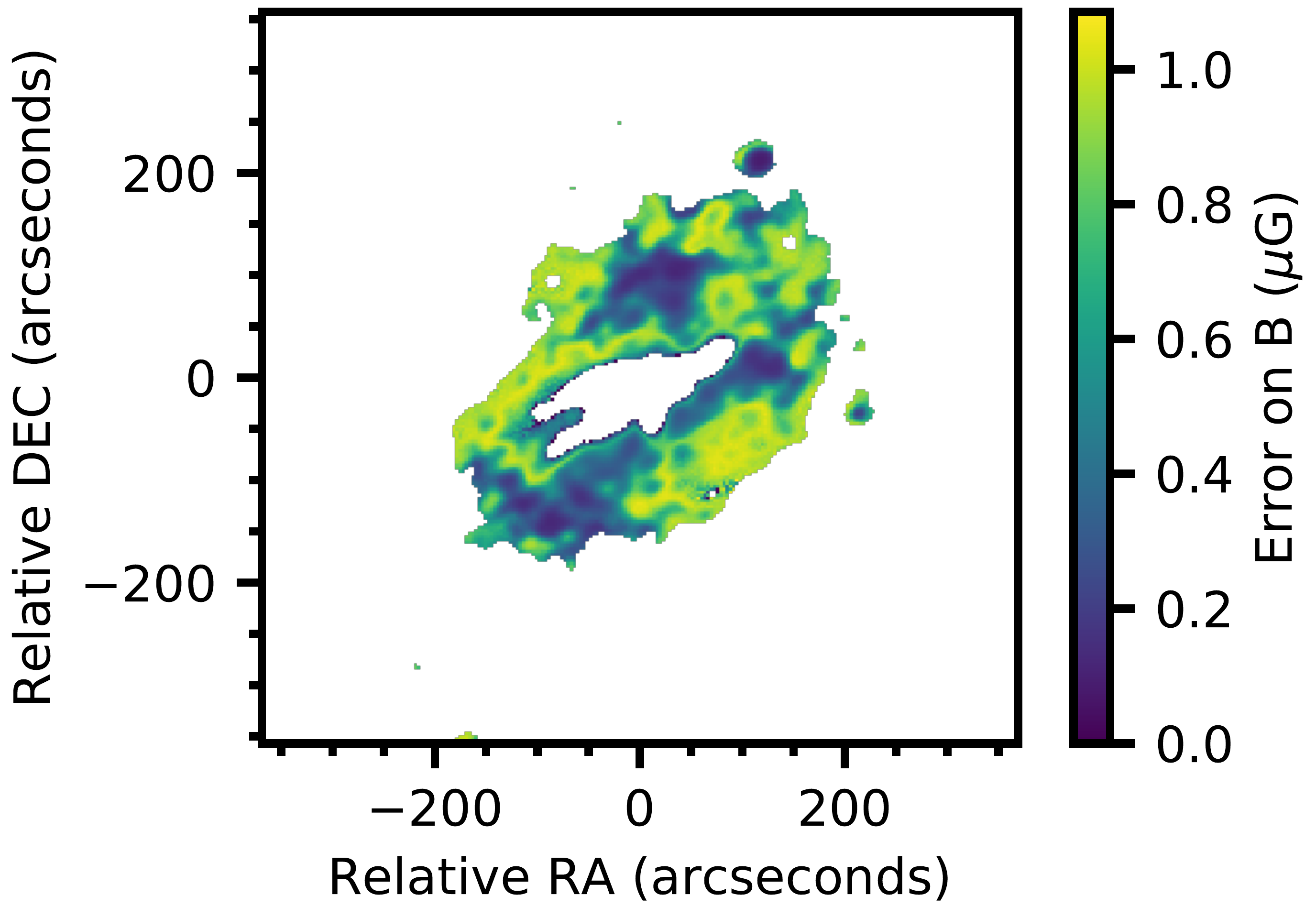}
 \includegraphics[width=0.31\linewidth]{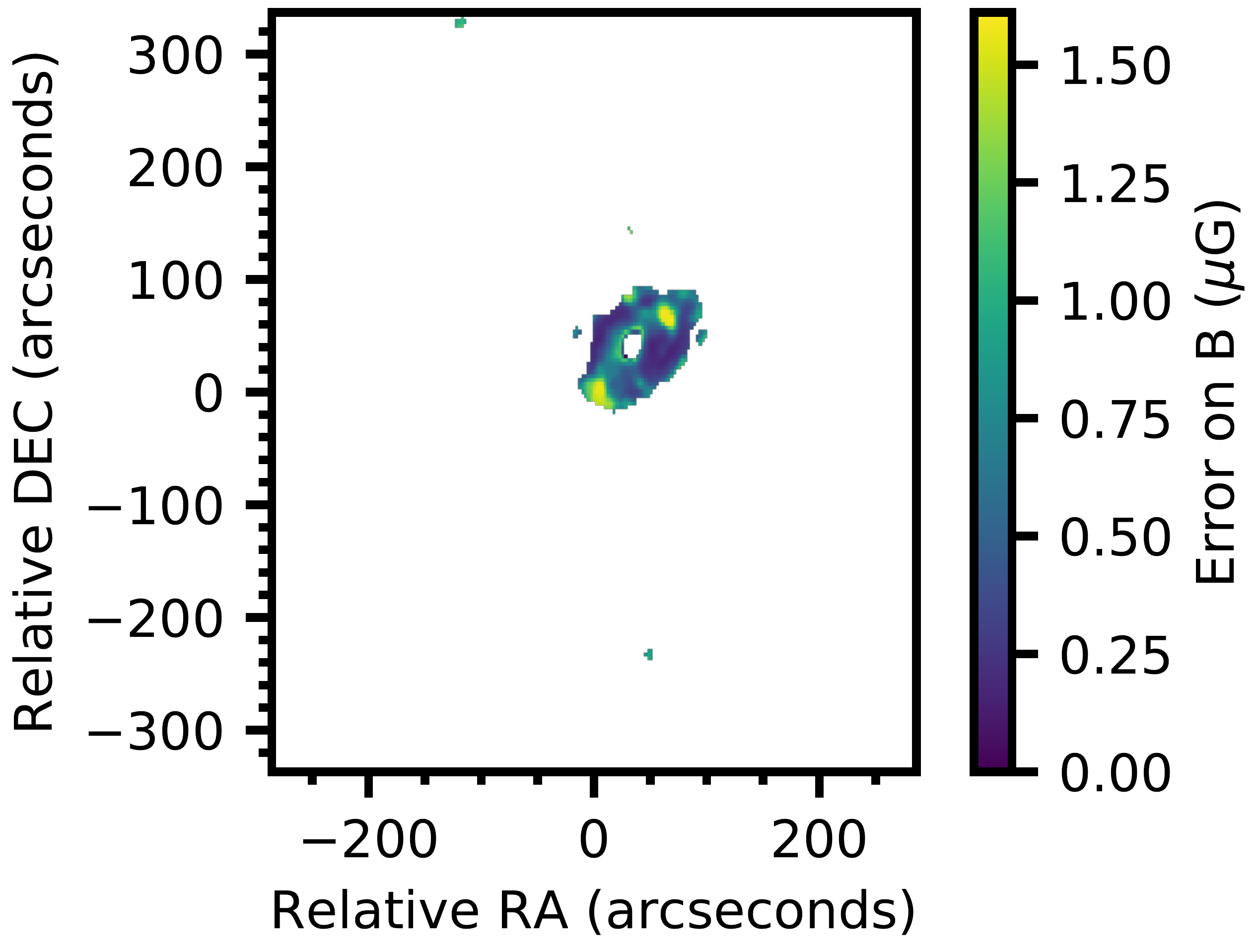}
 \includegraphics[width=0.32\linewidth]{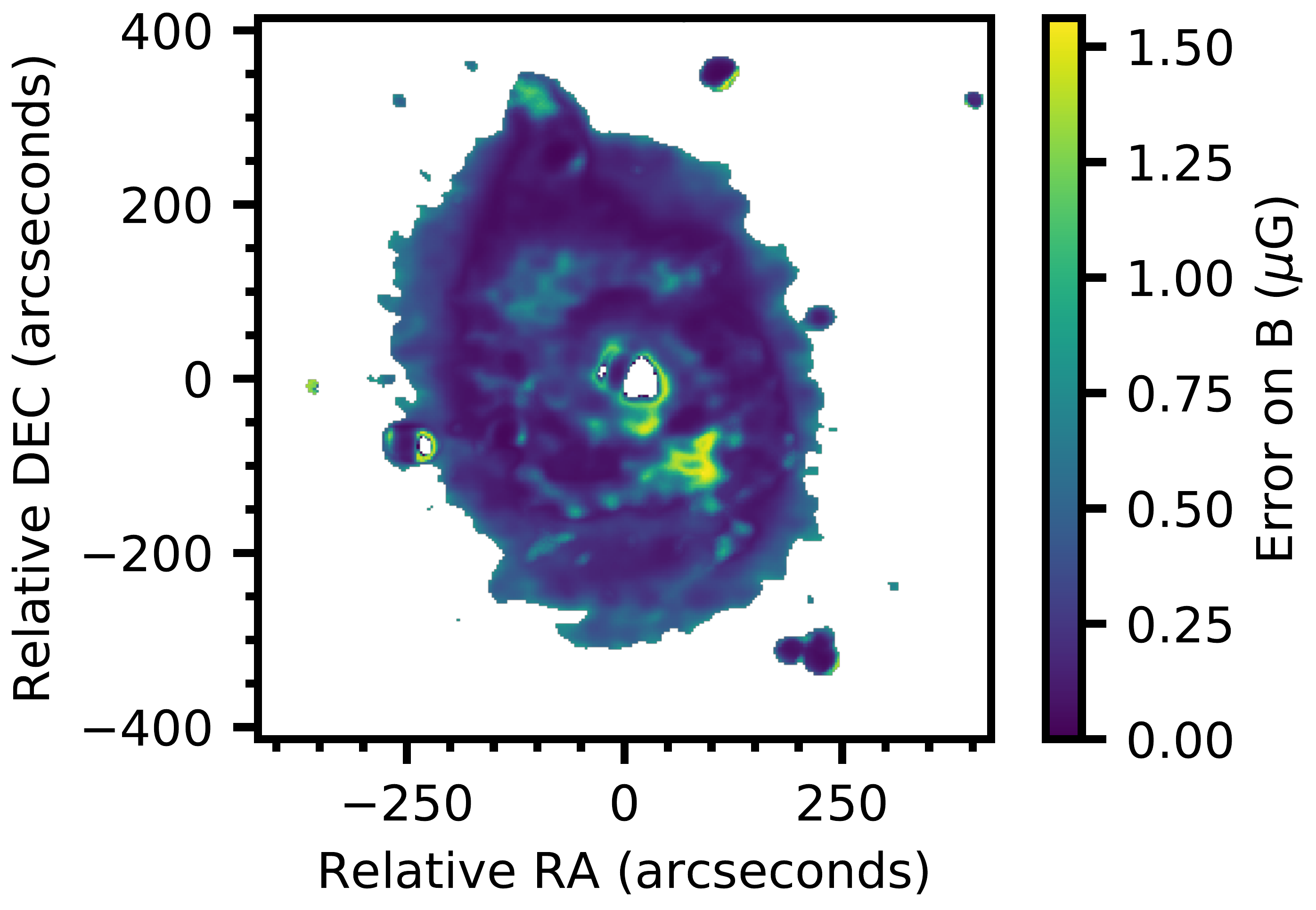}

\caption{The magnetic field uncertainty maps (in $\mu$G) of NGC 2683 (top left), NGC 3627 (top centre), NGC 4096 (top right), NGC 4449 (middle left), NGC 4490 (middle centre), NGC 4826 (middle right) and NGC 5194 (bottom) (Sample 1), shown in colour scale. Blanked regions (in white colour) in the centre of each galaxy correspond to regions with spectral index values $\leq$ 0.55.}
\label{B_map_err_1}
  \end{figure*}

We present here (Figure \ref{B_map_err_1}) magnetic field uncertainty maps of the galaxies in Sample 1, generated using the procedure described in Section~\ref{Error_Magnetic_Field}.

\section{Star Formation Rate Surface Density maps}
\begin{figure*}
 	\centering
 \includegraphics[width=0.45\linewidth]{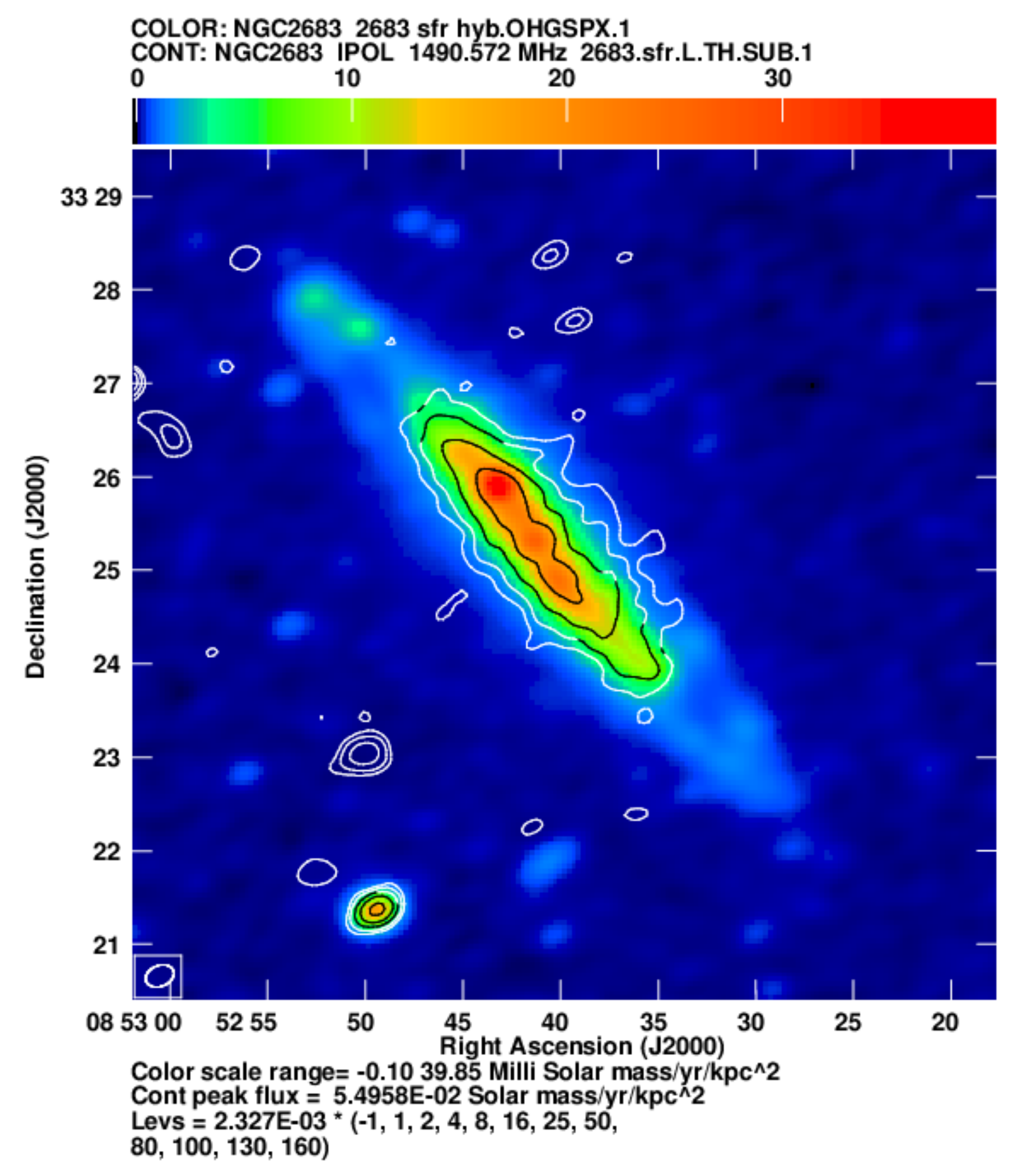}
 \includegraphics[width=0.45\linewidth]{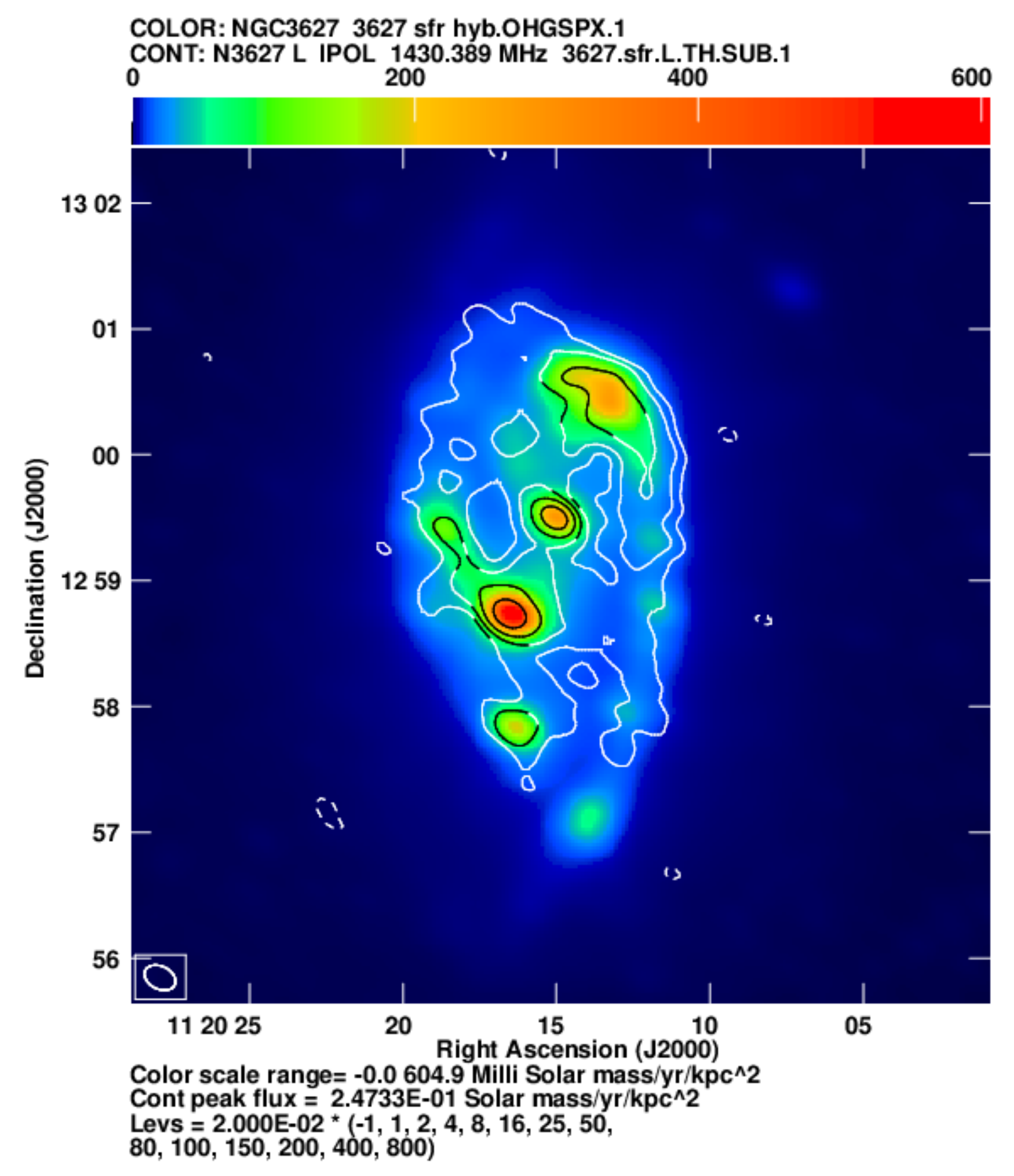}
 \includegraphics[width=0.45\linewidth]{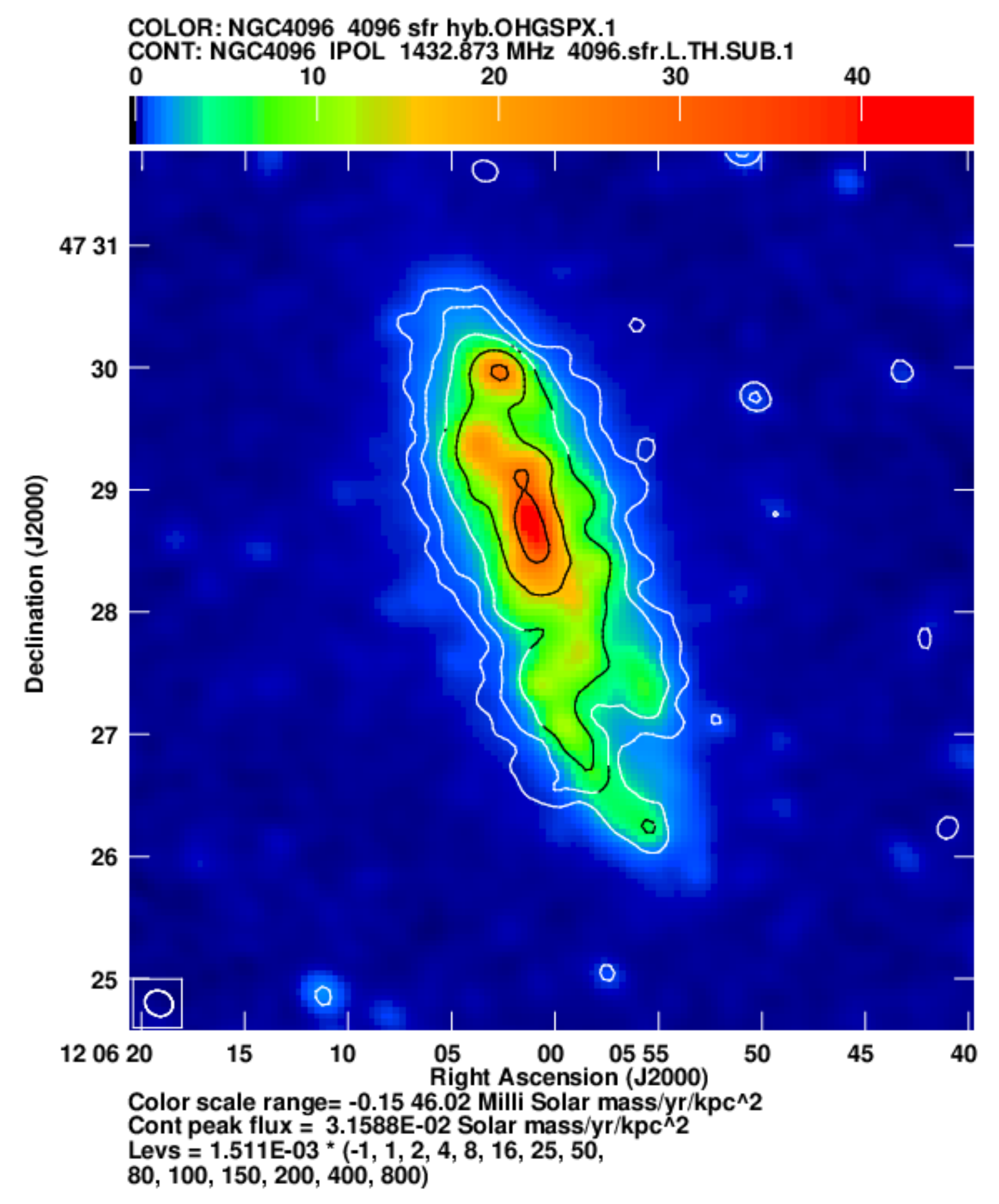}
 \includegraphics[width=0.45\linewidth]{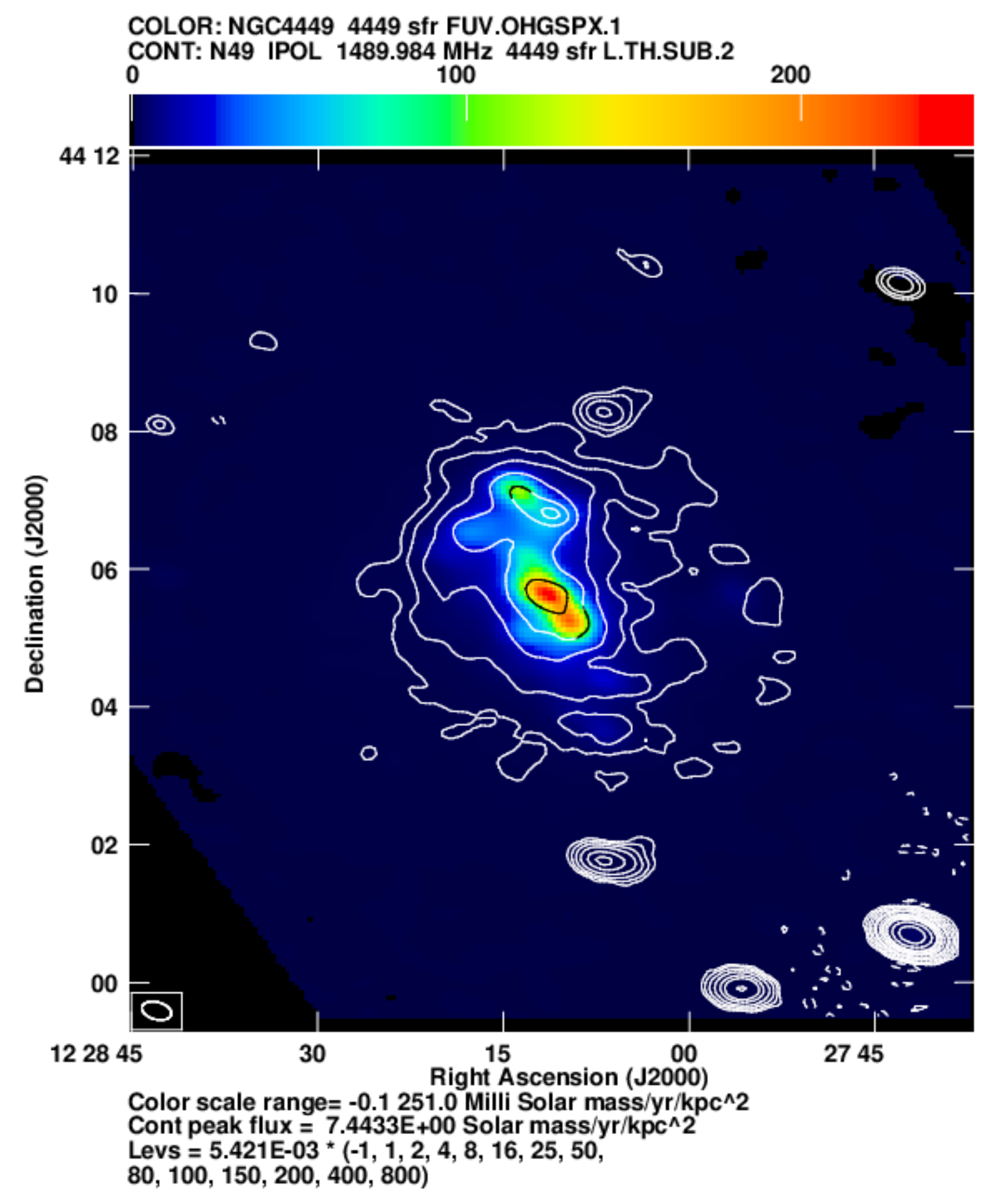}
 \caption{SFRSD (M$_\odot$yr$^{-1}$kpc$^{-2}$) maps of NGC 2683, NGC 3627, NGC 4449 and NGC 4096(clockwise from top left) (Sample 1). SFRSDs estimated using 1.4 GHz radio and FUV+24$\mu$m emission are shown in contours and colors, respectively. Contour levels are listed below each panel of the figure. The circle in the bottom-left corner of the images indicates the angular resolution of the maps.}
    \label{SFRSD_cont_1}
  \end{figure*}

\begin{figure*}
  	\centering
 \includegraphics[width=0.45\linewidth]{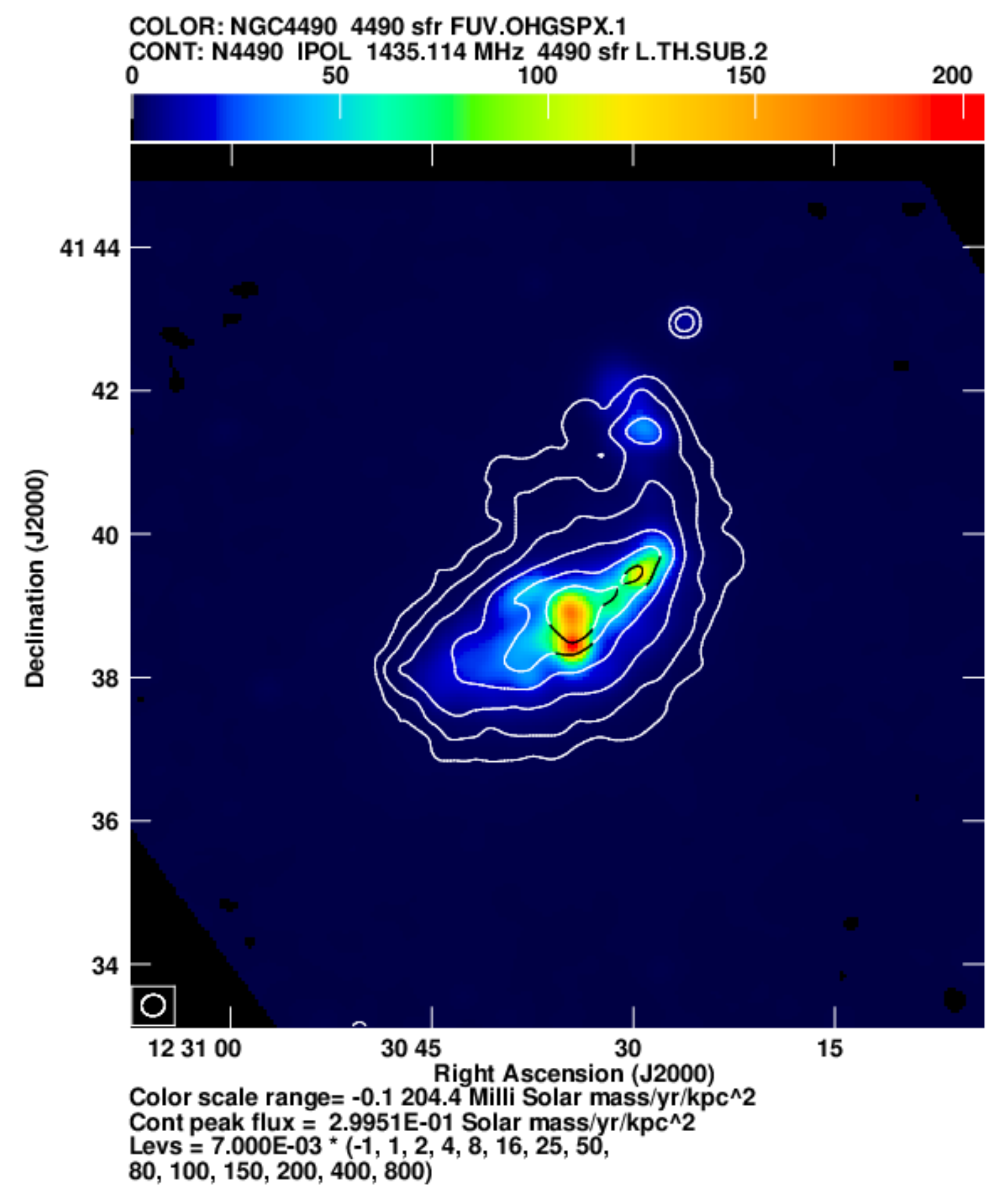}
 \includegraphics[width=0.45\linewidth]{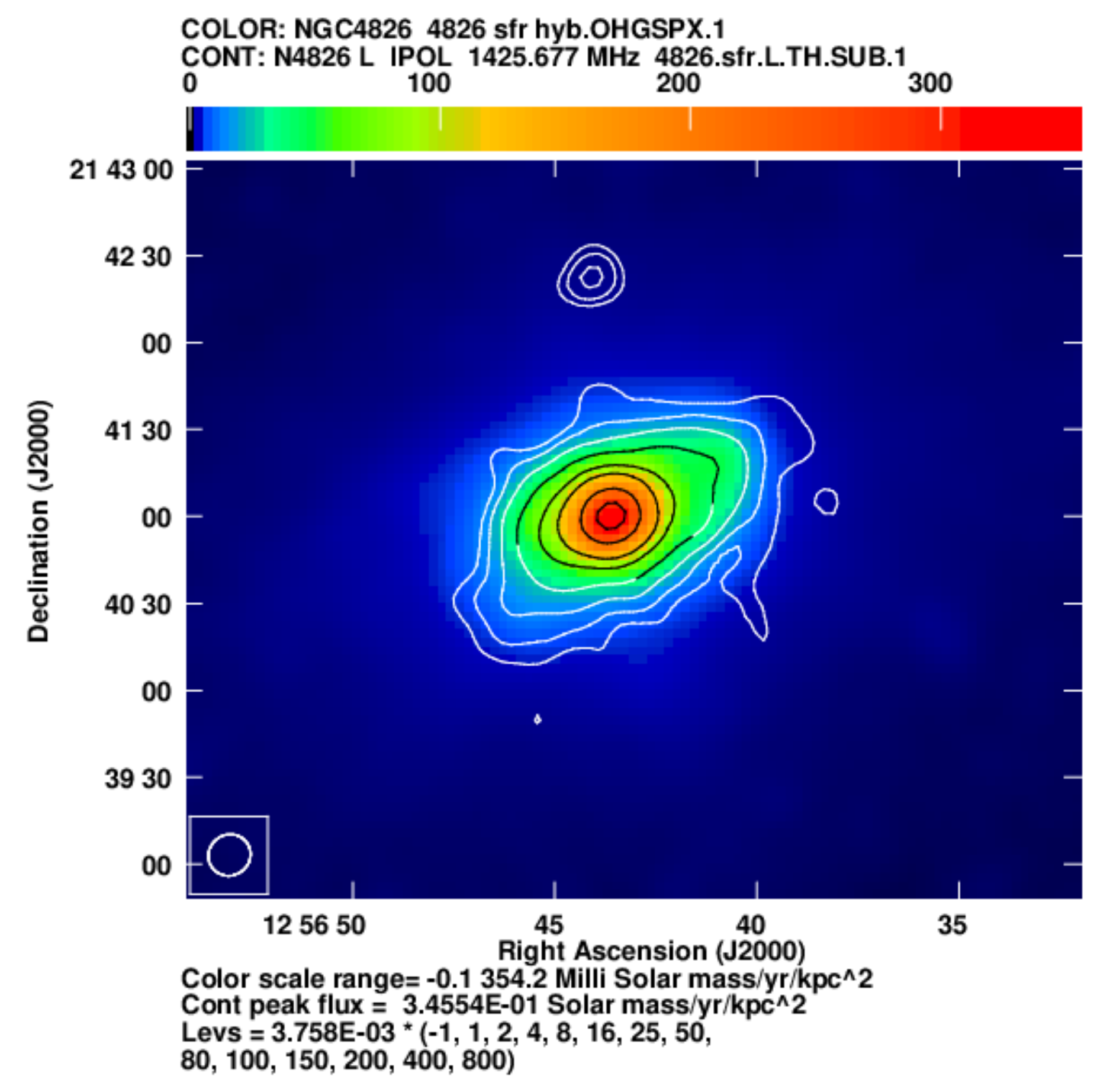}
 \includegraphics[width=0.45\linewidth]{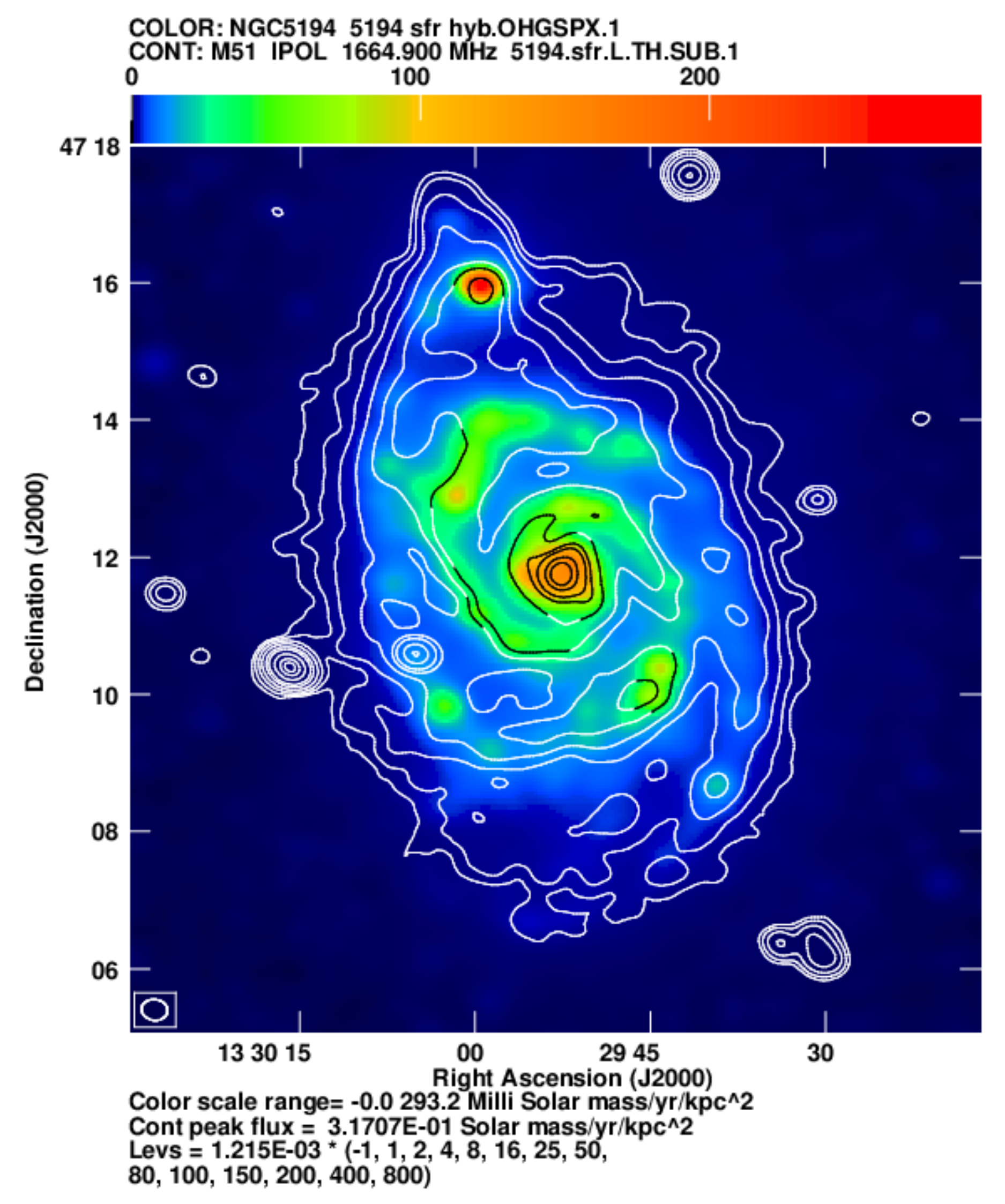}
 
  \caption{SFRSD (M$_\odot$yr$^{-1}$kpc$^{-2}$) maps of NGC 4490, NGC 4826 and NGC 5194 (clockwise from top left) (Sample 1). SFRSDs estimated using 1.4 GHz radio and FUV+24$\mu$m emission are shown in contours and colors, respectively. Contour levels are listed below each panel of the figure. The circle in the bottom-left corner of the images indicates the angular resolution of the maps.
 }
     \label{SFRSD_cont_2}
 \end{figure*}

\begin{figure*}
 	\centering
 \includegraphics[width=0.45\linewidth]{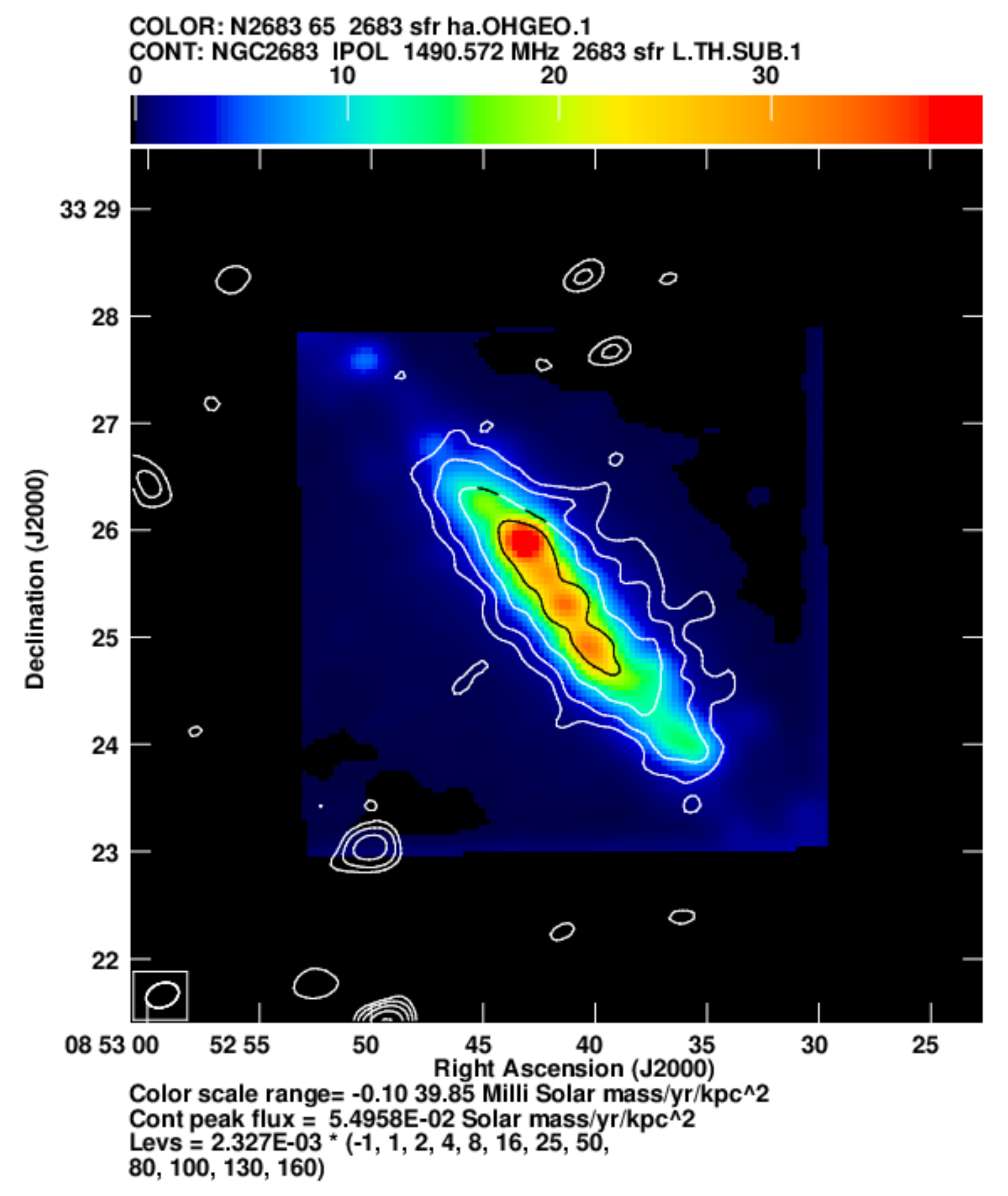}
 \includegraphics[width=0.45\linewidth]{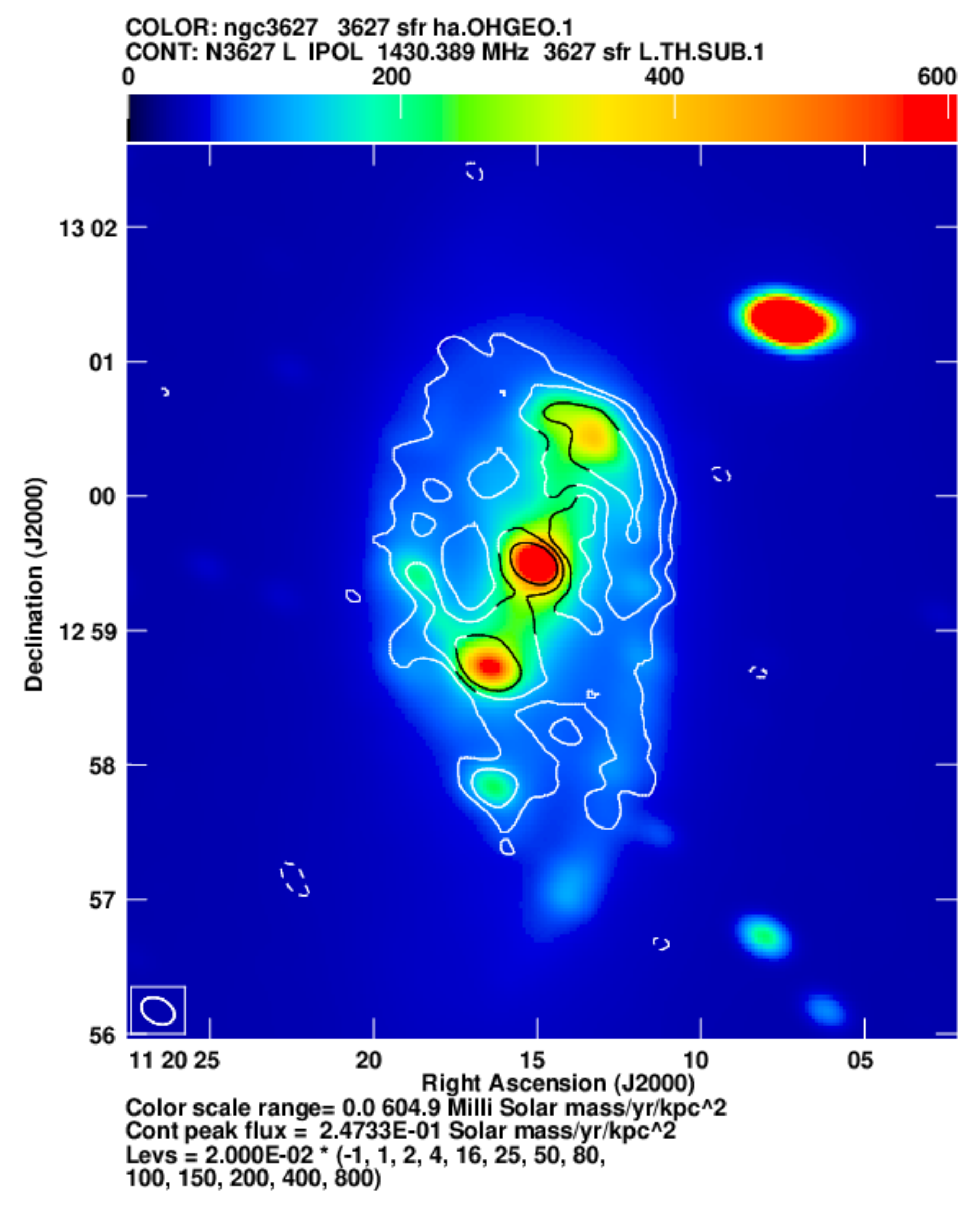}
 
 \includegraphics[width=0.45\linewidth]{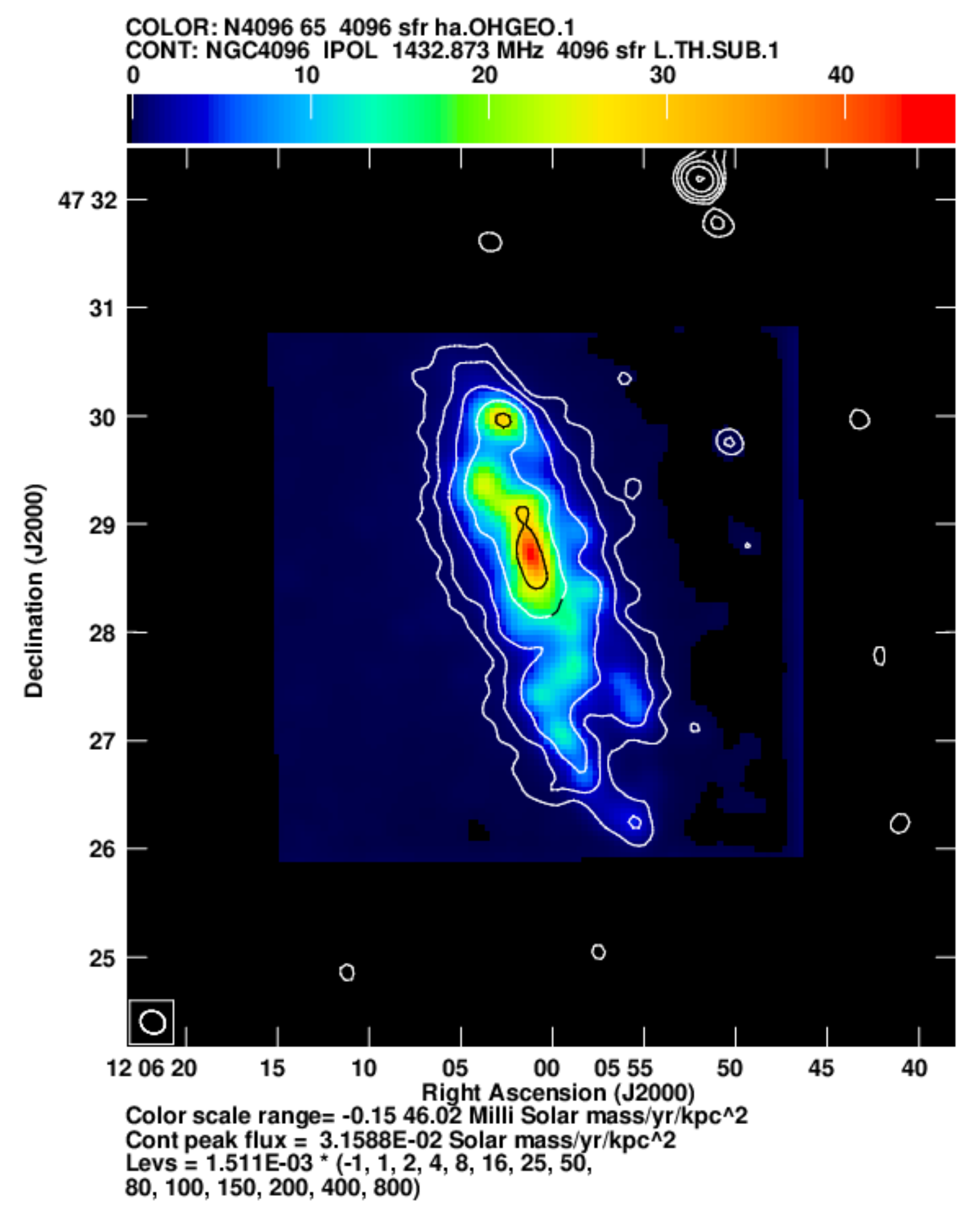}
 \includegraphics[width=0.45\linewidth]{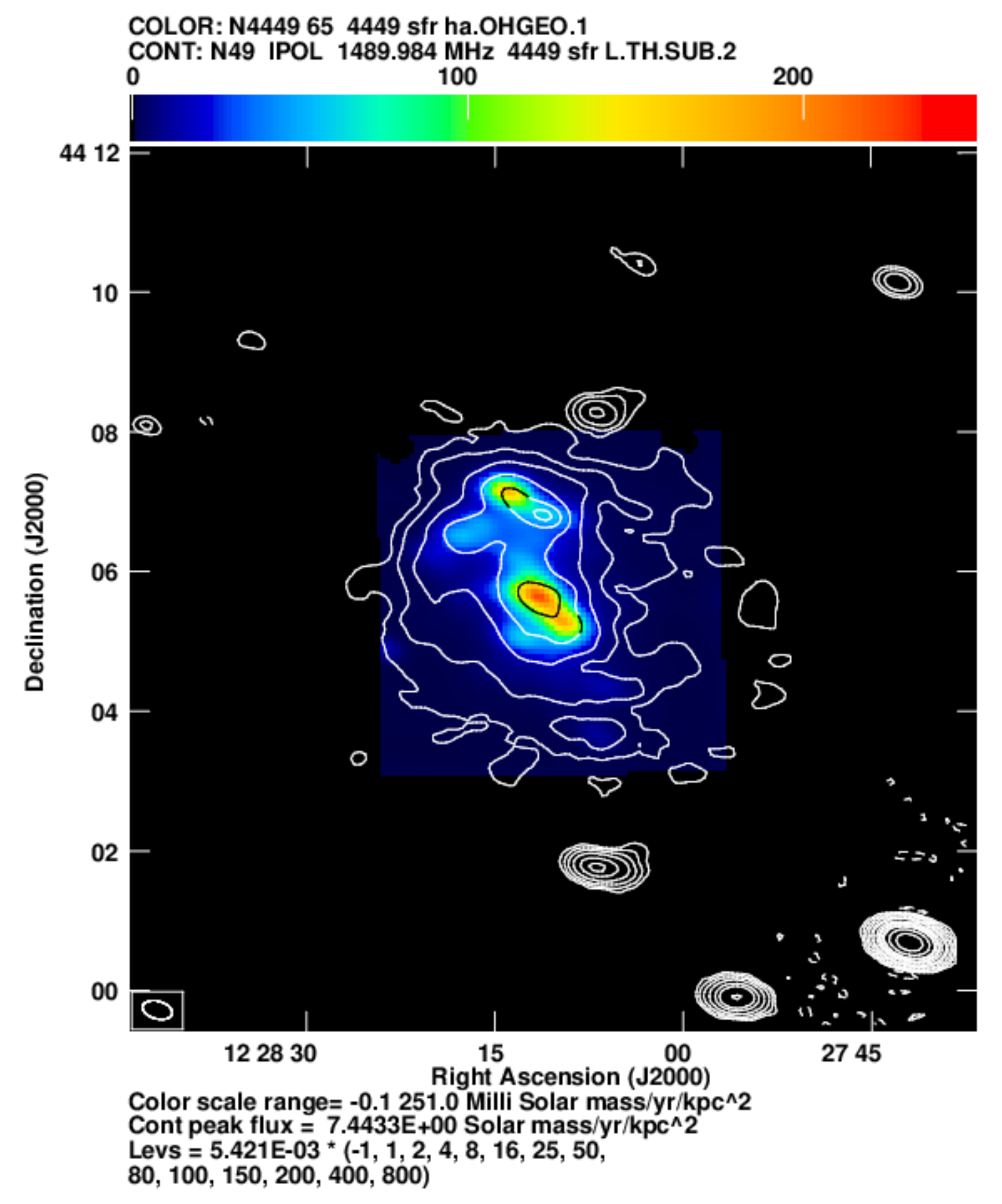}
 \caption{SFRSD (M$_\odot$yr$^{-1}$kpc$^{-2}$) maps of NGC 2683, NGC 3627, NGC 4449 and NGC 4096 (clockwise from top left) (Sample 1). SFRSDs estimated using 1.4 GHz radio and H$\alpha$+24$\mu$m emission are shown in contours and colors, respectively. Contour levels are listed below each panel of the figure. The circle in the bottom-left corner of the images indicates the angular resolution of the maps.}
  \label{SFRSD_halpha_1}
  \end{figure*}

\begin{figure*}
  	\centering
 \includegraphics[width=0.45\linewidth]{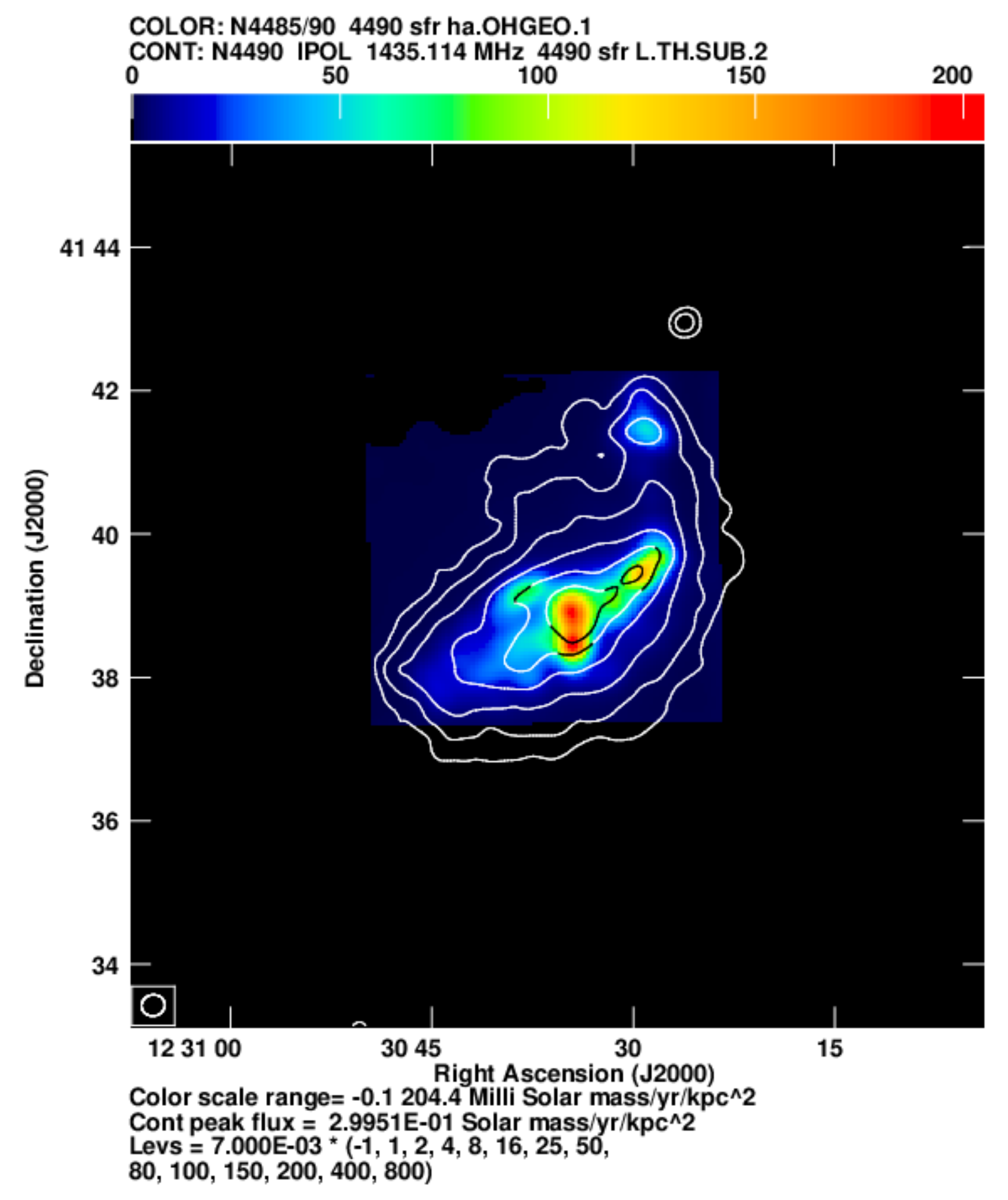}
 \includegraphics[width=0.45\linewidth]{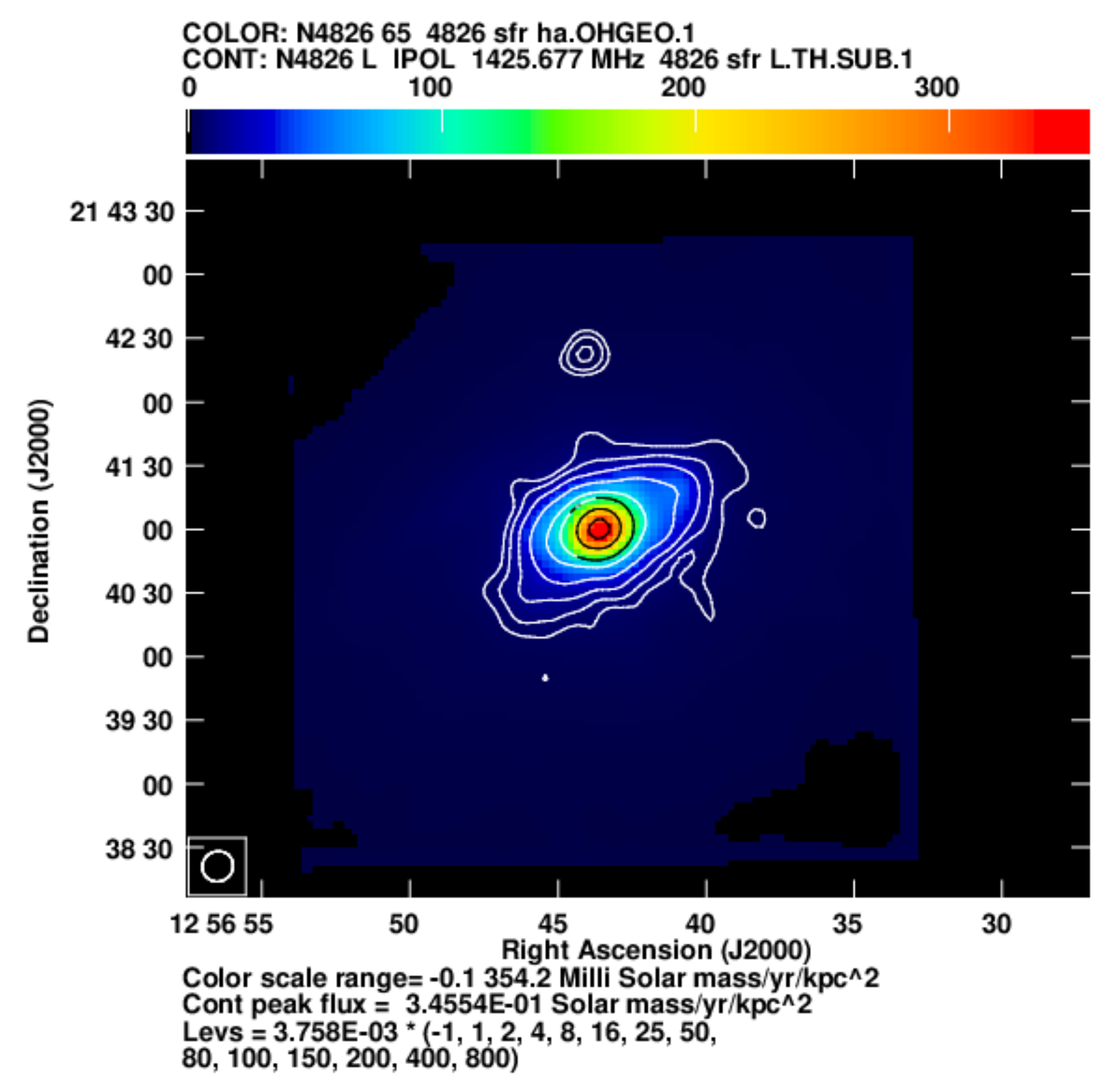}
 \includegraphics[width=0.45\linewidth]{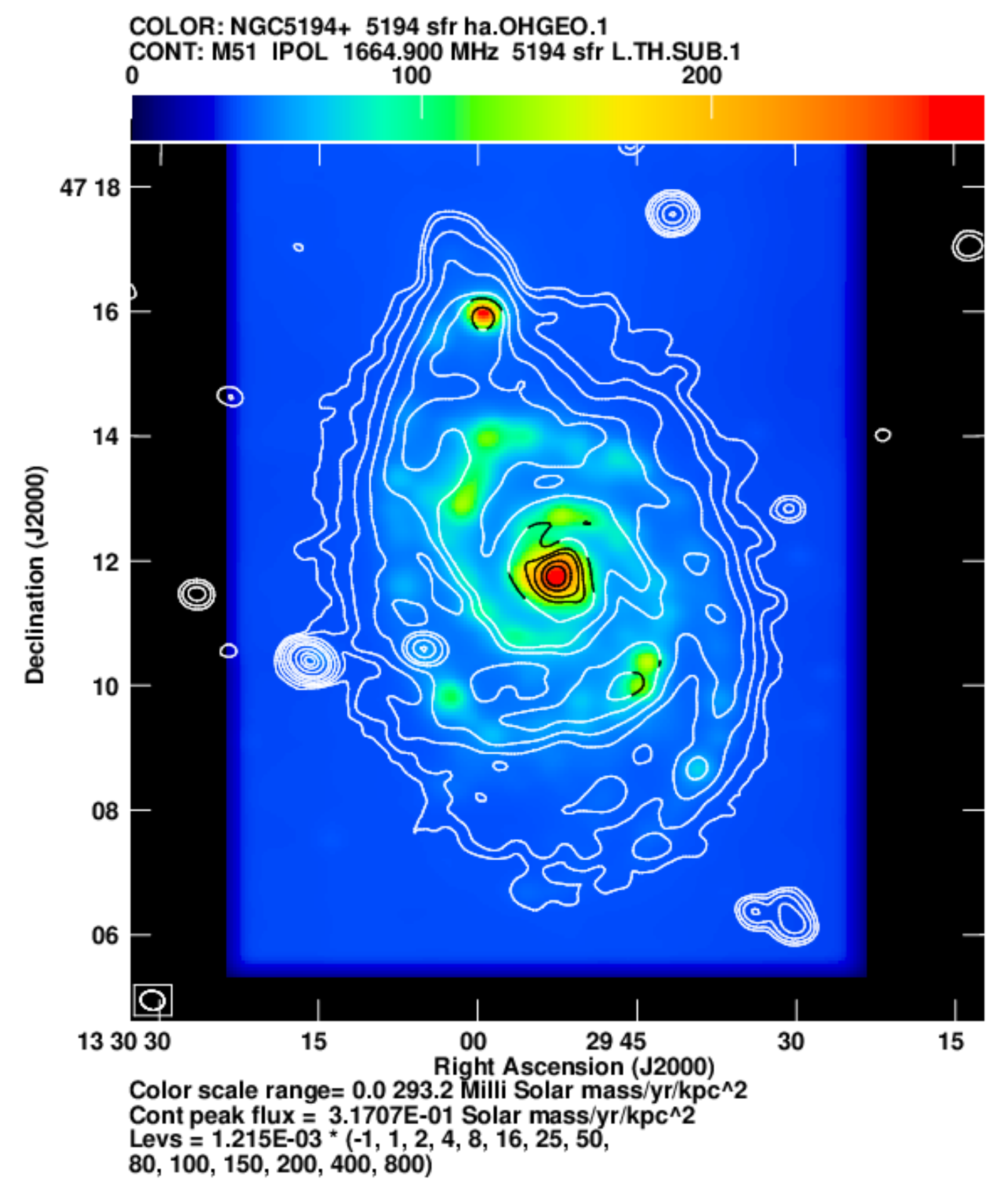}
 
  \caption{SFRSD (M$_\odot$yr$^{-1}$kpc$^{-2}$) maps of NGC 4490, NGC 4826 and NGC 5194  (clockwise from top left) (Sample 1). SFRSDs estimated using 1.4 GHz radio and H$\alpha$+24$\mu$m emission are shown in contours and colors, respectively. Contour levels are listed below each panel of the figure. The circle in the bottom-left corner of the images indicates the angular resolution of the maps.
 }
     \label{SFRSD_halpha_2}
 \end{figure*}

We show SFRSD maps of the seven galaxies (Sample 1) in Figures \ref{SFRSD_cont_1} and \ref{SFRSD_cont_2}, where SFRSDs estimated using 1.4 GHz and FUV+24$\mu$m emission are shown in contours and colors, respectively. In Figures \ref{SFRSD_halpha_1} and  \ref{SFRSD_halpha_2}, we have also shown the SFRSD maps estimated using H$\alpha$+24$\mu$m and 1.4GHz data in colors and contours, respectively. The SFRSD maps of each galaxy in Figures~\ref{SFRSD_cont_1}, \ref{SFRSD_cont_2}, \ref{SFRSD_halpha_1}, and \ref{SFRSD_halpha_2} have been shown in the same color scale and contours.


\end{document}